\documentclass[fleqn,usenatbib]{mnras}

\usepackage{newtxtext,newtxmath}
\usepackage[T1]{fontenc}

\DeclareRobustCommand{\VAN}[3]{#2}
\let\VANthebibliography\thebibliography
\def\thebibliography{\DeclareRobustCommand{\VAN}[3]{##3}\VANthebibliography}

\usepackage{graphicx}	% Including figure files
\usepackage{amsmath}	% Advanced maths commands
\usepackage{subfigure}
\usepackage{paralist}
\usepackage{bm}
\usepackage{bbding}
\usepackage{multirow}
\title[Brown Hamiltonian]{Extensions of Brown Hamiltonian--II. Analytical study on the modified von Zeipel--Lidov--Kozai effects}

\author[Lei \& Grishin]{
Hanlun Lei$^{1,2}$\thanks{leihl@nju.edu.cn},
Evgeni Grishin$^{3,4}$
\\
$^{1}$School of Astronomy and Space Science, Nanjing University, Nanjing 210023, China\\
$^{2}$Key Laboratory of Modern Astronomy and Astrophysics in Ministry of Education, Nanjing University, Nanjing 210023, China\\
$^{3}$School of Physics and Astronomy, Monash University, Clayton, VIC 3800, Australia\\
$^{4}$OzGrav: Australian Research Council Centre of Excellence for Gravitational Wave Discovery, Clayton, VIC 3800, Australia
}

\date{Accepted XXX. Received YYY; in original form ZZZ}

\pubyear{2025}

\begin{document}
\label{firstpage}
\pagerange{\pageref{firstpage}--\pageref{lastpage}}
\maketitle

% Abstract of the paper
\begin{abstract}
In triple systems of weak hierarchies, nonlinear perturbations arising from the periodic oscillations associated with the inner and outer binaries play a crucial role in shaping their long-term dynamical evolution. In this context, we have developed an extended Brown Hamiltonian in Paper I, which serves as a fundamental model for describing the modified von Zeipel–Lidov–Kozai (ZLK) oscillations. The present work aims to analyze the characteristics of ZLK oscillations within this extended framework, focusing on phase-space structures, the location of ZLK center, the maximum eccentricity reached, the boundaries of librating cycles, and the critical inclination required to trigger ZLK resonance. Under the extended Hamiltonian, we introduce the Lidov integral $C_{\rm ZLK}$, which is a combination of the Hamiltonian and the $z$-component of angular momentum, to characterize the modified ZLK properties. It is found that the librating and circulating cycles are separated by $C_{\rm ZLK}=0$, which is consistent with the classical theory. Furthermore, we derive analytical expressions of these ZLK properties using perturbation techniques. Analytical predictions are compared to numerical results, showing an excellent agreement between them. Notably, the results reveal that ZLK characteristics in prograde and retrograde regimes are no longer symmetric under the influence of Brown corrections. At last, we conduct $N$-body integrations about millions of orbits to generate dynamical maps, where the numerical structures are well captured by the analytical solutions derived from the extended model.    
\end{abstract}

% Select between one and six entries from the list of approved keywords.
% Don't make up new ones.

%%%%%%%%%%%%%%%%%%%%%%%%%%%%%%%%%%%%%%%%%%%%%%%%%%

%%%%%%%%%%%%%%%%% BODY OF PAPER %%%%%%%%%%%%%%%%%%

\section{Introduction}
\label{Sect1}

Triple and higher-order multiple systems are commonly found throughout the Universe, spanning a wide range of mass and physical scales: from high-altitude artificial satellites orbiting the Earth or Moon \citep{lidov1962evolution,wytrzyszczak2007regular,rosengren2016galileo,nie2019semi,ortore2023optimal,zhao2024analytical}, natural satellites of giant planets \citep{cuk2004secular,gaspar2013irregular,brozovic2022orbits,grishin2024irregularI,grishin2024irregularII}, and solar system asteroids \citep{kozai1962secular,margot2015quantitative}, to Kuiper Belt binaries \citep{noll2008evidence,grishin2020wide}, planets in circumbinary or multi-planet systems \citep{fabrycky2007shrinking,libert2009kozai,naoz2011hot,naoz2013secular,winn2015occurrence,luo2016double,lei2024dynamical}, and even stellar or black hole triple systems \citep{harrington1969stellar,ford2000secular,blaes2002kozai,li2015implications,kimpson2016hierarchical, mangipudi2022extreme}. For dynamical stability, such systems are typically hierarchical in structure, consisting of a close inner binary that is gravitationally perturbed by a more distant tertiary companion \citep{mardling2001tidal, tory2022empirical, vynateya2022}. In hierarchical systems, it is often appropriate to adopt a phase-averaged (or secular) approximation to explore their long-term dynamical evolution.

In a hierarchical three-body system under the secular and test-particle approximations, it is well established that, under the quadrupole-order truncation, long-term perturbations from the distant third body can induce large-amplitude, coupled oscillations in the eccentricity and inclination of the inner binary when the mutual inclination between the inner and outer orbits lies between approximately $39.2^{\circ}$ and $140.8^{\circ}$ \citep{von1910application,lidov1962evolution,kozai1962secular}. This phenomenon is now known as the von Zeipel--Lidov--Kozai (ZLK) oscillation \citep{ito2019lidov}. Recently, the ZLK cycles are shown to be
equivalent to the periodic solution in a simple pendulum \citep{basha2025kozai}. ZLK oscillations have been invoked in a wide range of astrophysical contexts to account for diverse dynamical phenomena, including the origin of high-inclination asteroids in the Solar System \citep{kozai1962secular,gomes2003origin,vinogradova2017amplitude,saillenfest2017non,bhaskar2020mildly,lei2022zeipel}, the inclination distribution of irregular satellites \citep{carruba2002inclination,kavelaars2004discovery,cuk2004secular,beauge2007proper,nesvorny2003orbital}, the production of sun-grazing asteroids \citep{farinella1994asteroids,vokrouhlicky2012sun,toliou2023resonant}, the formation of wide binaries in the Kuiper Belt \citep{parker2011characterization,grishin2020wide,gladman2021transneptunian,campbell2025non}, the formation of hot Jupiters \citep{wu2003planet,fabrycky2007shrinking,wu2007hot,naoz2011hot,petrovich2015steady,storch2014chaotic,munoz2016formation,dawson2018origins}, the origin of blue straggler stars \citep{perets2009triple,naoz2014mergers,antonini2016black, grishin2022wide}, the accelerated merger of black hole binaries \citep{blaes2002kozai,wen2003eccentricity,naoz2014mergers,vanlandingham2016role,kimpson2016hierarchical,antonini2017binary,hoang2018black}, the distribution of dark matter near supermassive black holes \citep{naoz2014formation,naoz2019dark}, and gravitational wave sources \citep{miller2002production,antonini2012secular,antonini2014black,belczynski2014formation,silsbee2017lidov,gondan2018accuracy,deme2020detecting,gupta2020gravitational,liu2021hierarchical,chandramouli2022ready}. For comprehensive reviews on the applications of ZLK oscillations to different astrophysical systems, see \citet{naoz2016eccentric}, \citet{shevchenko2016lidov}, \citet{ito2019lidov} and \citet{iye2025re}.

In the classical theory, ZLK oscillations are a type of secular phenomenon that occurs on timescales much longer than the orbital periods of both the inner and outer binaries \citep{antognini2015timescales}. However, the validity of the secular approximation relies on a clear separation between secular and orbital timescales \citep{tremaine2023hamiltonian}. Usually, the degree of hierarchy in a triple system is characterized by the single-averaging parameter $\epsilon_{\rm SA}$ \citep{luo2016double}. In particular, the condition $\epsilon_{\rm SA} \ll 1$ indicates a strongly hierarchical configuration, under which the secular approximation is well justified. In such cases, the classical double-averaged Hamiltonian formalism can be employed, wherein the Hamiltonian is averaged over the orbital periods of both the inner and outer orbits. At the lowest order in the semimajor axis ratio and in the test-particle limit, the resulting dynamical model is integrable, and the properties of ZLK oscillations have been extensively studied and are well understood \citep{kozai1962secular,lidov1962evolution,antognini2015timescales,kinoshita1999analytical,broucke2003long,kinoshita2007general,lubow2021analytic,basha2025kozai,hamers2021properties}.

However, as $\epsilon_{\rm SA}$ increases and the separation between orbital and secular timescales diminishes, the secular—and even quasi-secular—approximations may break down \citep{seto2013highly,luo2016double,liu2018black,naoz2016eccentric}. A classical example of this breakdown is Newton’s inability to account for the precession of lunar apsides \citep{brouwer1961methods,cuk2004secular,tremaine2023dynamics}.

In mildly hierarchical triple systems, additional nonlinear terms in the Hamiltonian play a crucial role in accurately describing long-term dynamical evolution. The second-order Hamiltonian, which accounts for the nonlinear effects of evection terms associated with the outer binary \citep{cuk2004secular}, was first introduced in a series of works by \citet{brown1936stellarI,brown1936stellarII,brown1936stellarIII,brown1937stellarIV}. This Hamiltonian is now commonly referred to as Brown Hamiltonian, as proposed by \citet{tremaine2023hamiltonian}. In the literature, more than three distinct formulations of Brown Hamiltonian have been developed \citep{soderhjelm1975three,cuk2004secular,breiter2015secular,luo2016double,lei2018modified,krymolowski1999studies,will2021higher,conway2024higher}. It is demonstrated by \citet{tremaine2023hamiltonian} that these different forms are related through a gauge freedom inherent in canonical transformations. Notably, the use of the third choice of fictitious time discussed by \citet{tremaine2023hamiltonian} leads to the simplest form of the Hamiltonian, which can be used to simplify the dynamical model.

Within the framework of the Brown Hamiltonian, several key dynamical features have been investigated, including the modified maximum eccentricity, the location of fixed points, and the critical inclination \citep{grishin2018quasi,mangipudi2022extreme,grishin2024irregularI}. The modified ZLK oscillations derived from this framework provide valuable insights into the Hill stability of inclined triple systems \citep{grishin2017generalized, vynateya2022, tory2022empirical}, the dynamical evolution of irregular satellites \citep{cuk2004secular,grishin2024irregularII}, the formation of wide binaries in the Kuiper Belt \citep{grishin2020wide,rozner2020wide}, and the mergers of stellar binaries accompanied by gravitational wave emission \citep{fragione2019black}.

In weak-hierarchy triple systems, the secular timescale can even become comparable to the orbital period of the inner binary. Under such conditions, the single-averaging dynamical model fails to capture the system's behavior, indicating a breakdown of the quasi-secular approximation. To address this issue, \citet{beauge2006high} and \citet{lei2019semi} developed long-term dynamical models based on elliptic expansions of the disturbing function. However, these expansion-based models present several limitations: (a) they are challenging to implement due to their algebraic complexity; (b) the Laplace convergence limit restricts their applicability to eccentricities below a critical value of $0.6627$ \citep{wintner1941analytical}; and (c) they often suffer from slow convergence, making them computationally expensive.

To overcome these limitations, we have developed a high-precision dynamical model that incorporates the nonlinear effects arising from both the inner and outer binaries, referred to as the extended Brown Hamiltonian model \citep{lei2025Extensions}. This framework expresses the Hamiltonian in an elegant and closed form with respect to the eccentricities of both the inner and outer orbits. The resulting model is particularly well suited for describing von Zeipel–Lidov–Kozai oscillations in weak-hierarchy three-body systems.

The aim of this study is to systematically investigate the dynamical properties of ZLK oscillations within the framework of the extended Brown Hamiltonian. Specifically, we analyze the associated phase-space structures, conduct a brief survey of the parameter space, identify the ZLK center, determine the maximum eccentricity attained, delineate the boundaries between libration and circulation regimes, and derive the critical inclination required to trigger ZLK resonance. Explicit expressions are formulated about these ZLK characteristics based on perturbation method \citep{Lei2024perturbation}. Notably, our analytical derivation can cover those classical results when the extended framework is reduced to the classical Brown Hamiltonian model or the classical ZLK model. Additionally, we conduct $N$-body simulations about millions of orbits to produce dynamical maps, where the numerical structures can be well captured by the analytical solutions derived from the extended model.  

The structure of this paper is organized as follows. In Section \ref{Sect2}, we provide a brief overview of the extended Brown Hamiltonian model, discuss the associated phase-space structures, and present a preliminary survey of the parameter space. In Section \ref{Sect3}, we develop two types of perturbative solutions to characterize the modified ZLK dynamics, including the location of the ZLK resonance center, the maximum eccentricity reached, and the critical inclination for the onset of resonance. Section \ref{Sect4} presents numerical investigations across various regions of parameter space to validate and complement the analytical results. Finally, the main conclusions of this work are summarized in Section \ref{Sect5}.

\begin{figure*}
\centering
\includegraphics[width=0.66\columnwidth]{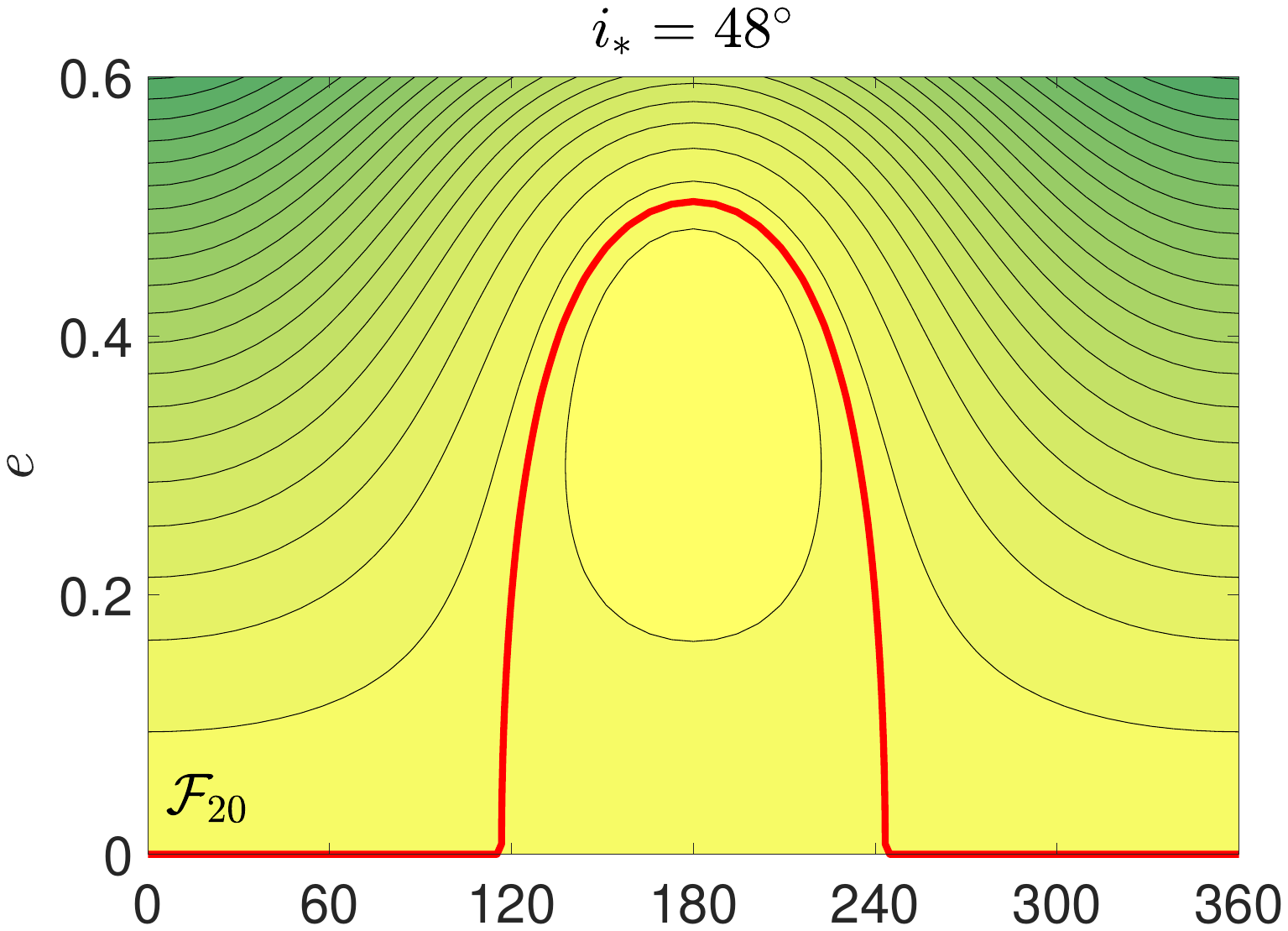}
\includegraphics[width=0.66\columnwidth]{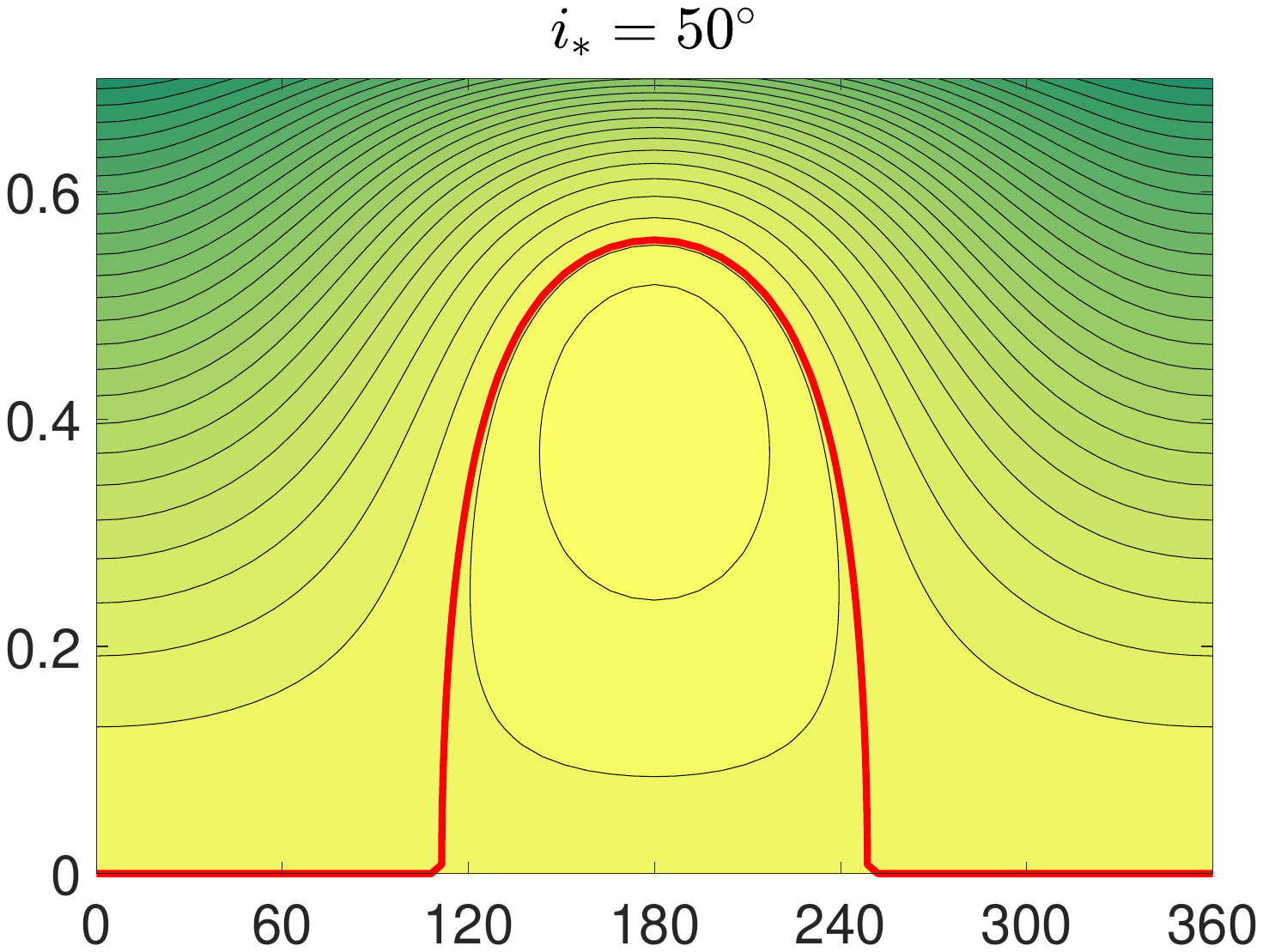}
\includegraphics[width=0.66\columnwidth]{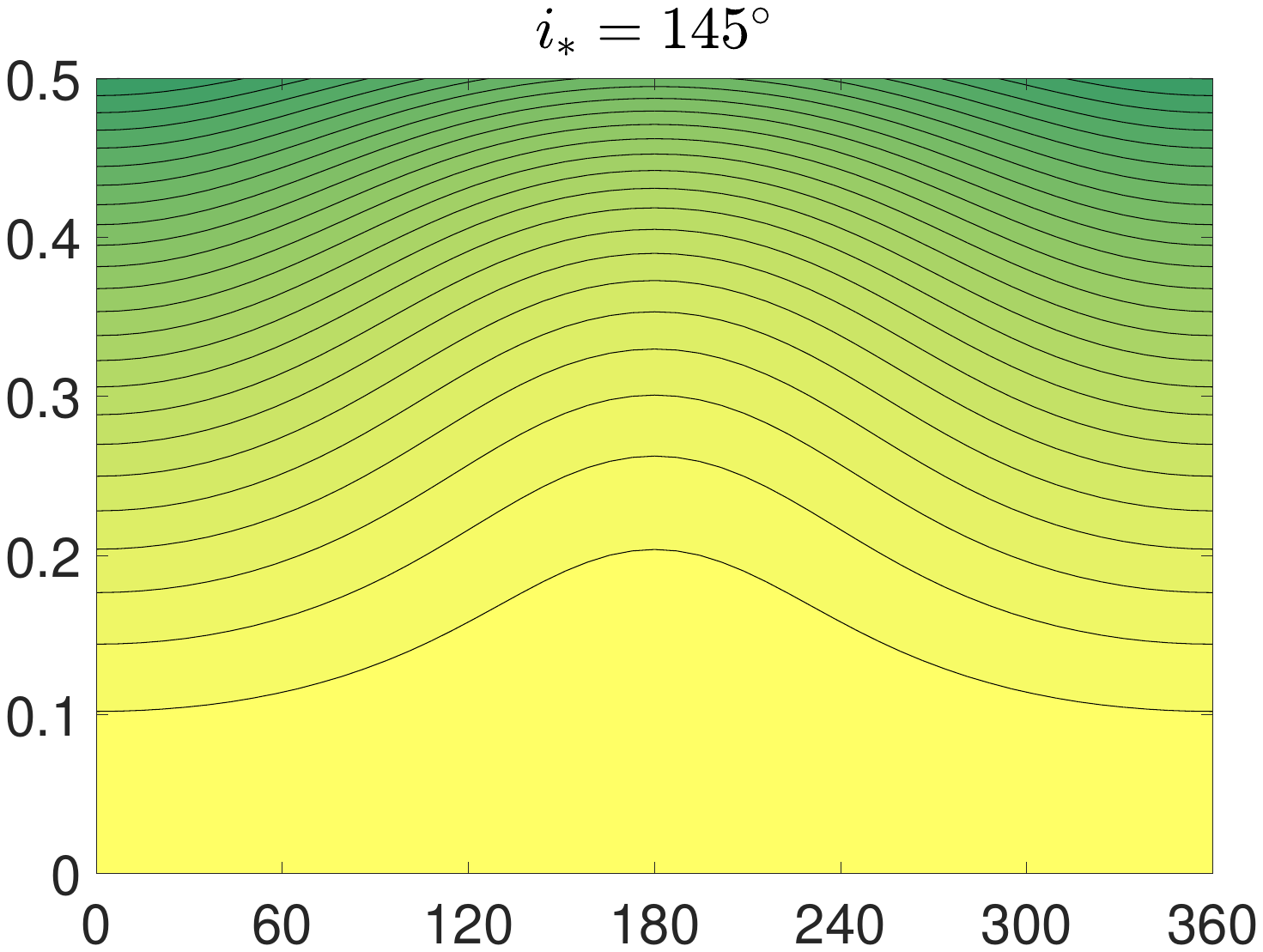}\\
\includegraphics[width=0.66\columnwidth]{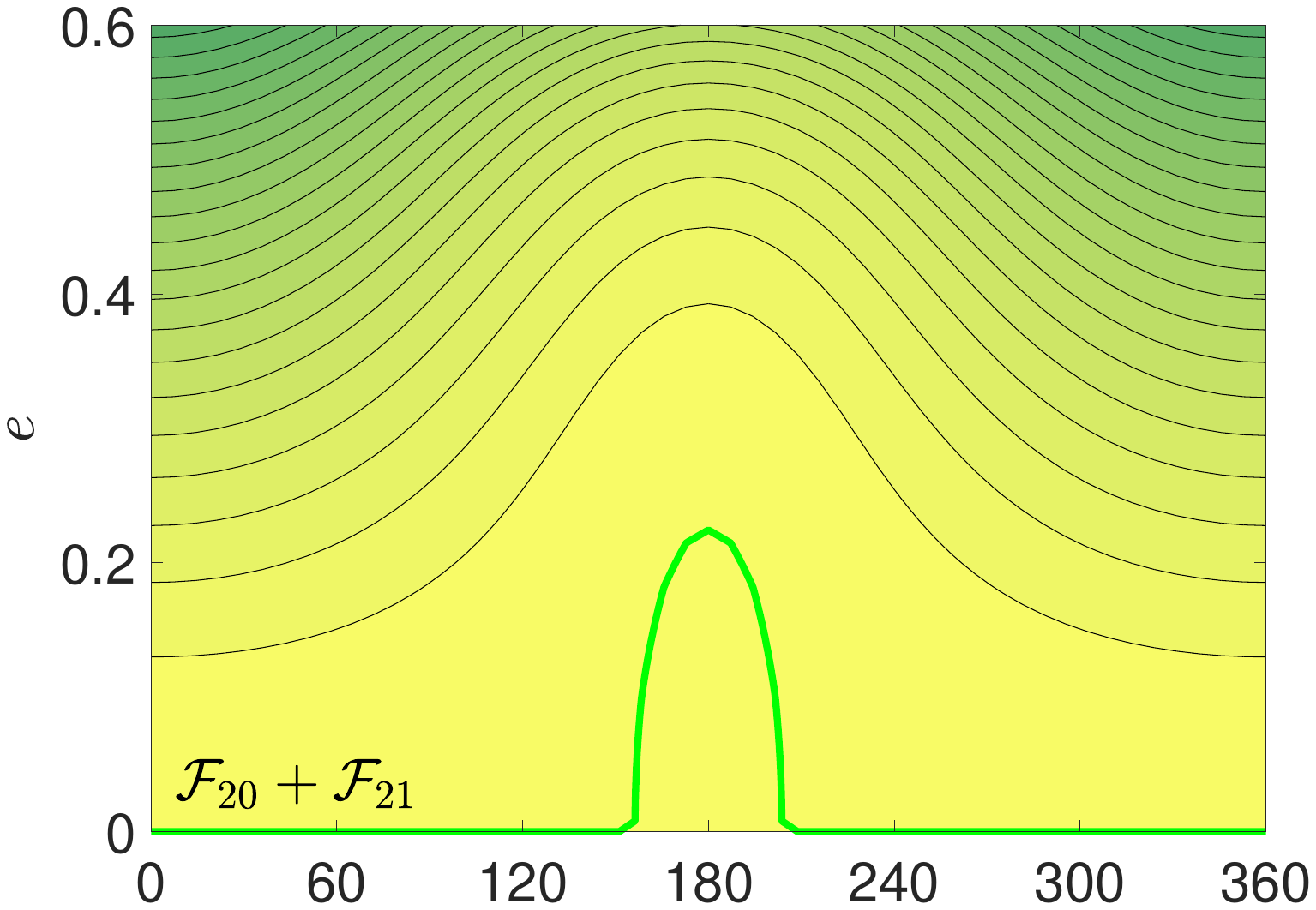}
\includegraphics[width=0.66\columnwidth]{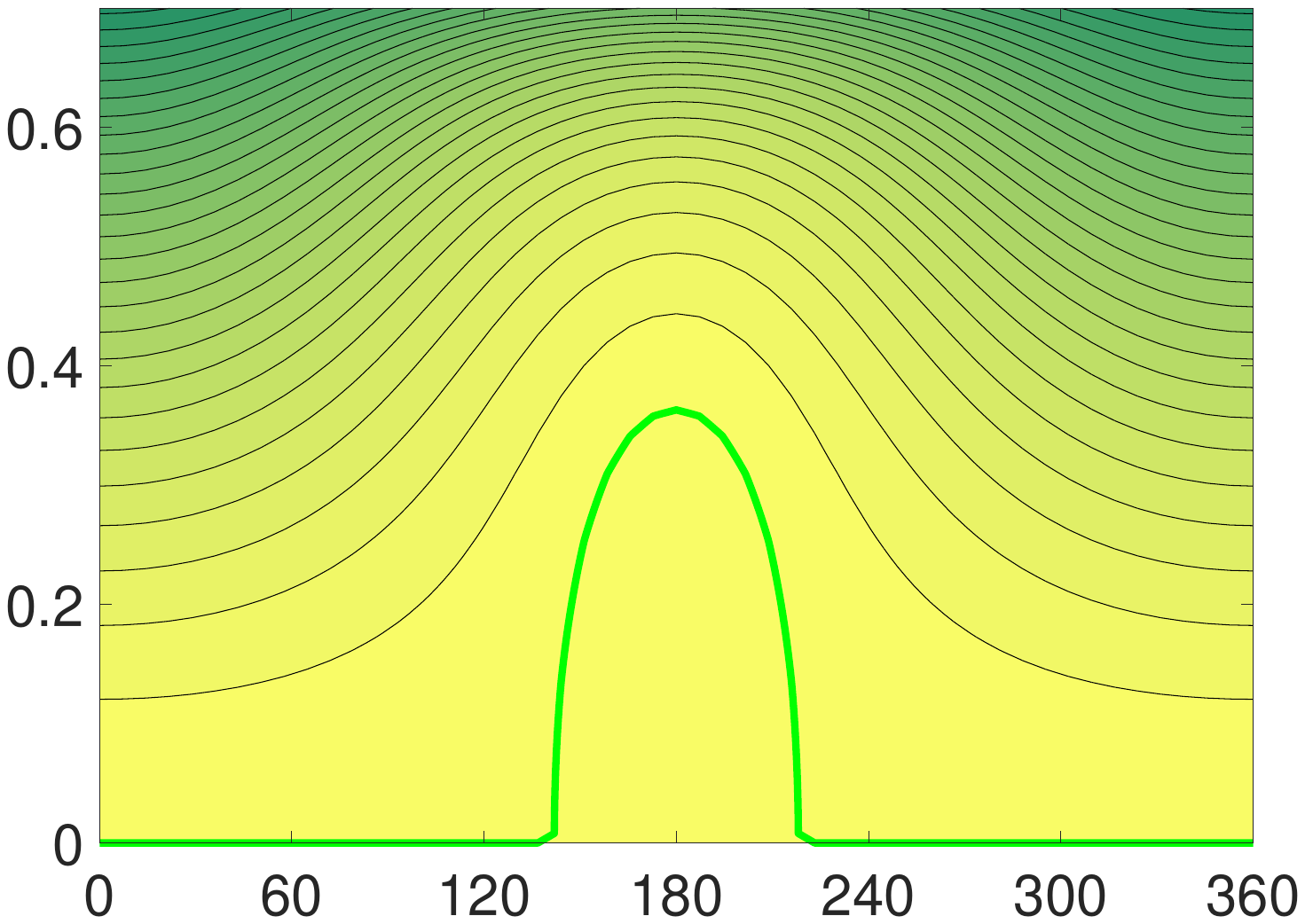}
\includegraphics[width=0.66\columnwidth]{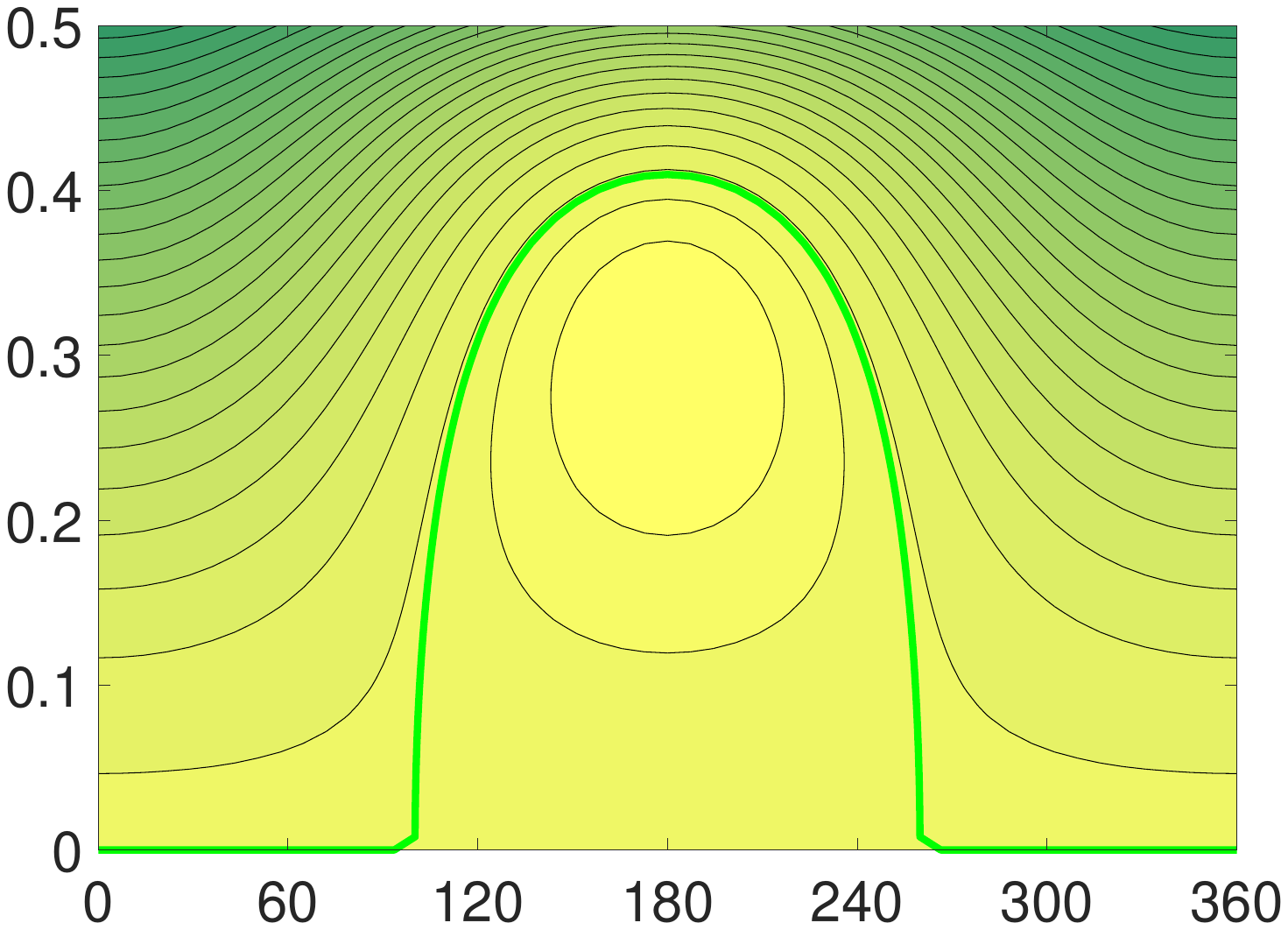}\\
\includegraphics[width=0.66\columnwidth]{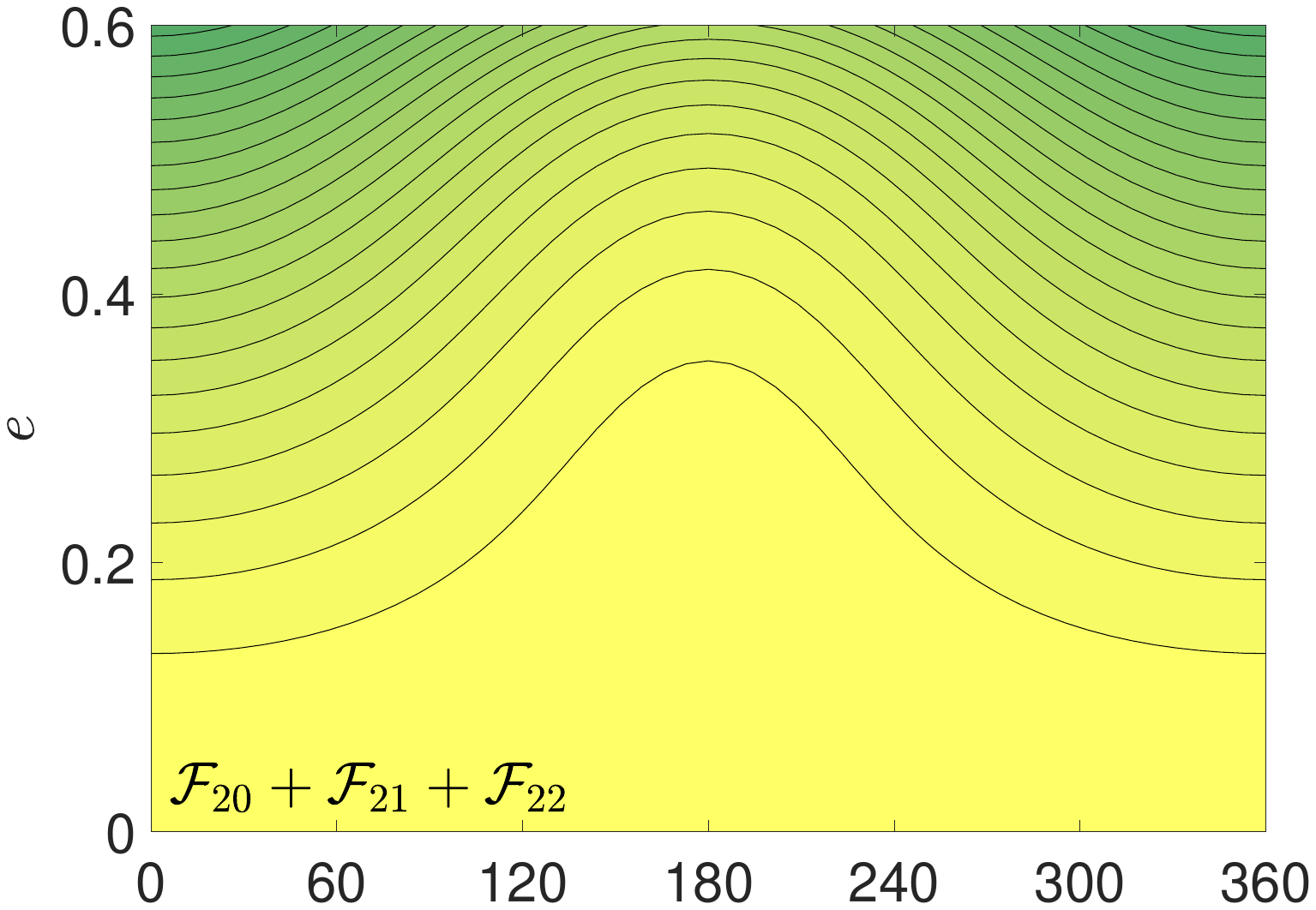}
\includegraphics[width=0.66\columnwidth]{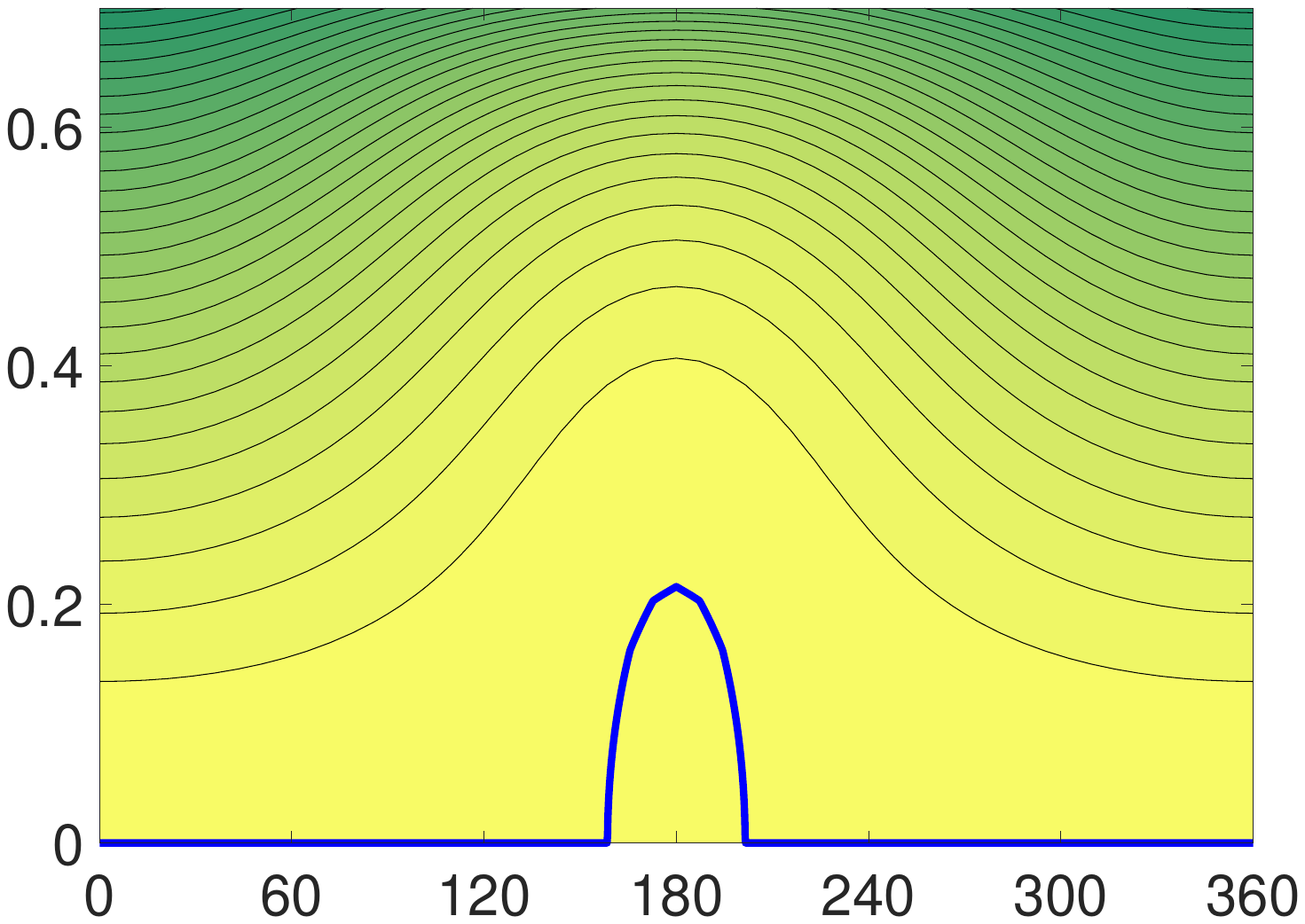}
\includegraphics[width=0.66\columnwidth]{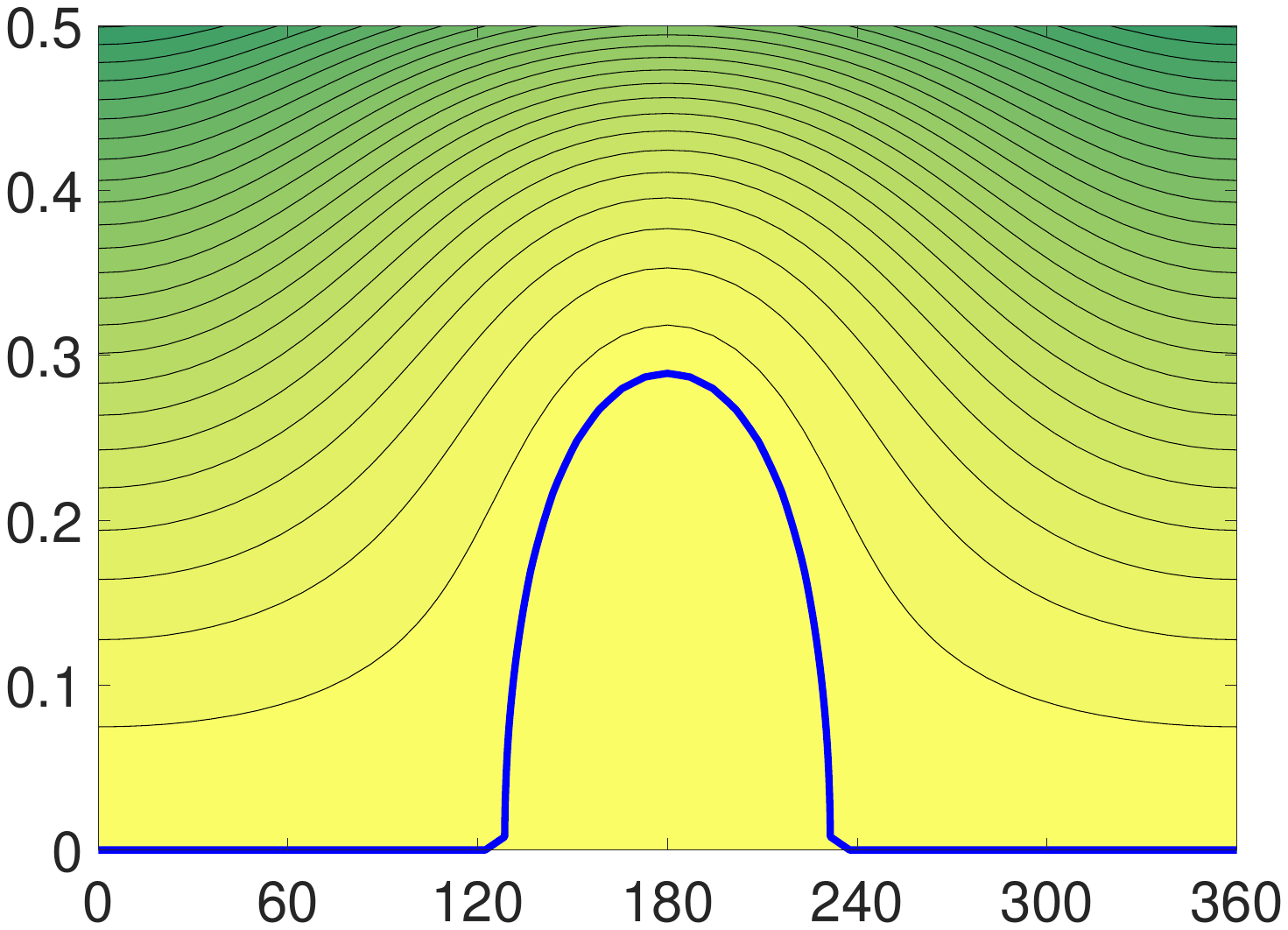}\\
\includegraphics[width=0.66\columnwidth]{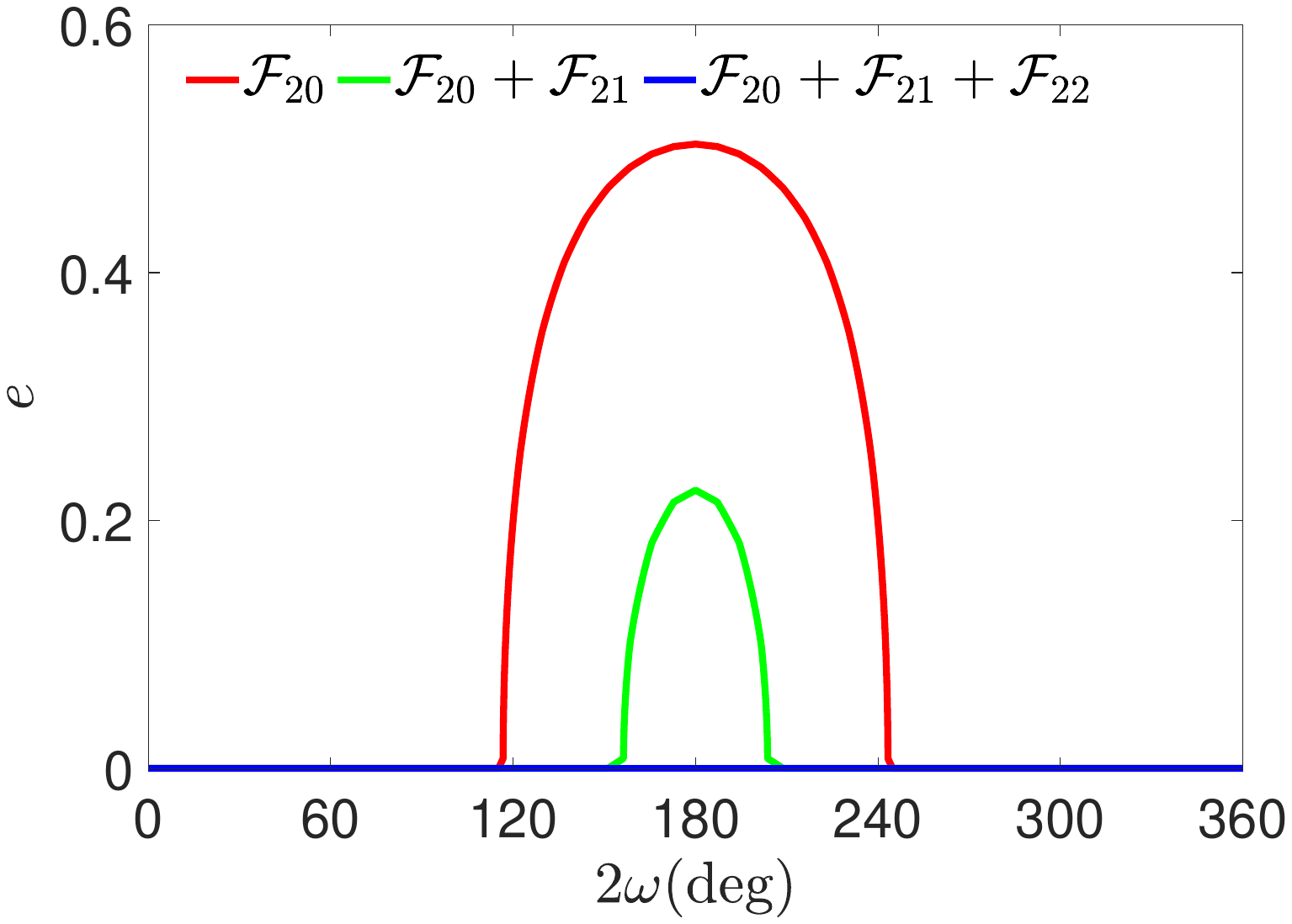}
\includegraphics[width=0.66\columnwidth]{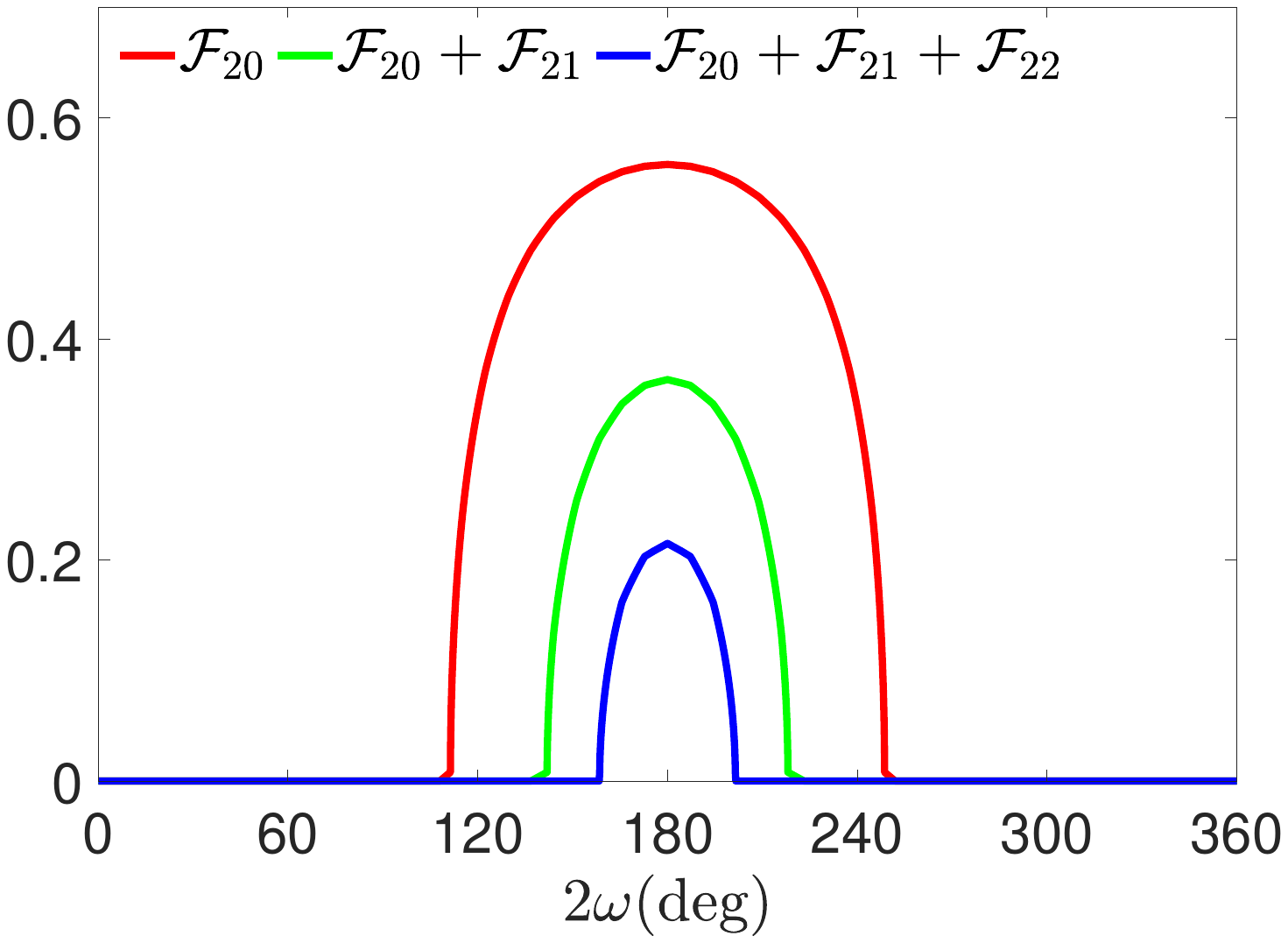}
\includegraphics[width=0.66\columnwidth]{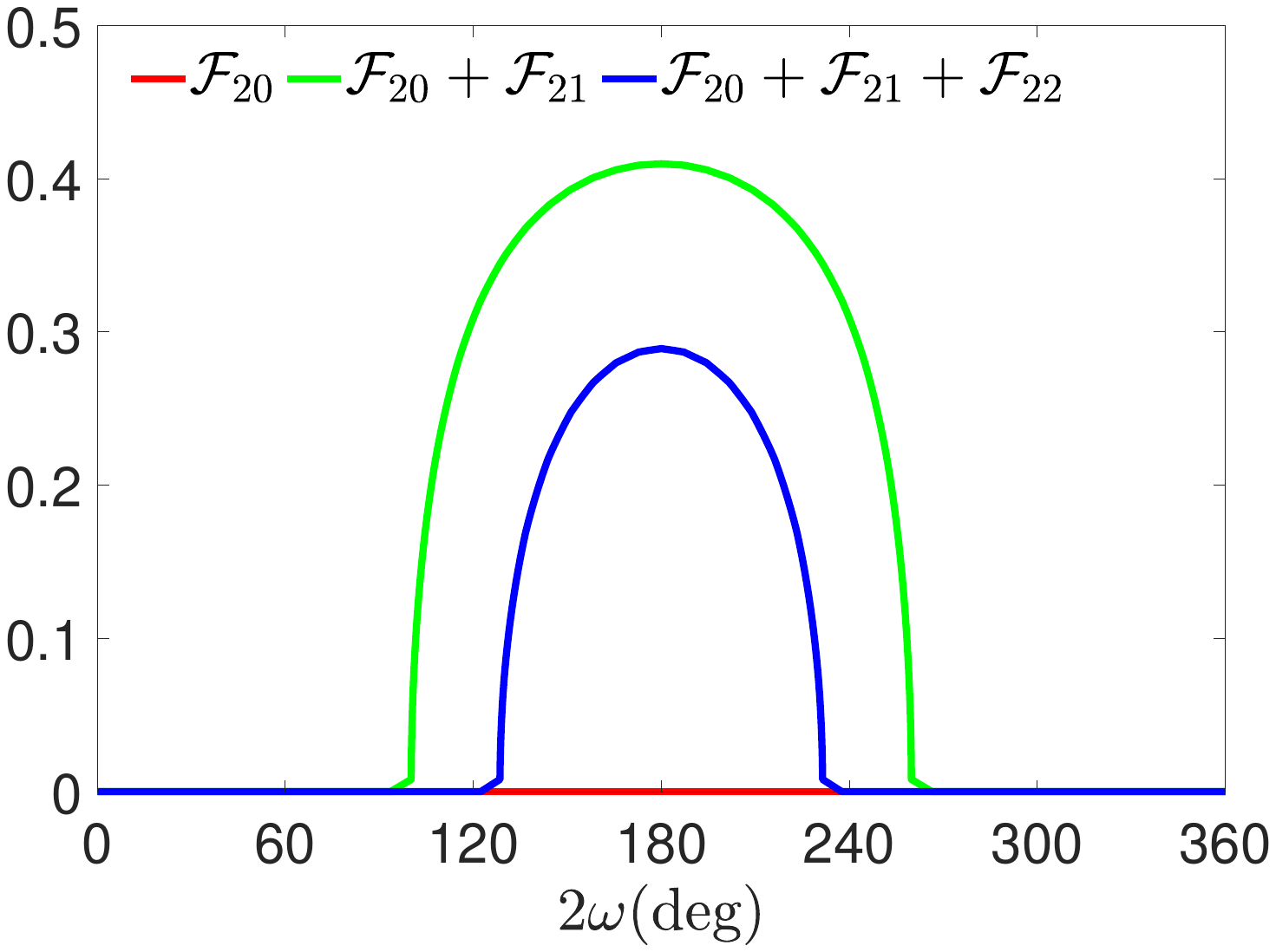}
\caption{Phase portraits (level curves of Hamiltonian) and the corresponding separatrix of ZLK resonance under the classical ZLK model (1st row, red lines), the classical Brown Hamiltonian model (2nd row, green lines) and the extended Brown Hamiltonian model (3rd row, blue lines) for the Kozai inclination at $i_*=48^{\circ}$ (\textit{left-column panels}), $i_*=50^{\circ}$ (\textit{middle-column panels}) and $i_*=145^{\circ}$ (\textit{right-column panels}). In the bottom-row panels, comparisons of ZLK separatrices are made for Hamiltonian models with different levels of correction. The orbit elements of the perturber is taken as $a_{\rm p} = 5.2\, {\rm au}$ and $e_{\rm p} = 0.049$, corresponding to the Jupiter--satellite--Sun system. The semimajor axis of satellite is fixed at $a = 0.15\, {\rm au}$, corresponding to $\alpha_{\rm H} = 0.423$ (the corresponding model parameters are $\varepsilon_{21} = 0.159$ and $\varepsilon_{22} = 0.026$).
}
\label{Fig1}
\end{figure*}

\section{The extended Brown Hamiltonian model}
\label{Sect2}

In the first paper of this series \citep{lei2025Extensions}, we have developed an extended Brown Hamiltonian model by taking advantage of the method of von Zeipel transformation \citep{brouwer1959solution}, providing a fundamental model for describing the ZLK oscillations in triple systems of weak hierarchies. In this section, we briefly introduce the extended Brown Hamiltonian, discuss its phase-space structures by analyzing phase portraits, and present a brief survey in parameter space. 

For convenience, we introduce an inertial coordinate system with the $z$-axis aligned with the angular momentum vector to describe the motion of interested objects by means of classical orbital elements, including the semimajor axis $a$, the eccentricity $e$, the inclination $i$, longitude of ascending node $\Omega$, argument of pericenter $\omega$ and mean anomaly $M$. Without otherwise specified, the variables with subscript `${\rm p}$' are for the perturber and the ones without subscripts are for inner test particles. The mass of the central object and the perturber are denoted by $m_0$ and $m_{\rm p}$, respectively.

Without otherwise stated, in the entire work we take Jupiter--satellite--Sun system to perform practical simulations.

\subsection{The extended framework of Brown Hamiltonian}
\label{Sect2-1}

Considering the nonlinear effects of the quadrupole-order potential arising from both the inner and outer bodies, we formulated the extended Brown Hamiltonian in a closed form with respect to the eccentricities of the inner and outer orbits as follows \citep{lei2025Extensions}:
\begin{equation}\label{Eq1}
{\cal F} = {{\cal F}_{20}} + {\varepsilon _{{\rm{21}}}}{{\cal F}_{21}} + {\varepsilon _{{\rm{22}}}}{{\cal F}_{22}},
\end{equation}
where
\begin{equation}\label{Eq2-1}
{{\cal F}_{20}} = \frac{1}{6}\left( {2 + 3{e^2}} \right)\left( {3{{c}^2} - 1} \right) + \frac{5}{2}{e^2}{s^2} \cos 2\omega,
\end{equation}
\begin{equation}\label{Eq2-2}
{{\cal F}_{21}} = \frac{{3}}{{16}}\eta c \left[ {2{{s}^2} + {e^2}\left( {33 + 17{{c}^2} + 15{{s}^2} \cos 2\omega } \right)} \right],
\end{equation}
and
\begin{equation}\label{Eq2-3}
\begin{aligned}
{{\cal F}_{22}} = & \frac{3}{{512}}\left\{ {227{e^2}\left( {8 + {e^2}} \right) - {{c}^4} \left( {56 - 472{e^2} + 701{e^4}} \right)} \right.\\
&- 2{c^2}\left( {\frac{{376}}{3} - 360{e^2} - 305{e^4}} \right) - 95{e^4}{s^4}\cos 4\omega \\
&+ \left. {  4{e^2}{{s}^2}\left[ {186 + \frac{{109}}{3}{e^2} + 5\left( {18 - 37{e^2}} \right){{c}^2}} \right]\cos 2\omega } \right\}.
\end{aligned}
\end{equation}
with $\eta = \sqrt{1-e^2}$, $c=\cos{i}$ and $s=\sin{i}$. In terms of the eccentricity vector $\bm e$ and normalized angular momentum vector $\bm j$, the extended Brown Hamiltonian model can be alternatively written in a much more elegant form \citep{lei2025Extensions}:
\begin{equation}\label{Eq8-3}
\begin{aligned}
{{\cal F}_{20}} =&  2{e^2}- 5{e_z^2}  + {j_z^2} - \frac{1}{3},\\
{{\cal F}_{21}} =& \frac{3}{8}{j_z}\left( {1 - j_z^2 + 24{e^2} - 15e_z^2} \right),\\
{{\cal F}_{22}} =& \frac{1}{{64}}\left\{{ 8e^2\left(13{e^2} + {22e_z^2 + 4j_z^2 + 120}\right) - 94j_z^2} \right.\\
&\left. { - 3\left[ {95e_z^4 + 6e_z^2\left( {15j_z^2 + 31} \right) + 7j_z^4} \right]} \right\},
\end{aligned}
\end{equation}
where
\begin{equation}\label{8-1}
{\bm j} = \left( {\begin{array}{*{20}{c}}
{{j_x}}\\
{{j_y}}\\
{{j_z}}
\end{array}} \right) = \sqrt {1 - {e^2}} \left( {\begin{array}{*{20}{c}}
{\sin i\sin \Omega }\\
{ - \sin i\cos \Omega }\\
{\cos i}
\end{array}} \right),
\end{equation}
and
\begin{equation}\label{8-2}
{\bm e} = \left( {\begin{array}{*{20}{c}}
{{e_x}}\\
{{e_y}}\\
{{e_z}}
\end{array}} \right) = e\left( {\begin{array}{*{20}{c}}
{\cos \Omega \cos \omega  - \cos i\sin \Omega \sin \omega }\\
{\sin \Omega \cos \omega  + \cos i\cos \Omega \sin \omega }\\
{\sin i\sin \omega }
\end{array}} \right).
\end{equation}
In particular, ${\cal F}_{20}$ corresponds to the classical ZLK Hamiltonian \citep{kozai1962secular}, ${\cal F}_{21}$ represents the classical Brown Hamiltonian correction, standing for the nonlinear effects arising from the short-period terms associated with the outer orbits \citep{tremaine2023hamiltonian}, and ${\cal F}_{22}$ is the extended Brown Hamiltonian, representing the nonlinear effects arising from the short-period terms associated with the inner orbits \citep{lei2025Extensions}. From the explicit expressions, we can see that ${\cal F}_{20}$ and ${\cal F}_{22}$ are even functions of $j_z$, but ${\cal F}_{21}$ is an odd function of $j_z$. This shows that the asymmetry of ZLK dynamical structures is mainly caused by the introduction of the classical Brown correction (${\cal F}_{21}$), instead of the extended Brown correction (${\cal F}_{22}$).

The contributions of the classical and extended Brown Hamiltonian corrections are, respectively, controlled by the associated coefficients $\varepsilon_{21}$ and $\varepsilon_{22}$, given by \citep{lei2025Extensions}
\begin{equation}\label{Eq5}
\begin{aligned}
{\varepsilon _{21}} &= \left( {\frac{{{n_{\rm p}}}}{n}} \right)\left( {\frac{{{m_{\rm p}}}}{{{m_0} + {m_{\rm p}}}}} \right)\frac{1}{{{{\left( {1 - e_{\rm p}^2} \right)}^{3/2}}}}\left( {1 + \frac{2}{3}e_{\rm p}^2} \right),\\
{\varepsilon _{22}} &= {\left( {\frac{{{n_{\rm p}}}}{n}} \right)^2}\left( {\frac{{{m_{\rm p}}}}{{{m_0} + {m_{\rm p}}}}} \right)\frac{1}{{{{\left( {1 - e_{\rm p}^2} \right)}^3}}}\left( {1 + 3e_{\rm p}^2 + \frac{3}{8}e_{\rm p}^4} \right),
\end{aligned}
\end{equation}
where $n$ and $n_{\rm p}$ are the mean motions of the test particle and the perturber, respectively. It is noted that the coefficient $\varepsilon_{21}$ is related to the single-averaging parameter $\epsilon_{\rm SA}$ introduced in \citet{luo2016double} by
\begin{equation*}
\varepsilon_{21} = \epsilon_{\rm SA} \left( {1 + \frac{2}{3}e_{\rm p}^2} \right).
\end{equation*}
In the limit of $m_0 \ll m_{\rm p}$ and $e_{\rm p} \ll 1$, the coefficients $\varepsilon_{21}$ and $\varepsilon_{22}$ can reduce to
\begin{equation}\label{Eq6}
\begin{aligned}
{\varepsilon _{21}} &= \epsilon_{\rm SA} \simeq \left(\frac{n_{\rm p}}{n}\right)  \simeq \frac{8}{15\pi} \frac{P_{\rm out}}{t_{\rm ZLK}},\\
{\varepsilon _{22}} &= \epsilon_{\rm SA}^2 \simeq \left(\frac{n_{\rm p}}{n}\right)^2 \simeq \frac{8}{15\pi} \frac{P_{\rm in}}{t_{\rm ZLK}},
\end{aligned}
\end{equation}
where $P_{\rm in}$ and $P_{\rm out}$ are the inner and outer binary periods, $t_{\rm ZLK}$ is the period (or timescale) of ZLK oscillations, given by \citep{antognini2015timescales}
\begin{equation}\label{Eq7}
t_{\rm ZLK}  \simeq  \frac{16}{15} \left(\frac{1}{n}\right) \left(\frac{m_0}{m_{\rm p}}\right)\left(\frac{a_{\rm p}}{a}\right)^3 \left(1-e_{\rm p}^2\right)^{3/2}.
\end{equation}
It should be noted that the approximate timescale of ZLK oscillation is derived from the classical ZLK model (only ${\cal F}_{20}$)\footnote{The exact ZLK timescale under the ${\cal F}_{20}$ model can be evaluated using elliptic integrals \citep{vashkov1999evolution,kinoshita2007general,antognini2015timescales,sidorenko2018eccentric,basha2025kozai}.}. Accordingly, the timescale of ZLK oscillation is modified with inclusion of Brown corrections. This topic is interesting and it can be discussed in a similar manner to that of \citet{antognini2015timescales}.

The single-averaging parameter $\epsilon_{\rm SA}$ is usually used to measure the hierarchy of triple systems \citep{luo2016double,liu2018black}. In particular, when the single-averaging parameter ${\epsilon _{\rm SA}}$ is much smaller than unity (strong-hierarchy configurations), the Brown Hamiltonian corrections can be ignored, meaning that the classical double-averaged model can work well. However, when ${\epsilon _{\rm SA}}$ increases, the orbital and secular timescales are not well separated, leading to the fact that the Brown Hamiltonian corrections have significant influences on the ZLK oscillations \citep{cuk2004secular,beauge2006high,luo2016double,lei2018modified,grishin2018quasi,lei2019semi,lei2025Extensions}. Thus, for those weak- and mild-hierarchy triple systems, Brown Hamiltonian corrections need to be considered.

\subsection{Phase-space structures (phase portraits)}
\label{Sect2-2}

We can see that the mean anomalies of the test particle and the perturber $(M,M_{\rm p})$ have been eliminated from the formulated Hamiltonian due to averaging approximation. Thus, the semimajor axis of the test particle remains constant in the long-term evolution, showing that the action $L=\sqrt{\mu a}$ is a motion integral. For convenience, we adopt the following set of (normalized) Delaunay variables to formulate the long-term Hamiltonian model,
\begin{equation}\label{Eq7-1}
\begin{aligned}
g &= \omega,\quad h = \Omega,\\
G &= \sqrt {1 - {e^2}},\quad H = G\cos i,
\end{aligned}
\end{equation}  
where $H=j_z$ is the $z$ component of orbital angular momentum. It is further observed that, under the extended Brown Hamiltonian model, the angle $h (=\Omega)$ is a cyclic coordinate, meaning that its conjugate momenta $H=\sqrt{1-e^2}\cos{i}$ is a motion integral. During the long-term evolution, the eccentricity exchanges with inclination in order to conserve $H$, which is consistent to the classical ZLK theory \citep{kozai1962secular}. For convenience, we adopt the Kozai inclination $i_*$ to characterize the motion integral $H$ by \citep{kozai1962secular}
\begin{equation}\label{Eq8}
H = \sqrt{1-e^2}\cos{i} = \cos{i_*},
\end{equation}
which indicates that ${i_*}$ corresponds to the inclination evaluated at zero eccentricity. As a result, there is a one-to-one correspondence between ${i_*}$ and $H$. Since there is only one angular coordinate $g (=\omega)$, the extended Brown Hamiltonian is of one degree of freedom. Conservation of the Hamiltonian makes it be an integral model. Similar to the classical ZLK oscillations discussed in \citet{kinoshita1999analytical, kinoshita2007general} and \citet{lubow2021analytic}, the trajectories under our novel extended model can be expressed analytically as Jacobian elliptic functions (but in a much more complicated form).

When the motion integrals $a$ and $i_*$ are given (i.e., the motion integrals $L$ and $H$ are provided), the solution of the extended Brown Hamiltonian model in the phase space $(g,G)$ corresponds to level curves of Hamiltonian (i.e., phase portraits). To this end, in Figure \ref{Fig1} we present a series of phase portraits with different motion integrals characterized by the Kozai inclinations at $i_* = 48^{\circ}$, $i_* = 50^{\circ}$ and $i_* = 145^{\circ}$ under different Hamiltonian models, including the standard double-averaged model (${\cal F}_{20}$ only), the Brown Hamiltonian model (turning on $\varepsilon_{21}$), and our novel extended model (turning on both $\varepsilon_{21}$ and $\varepsilon_{22}$). In the bottom-row panels, ZLK separatrices, dividing the librating region from the circulating region, are shown. 

As Jupiter--irregular satellite--Sun model is a weak-hierarchy three-body system, there are evident discrepancies of phase-space structures under Hamiltonian models with different levels of correction. In the case of $i_* = 48^{\circ}$, ZLK resonance can occur under the ${\cal F}_{20}$ and ${\cal F}_{20} + {\cal F}_{21}$ models, but it is not in the ${\cal F}_{20} + {\cal F}_{21} + {\cal F}_{22}$ model, indicating that, in the prograde space, the Kozai inclination to trigger ZLK resonance in the extended Brown Hamiltonian model is higher than that of the other two models. In the case of $i_* = 50^{\circ}$, ZLK resonance can happen under three Hamiltonian models. In the case of $i_* = 145^{\circ}$, the ZLK resonance can occur under both the ${\cal F}_{20} + {\cal F}_{21}$ and ${\cal F}_{20} + {\cal F}_{21} + {\cal F}_{22}$ models, but it is not in the ${\cal F}_{20}$ (the classical ZLK) model. Among the three Hamiltonian models, it is further observed that, for the prograde cases, the maximum eccentricity excited by the ZLK effect is the largest in the ${\cal F}_{20}$ (the classical ZLK) model, while for the retrograde case it is the largest in the ${\cal F}_{20} + {\cal F}_{21}$ (the classical Brown) model. 

\begin{figure*}
\centering
\includegraphics[width=2\columnwidth]{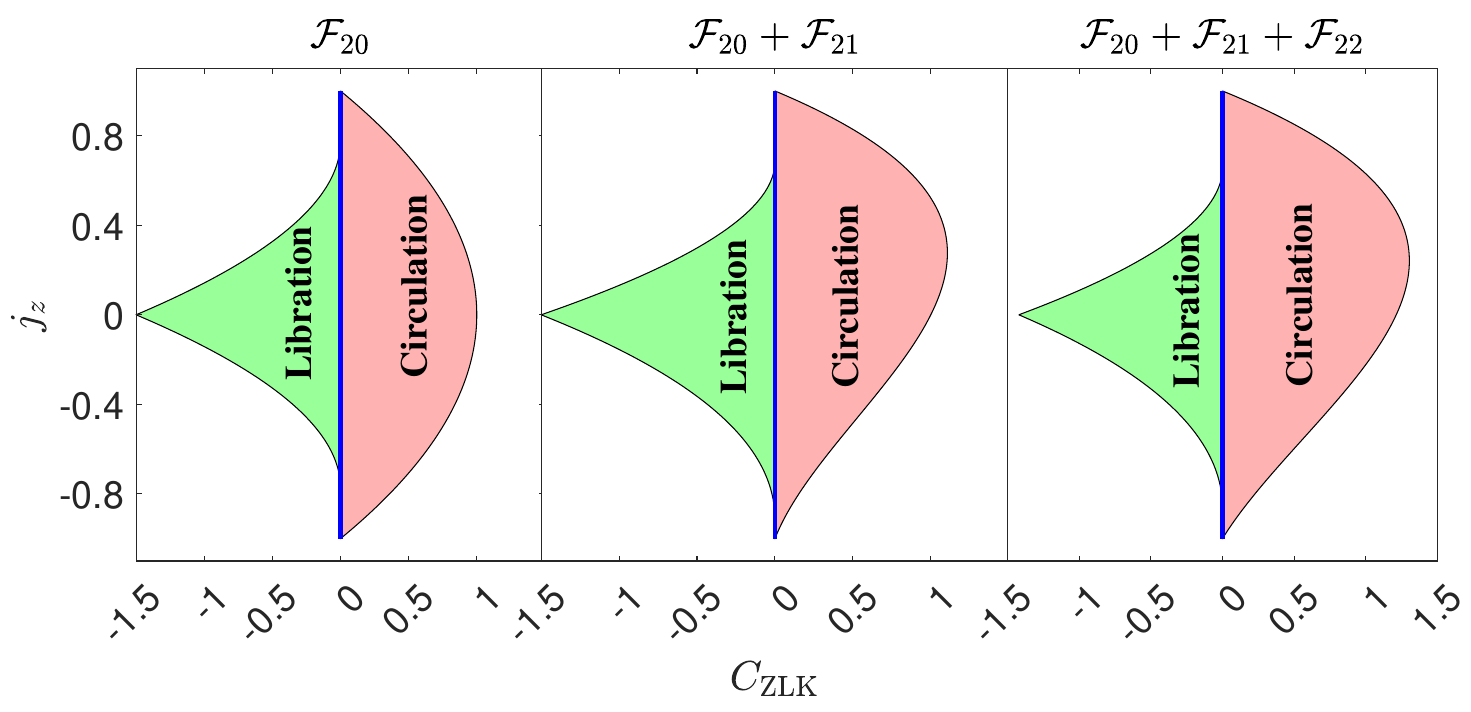}
\caption{Distribution of libration and circulation regimes in the $(C_{\rm ZLK},j_z)$ space for the case of $a = 0.15\,{\rm au}$, under the ${\cal F}_{20}$ (classical ZLK) model (\textit{left panel}), the ${\cal F}_{20} + {\cal F}_{21}$ (classical Brown Hamiltonian) model (\textit{middle panel}) and the ${\cal F}_{20} + {\cal F}_{21} + {\cal F}_{22}$ (extended Brown Hamiltonian) model (\textit{right panel}). The separatrix between the regions of libration and circulation, corresponding to $C_{\rm ZLK}= 0$, is marked by blue lines. In each panel, the white space stands for the forbidden region.}
\label{Fig1-1}
\end{figure*}

\subsection{A brief survey in parameter space}
\label{Sect2-3}

In the classical ZLK theory ($\varepsilon_{21} =\varepsilon_{22} =0$), the Lidov integral is introduced as \citep{lidov1962evolution}
\begin{equation}\label{Eq8-8-1}
c_2 = {e^2} - \frac{5}{2}e_z^2 = e^2\left(1-\frac{5}{2}\sin^2{i}\sin^2{\omega}\right),
\end{equation}
which corresponds to a combination of ${\cal F}_{20}$ and $j_z$ in the following manner,
\begin{equation}\label{Eq8-8-2}
c_2 = \frac{1}{2}\left({\cal F}_{20} - j_z^2 + \frac{1}{3}\right).
\end{equation}
Discussions about the Lidov integral can also be found in \citet{broucke2003long}, \citet{antognini2015timescales} and \citet{shevchenko2016lidov}.

Similarly, let us introduce the Lidov integral under our novel extended model as follows:
\begin{equation}\label{Eq8-6}
{C_{\rm ZLK}} = C_{\rm ZLK}^{20} + {\varepsilon _{21}}C_{\rm ZLK}^{21} + {\varepsilon _{22}}C_{\rm ZLK}^{22},
\end{equation}
where
\begin{equation}\label{Eq8-7}
C_{\rm ZLK}^{20} = {e^2} - \frac{5}{2}e_z^2,
\end{equation}
\begin{equation}\label{Eq8-7-1}
C_{\rm ZLK}^{21} = \frac{9}{2}{j_z}\left( {{e^2} - \frac{5}{8}e_z^2} \right),
\end{equation}
and
\begin{equation}\label{Eq8-8}
\begin{aligned}
C_{\rm ZLK}^{22} =& \frac{1}{{64}}\left\{ {8{e^2}\left( {\frac{{13}}{2}{e^2} + 11e_z^2 + 2j_z^2 + 60} \right)} \right.\\
&\left. { - \frac{3}{2} e_z^2\left[ {95e_z^2 + 6\left( {15j_z^2 + 31} \right)} \right]} \right\}.
\end{aligned}
\end{equation}
Based on the introduced parameter $C_{\rm ZLK}$, the extended Brown Hamiltonian becomes
\begin{equation}\label{Eq8-9}
\begin{aligned}
{\cal F} =& 2{C_{\rm ZLK}} + \left( {j_z^2 - \frac{1}{3}} \right) + \frac{3}{8}{\varepsilon _{21}}{j_z}\left( {1 - j_z^2} \right)\\
& - \frac{1}{{64}}{\varepsilon _{22}}j_z^2\left( {94 + 21j_z^2} \right).
\end{aligned}
\end{equation}

In particular, under the classical Brown Hamiltonian model (here it holds $\varepsilon_{22} =0$) the parameter becomes
\begin{equation}\label{Eq8-8-3}
{C_{\rm ZLK}} = {e^2}\left[ {1 + \frac{9}{2}{\varepsilon _{21}}{j_z} - \frac{5}{2}\left( {1 + \frac{9}{8}{\varepsilon _{21}}{j_z}} \right){{\sin }^2}i{{\sin }^2}\omega } \right].
\end{equation}

Recall that, under the extended Brown Hamiltonian model, there are two conserved quantities, namely the Hamiltonian ${\cal F}$ and the $z$ component of angular momentum $j_z (=H)$. As a result, equation (\ref{Eq8-9}) shows that the introduced parameter $C_{\rm ZLK}$ is also a conserved quantity during the long-term evolution. It means that $C_{\rm ZLK}$ can be used to parameterize the dynamical property of a triple system. Usually, it is convenient to work with $C_{\rm ZLK}$ instead of the Hamiltonian ${\cal F}$ \citep{lidov1962evolution}. It is mentioned that, under the ${\cal F}_{20}$ (classical ZLK) model, the distribution of libration and circulation regimes in the space of $(c_2 = C_{\rm ZLK}^{20}, c_1 = j_z^2)$ is now known as the Lidov diagram \citep{shevchenko2016lidov,sidorenko2018eccentric}, which is quite helpful in understanding the dynamical nature. However, in the extended Brown Hamiltonian model, ${\cal F}$ is an odd function of $j_z$, showing that it is not convenient to introduce the parameter $c_1 = j_z^2$. Thus, we directly take $j_z$, instead of $c_1$, as the associated parameter. Here, we aim to present a brief survey in the parameter space $(C_{\rm ZLK},j_z)$, following a similar method given in \citet{antognini2015timescales}. 

According to the phase portraits shown in Figure \ref{Fig1}, we can see that the separatrix between librating cycles and circulating cycles corresponds to the level curve of Hamiltonian passing through the zero-eccentricity point (saddle point). It means that, along the separatrix, $C_{\rm ZLK}$ is a constant, as shown by equation (\ref{Eq8-9}). As a result, $C_{\rm ZLK}$ of the separatrix can be evaluated at the saddle point ($e=0$) as
\begin{equation}\label{Eq8-10}
C_{\rm ZLK}^{\rm Sep} = 0.
\end{equation}
It provides a criterion for separating libration from circulation\footnote{It becomes possible for us to apply the criterion of $C_{\rm ZLK} < 0$ to quickly identify ZLK librating candidates among irregular satellites of giant planets, which will be discussed in the forthcoming paper of this series.}: (a) when $C_{\rm ZLK}<0$, the argument of pericenter of the inner binary $\omega$ is librating around $\pi/2$ or $3\pi/2$ (librating cycles); (b) when $C_{\rm ZLK}>0$, the angle $\omega$ may sweep through the full range of $[0,2\pi]$ (circulating cycles). Without Brown corrections, this criterion can reduce to the classical ZLK theory \citep{lidov1962evolution,broucke2003long,antognini2015timescales,shevchenko2016lidov,klein2024hierarchicalI,klein2024hierarchical,basha2025kozai}.

With a given $j_z$, it is not difficult to demonstrate that, in the case of $C_{\rm ZLK}>0$ (circulation), $C_{\rm ZLK}$ takes the maximum value at the point of $i=0$ if $j_z>0$ (prograde space) or at $i=\pi$ if $j_z<0$ (retrograde space), and in the case of $C_{\rm ZLK}<0$ (libration), it takes the minimum value at the ZLK resonance center. ZLK center corresponds to the stable stationary point, which will be discussed in Section \ref{Sect3}.

In the configuration of $a = 0.15\,{\rm au}$ (corresponding to $\alpha_{\rm H} = 0.423$), Figure \ref{Fig1-1} shows a distribution of ZLK librating and circulating orbits in the $(C_{\rm ZLK}, j_z)$ space, under the ${\cal F}_{20}$ (classical ZLK) model, the ${\cal F}_{20} + {\cal F}_{21}$ (classical Brown Hamiltonian) model and the ${\cal F}_{20} + {\cal F}_{21} + {\cal F}_{22}$ (extended Brown Hamiltonian) model. The separatrix between the regions of libration and circulation, determined by $C_{\rm ZLK}= 0$, is marked by blue lines. It is observed from Figure \ref{Fig1-1} that the distribution of librating (or circulating) cycles in the ${\cal F}_{20}$ model is symmetric with respect to $j_z = 0$, while they are no longer symmetric under the Hamiltonian models with Brown corrections. According to the discussion made in Section \ref{Sect2-1}, we know that the asymmetry is caused by the introduction of the classical Brown correction (here it corresponds to $C_{\rm ZLK}^{21}$).

\section{Analytical study of the modified ZLK oscillations}
\label{Sect3}

In this section, we aim to discuss the location of modified fixed points or ZLK resonance centers (Section \ref{Sect3-1}), the maximum eccentricity excited by ZLK effects (Section \ref{Sect3-2}), and the critical Kozai inclination for triggering ZLK resonances (Section \ref{Sect3-3}) under the extended Brown Hamiltonian model. Their perturbative solutions are formulated with $\varepsilon_{21}$ and $\varepsilon_{22}$ as small parameters. According to equation (\ref{Eq6}), we can see that $\varepsilon_{21}$ is a first-order small quantity of mean motion ratio $n_{\rm p}/n$, and $\varepsilon_{22}$ is a second-order small quantity. 

In the following we utilize a perturbation theory either around ${\cal F}_{20}$ (approach I), or ${\cal F}_{20} + {\cal F}_{21}$ (approach II) as a starting point. Approach I will usually lead to $5$ unknown coefficients in the expansion, while approach II will lead to $3$ unknown coefficients in the expansion.

\subsection{The ZLK resonance center (fixed point)}
\label{Sect3-1}

Substituting equation (\ref{Eq1}) into the Lagrange planetary equations (or the Hamiltonian canonical relations) leads to
\begin{equation}\label{Eq9}
\frac{{{\rm d}e}}{{{\rm d}t}} =  - \frac{\eta }{{n{a^2}e}}\frac{{\partial {\cal F}}}{{\partial \omega }},\quad \frac{{{\rm d}\omega }}{{{\rm d}t}} = \frac{\eta }{{n{a^2}}}\left( {\frac{1}{e}\frac{{\partial {\cal F}}}{{\partial e}} - \frac{{\cot i}}{{{\eta ^2}}}\frac{{\partial {\cal F}}}{{\partial i}}} \right),
\end{equation}
where $\eta = \sqrt{1-e^2}$. The stationary solution of equation (\ref{Eq9}) corresponds to the fixed points under the extended Hamiltonian model, defined by
\begin{equation}\label{Eq10}
\frac{{{\rm d}e}}{{{\rm d}t}}=0,\quad \frac{{{\rm d}\omega }}{{{\rm d}t}} = 0.
\end{equation}
The first condition ${\rm d}e/{\rm d}t=0$ %$\frac{{{\rm d}e}}{{{\rm d}t}} = 0$
requires that $\partial {\cal F}/\partial \omega = 0$, %$\frac{{\partial {\cal F}}}{{\partial \omega }} = 0$,
leading to $2\omega = 0$ or $2\omega = \pi$, which is consistent to the classical ZLK theory \citep{lidov1962evolution,kozai1962secular}. In particular, the fixed points at $2\omega = 0$ are unstable, and the ones at $2\omega = \pi$ are stable, as shown by the phase portraits in Figure \ref{Fig1}. Usually, the stable fixed points are referred to as the ZLK resonance center. Replacing $2\omega = \pi$ in the second condition of equation (\ref{Eq10}), we can get the equation of ZLK center as follows: %\eg{[I think the different terms need to be normalised with different powers of $L$ to have the same units.]}
\begin{equation}\label{Eq11}
\begin{aligned}
&\frac{5}{4}{\varepsilon _{22}}\left( {{G_{\rm fix}^8} - 57{H^4}} \right) - {G_{\rm fix}^6}\left( {24 - 27{\varepsilon _{21}}H - 49{\varepsilon _{22}} + 79{\varepsilon _{22}}{H^2}} \right)  \\
& + {G_{\rm fix}^2}{H^2}\left( {40 + 45{\varepsilon _{21}}H + 119{\varepsilon _{22}} + 105{\varepsilon _{22}}{H^2}} \right) = 0,
\end{aligned}
\end{equation}
where $H=\cos{i_*}$ is the motion integral determined by the initial condition and $G_{\rm fix}^2=1-e_{\rm fix}^2$ represents the location of ZLK center to be determined. Recall that $e_{\rm fix}$ is the eccentricity of ZLK center and $i_{\rm fix}$ denotes the associated inclination. Assuming $x= G_{\rm fix}^2=1-e_{\rm fix}^2 $, we can get the equation of ZLK center as follows:
\begin{equation}\label{Eq12}
\begin{aligned}
&  285{H^4}{\varepsilon _{22}} - 4{H^2}\left( {40 + 45H{\varepsilon _{21}} + 119{\varepsilon _{22}} + 105{H^2}{\varepsilon _{22}}} \right)x\\
& + 4\left( {24 - 27H{\varepsilon _{21}} - 49{\varepsilon _{22}} + 79{H^2}{\varepsilon _{22}}} \right){x^3} - 5{\varepsilon _{22}}{x^4} = 0,
\end{aligned}
\end{equation}
which is a quartic equation of $x$.  Considering the fact that there are two small parameters (including $\varepsilon_{21}$ and $\varepsilon_{22}$) in equation (\ref{Eq12}), we have two choices to solve it: (a) taking the solution of ${\cal F}_{20}$ model as the starting point, and (b) taking the solution of ${\cal F}_{20} + {\cal F}_{21}$ model as starting point. In case (a), both $\varepsilon_{21}$ and $\varepsilon_{22}$ are taken as small parameters. In case (b), $\varepsilon_{21}$ is included in the starting solution and thus only $\varepsilon_{22}$ is a small parameter in deriving the perturbation solution.

\textbf{Approach I: The ${\cal F}_{20}$ model as the starting point.}

The ${\cal F}_{20}$ model corresponds to the classical ZLK Hamiltonian, where the ZLK center is known at \citep{kozai1962secular}
\begin{equation}\label{Eq13}
x_0 = 1-e_{\rm fix}^2 = \sqrt {\frac{5}{3}} \left| H \right|.
\end{equation}
Under the extended Brown Hamiltonian model, the terms $\varepsilon_{21} {\cal F}_{21}$ and $\varepsilon_{22}{\cal F}_{22}$ can be considered as the perturbations to the classical ZLK dynamical model, thus the solution can be expressed as a two-parameter series solution (up to order 3) in the following form:
\begin{equation}\label{Eq14}
x = {x_0} + {\varepsilon _{21}}{x_1} + \varepsilon _{21}^2{x_2} + \varepsilon _{21}^3{x_3} + {\varepsilon _{22}}{x_4} + {\varepsilon _{21}}{\varepsilon _{22}}{x_5},
\end{equation}
where $x_i$ ($i=\{1,...,5\}$) are unknown coefficients to be determined. Replacing equation (\ref{Eq14}) in equation (\ref{Eq11}), we can obtain the perturbation solution of ZLK center as follows:
\begin{equation}\label{Eq15}
\begin{aligned}
x =& \sqrt {\frac{5}{3}} \left| H \right| + \frac{3}{8}\sqrt {15}{\varepsilon _{21}} H\left| H \right|\left( {1 + \frac{9}{{16}}{\varepsilon _{21}}H + \frac{{81}}{{128}}\varepsilon _{21}^2{H^2}} \right)\\
 &+ \frac{1}{{360}}{\varepsilon _{22}}\left| H \right|\left( {301\sqrt {15}  - 305\left| H \right| - 40\sqrt {15} {{\left| H \right|}^2}} \right)\\
 &+ \frac{1}{{128}}{\varepsilon _{21}}{\varepsilon _{22}}H\left| H \right|\left( {98\sqrt {15}  + 147\left| H \right| - 158\sqrt {15} {{\left| H \right|}^2}} \right).
\end{aligned}
\end{equation}
This perturbation solution explicitly shows how the classical and extended Brown Hamiltonian corrections influence the location of ZLK center.

\textbf{Approach II: The ${\cal F}_{20}+{\cal F}_{21}$ model as the starting point.}

The ${\cal F}_{20} + {\cal F}_{21}$ model corresponds to the classical Brown Hamiltonian model \citep{tremaine2023hamiltonian}, where the ZLK center is known at \citep{grishin2024irregularI}
\begin{equation}\label{Eq16}
x_0 = 1-e_{\rm fix}^2 = \sqrt {\frac{{5\left( {8 + 9{\varepsilon _{21}}H} \right)}}{{3\left( {8 - 9{\varepsilon _{21}}H} \right)}}} \left| H \right|,
\end{equation}
where the parameter $\varepsilon_{21}$ is included in the solution. Under the extended Brown Hamiltonian model, the term $\varepsilon_{22}{\cal F}_{22}$ plays a role of perturbation to the classical Brown model, thus the solution can be expressed as (up to order 3 in $\varepsilon_{22}$, corresponding to order 6 in $\varepsilon_{21}$)
\begin{equation}\label{Eq17}
x = {x_0} + {\varepsilon _{22}}{x_1} + \varepsilon _{22}^2{x_2} + \varepsilon _{22}^3{x_3},
\end{equation}
where the coefficients $x_i (i=1,2,3)$ are to be determined. Similarly, replacing equation (\ref{Eq17}) in equation (\ref{Eq11}) leads to
\begin{equation}\label{Eq18}
\begin{aligned}
{x_1} =& \frac{1}{{4{c_0}}}\left[ {5\left( {84{x_0} - 57} \right){H^4} + x_0^3\left( {196 + 5{x_0}} \right)} \right.\\
& + \left. {4{H^2}\left( {119{x_0} - 79x_0^3} \right)} \right],
\end{aligned}
\end{equation}
\begin{equation}\label{Eq19}
\begin{aligned}
{x_2} =& \frac{{{x_1}}}{{{c_0}}}\left[ {105{H^4} + {H^2}\left( {119 - 237x_0^2} \right) + x_0^2\left( {147 + 5{x_0}} \right)} \right.\\
&- \left. {  9{x_0}{x_1}\left( {8 - 9{\varepsilon _{21}}H} \right)} \right],
\end{aligned}
\end{equation}
and
\begin{equation}\label{Eq20}
\begin{aligned}
{x_3} =& \frac{1}{{2{c_0}}}\left\{ {14{H^2}{x_2}\left( {17 + 15{H^2}} \right) + 10x_0^3{x_2}} \right.\\
& + 3x_0^2\left[ {5x_1^2 + 2{x_2}\left( {49 - 79{H^2}} \right)} \right] - 6x_1^3\left( {8 - 9{\varepsilon _{21}}H} \right)\\
& - \left. {  6{x_0}{x_1}\left[ {{x_1}\left( {79{H^2} - 49} \right) + 6{x_2}\left( {8 - 9{\varepsilon _{21}}H} \right)} \right]} \right\},
\end{aligned}
\end{equation}
where ${c_0} = 10{H^2}\left( {8 + 9{\varepsilon _{21}}H} \right)$ and $x_0$ is provided by equation (\ref{Eq16}).

In summary, the location of modified ZLK center in the eccentricity--inclination space can be written as
\begin{equation}\label{Eq21}
e_{\rm fix} = \sqrt{1-x},\quad i_{\rm fix} = \arccos\left({\frac{\cos{i_*}}{\sqrt{x}}}\right),
\end{equation}
where $x$ is provided by equations (\ref{Eq14}) and (\ref{Eq17}), corresponding to approaches I and II, respectively.

\begin{figure*}
\centering
\includegraphics[width=\columnwidth]{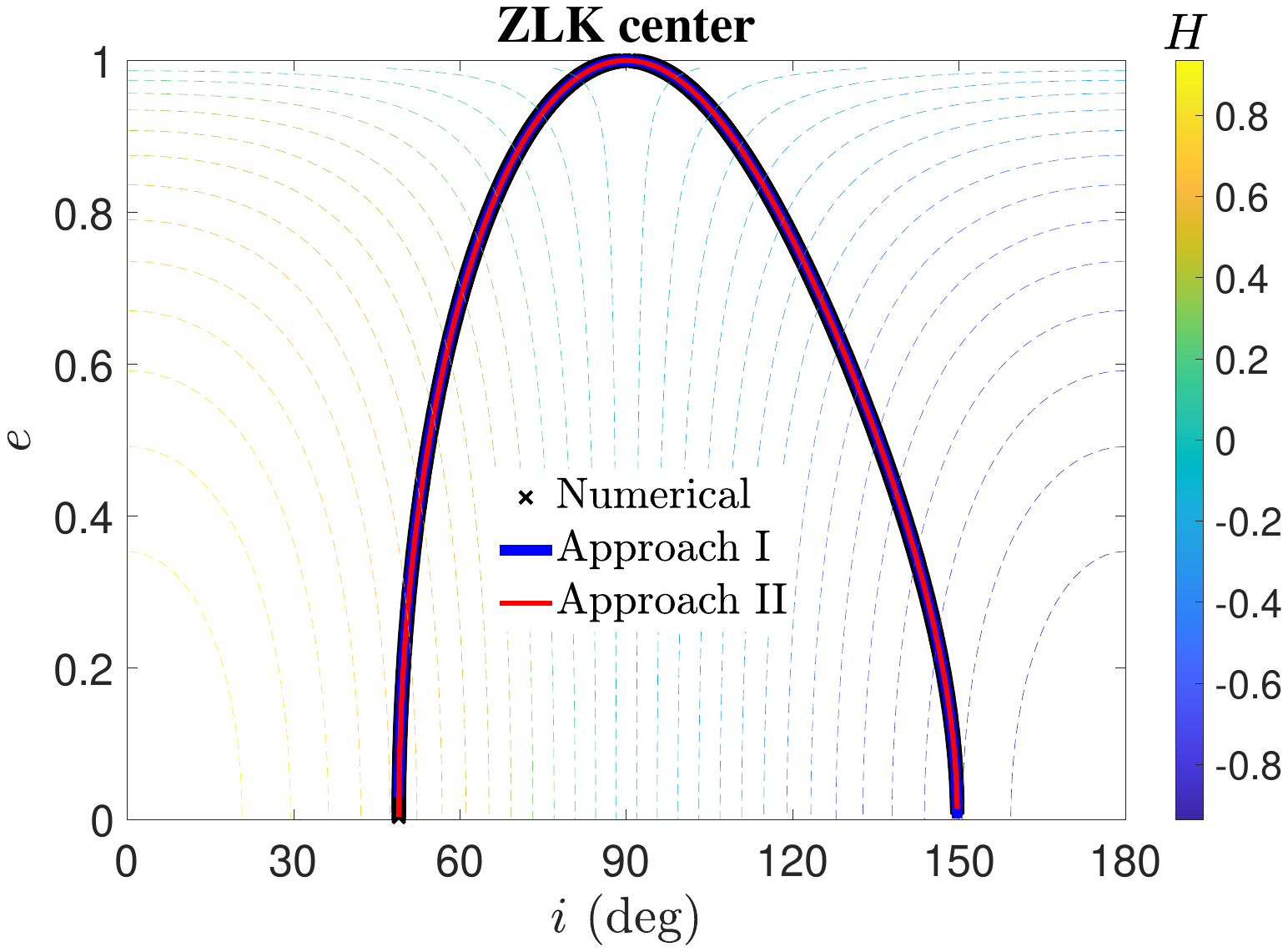}
\includegraphics[width=\columnwidth]{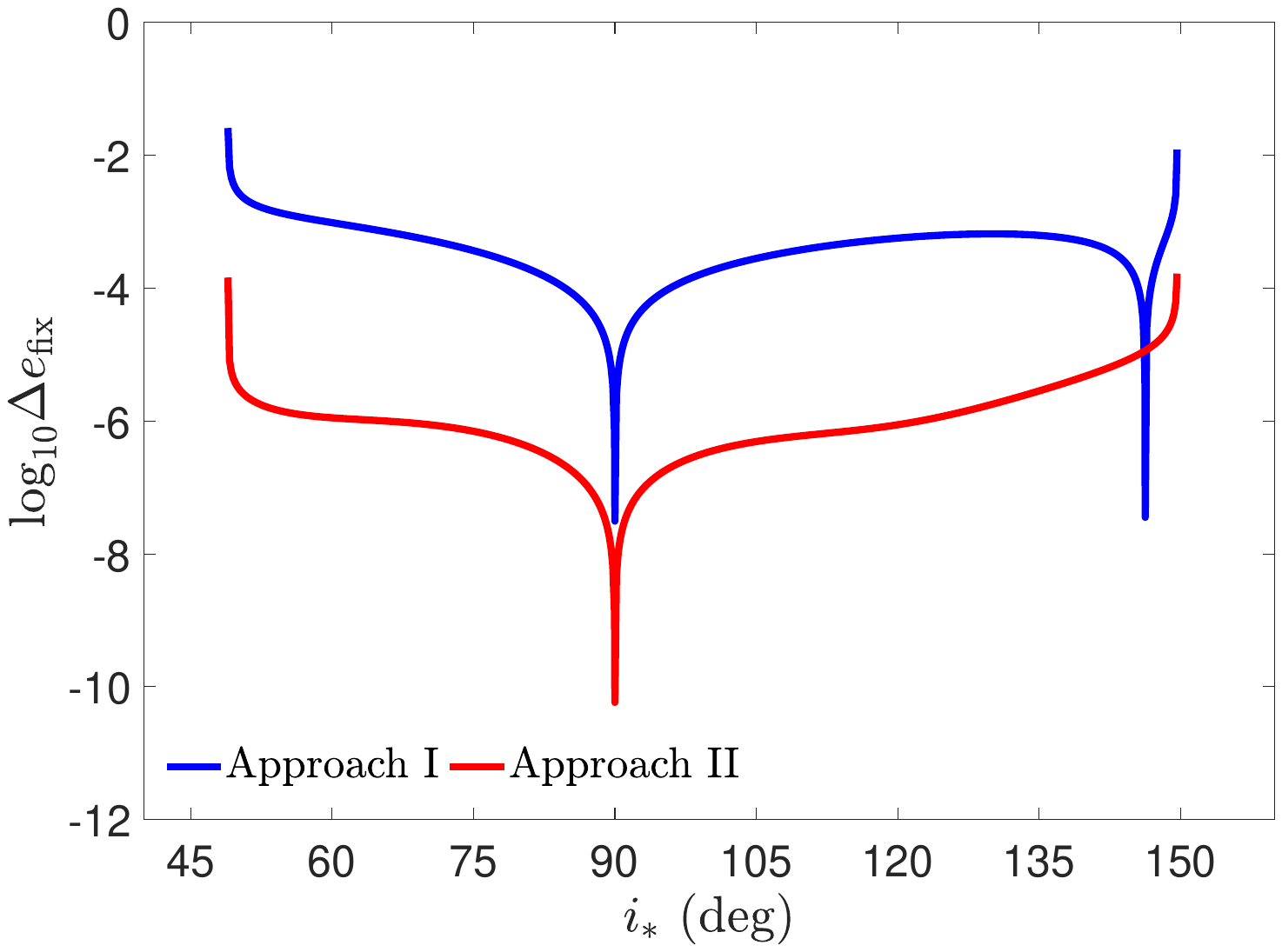}
\caption{Location of ZLK center in the eccentricity--inclination space under the extended Brown Hamiltonian model, determined by means of numerical and analytical methods (\textit{left panel}), as well as the deviation of $e_{\rm fix}$ as a function of $i_*$ for analytical expressions (\textit{right panel}), in the configuration of $a = 0.15\, {\rm au}$ (corresponding to $\alpha_{\rm H} = 0.423$). Note that the location of ZLK center $(e_{\rm fix},i_{\rm fix})$ is measured at the argument of $2\omega=\pi$. In the left panel, level curves of the motion integral $H$ are presented as background (the magnitude of $H$ is evaluated from the inclination $i_*$ at zero eccentricity by $H = \cos{i_*}$).}
\label{Fig2}
\end{figure*}

Figure \ref{Fig2} shows the location of ZLK center under the extended Brown Hamiltonian model determined by means of numerical method\footnote{In the entire work we refer to the Newton--Raphson method where the solution of the classical ZLK model is taken as an initial guess.} and analytical methods (including approaches I and II), together with the error of $e_{\rm fix}$ as a function of the Kozai inclination $i_*$ for perturbation solutions. In the left panel of Figure \ref{Fig2}, level curves of the motion integral $H$ are presented as background. It is noted that, with a given $H$, there exists only one physical solution of $x \in [0,1]$ by solving equation (\ref{Eq12}). For convenience, the base 10 logarithm of the deviation between the analytical and numerical results is taken for error analysis. It is observed that the analytical curve of ZLK center is in good agreement with numerical result. In particular, the perturbation solution of approach II has a higher precision in comparison to that of approach I. This is because in approach I the series solution is up to the third order in $\varepsilon_{21}$, while the solution of approach II is up to the third order in $\varepsilon_{22}$ (corresponding to the sixth order in $\varepsilon_{21}$).

\begin{figure*}
\centering
\includegraphics[width=\columnwidth]{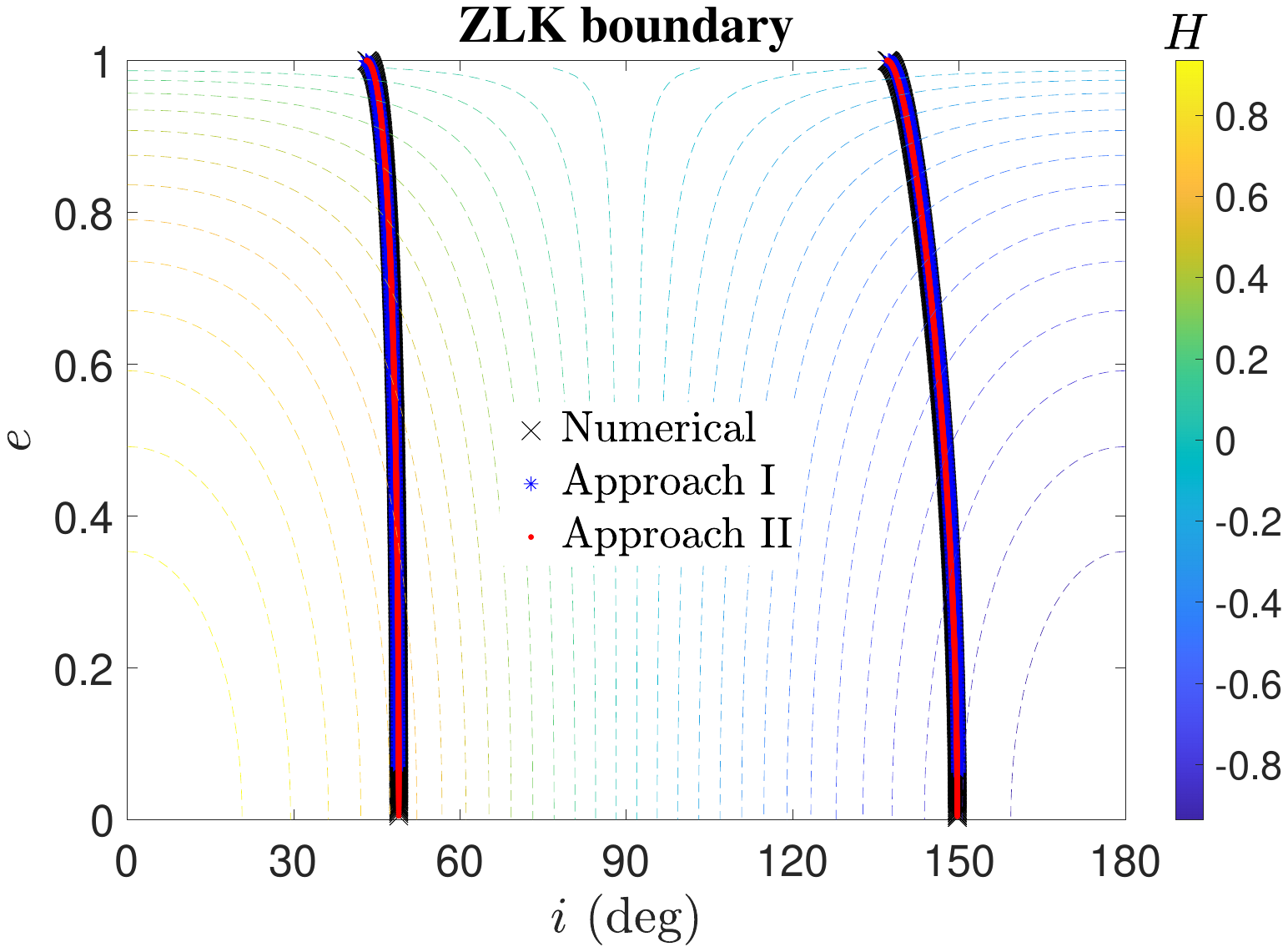}
\includegraphics[width=\columnwidth]{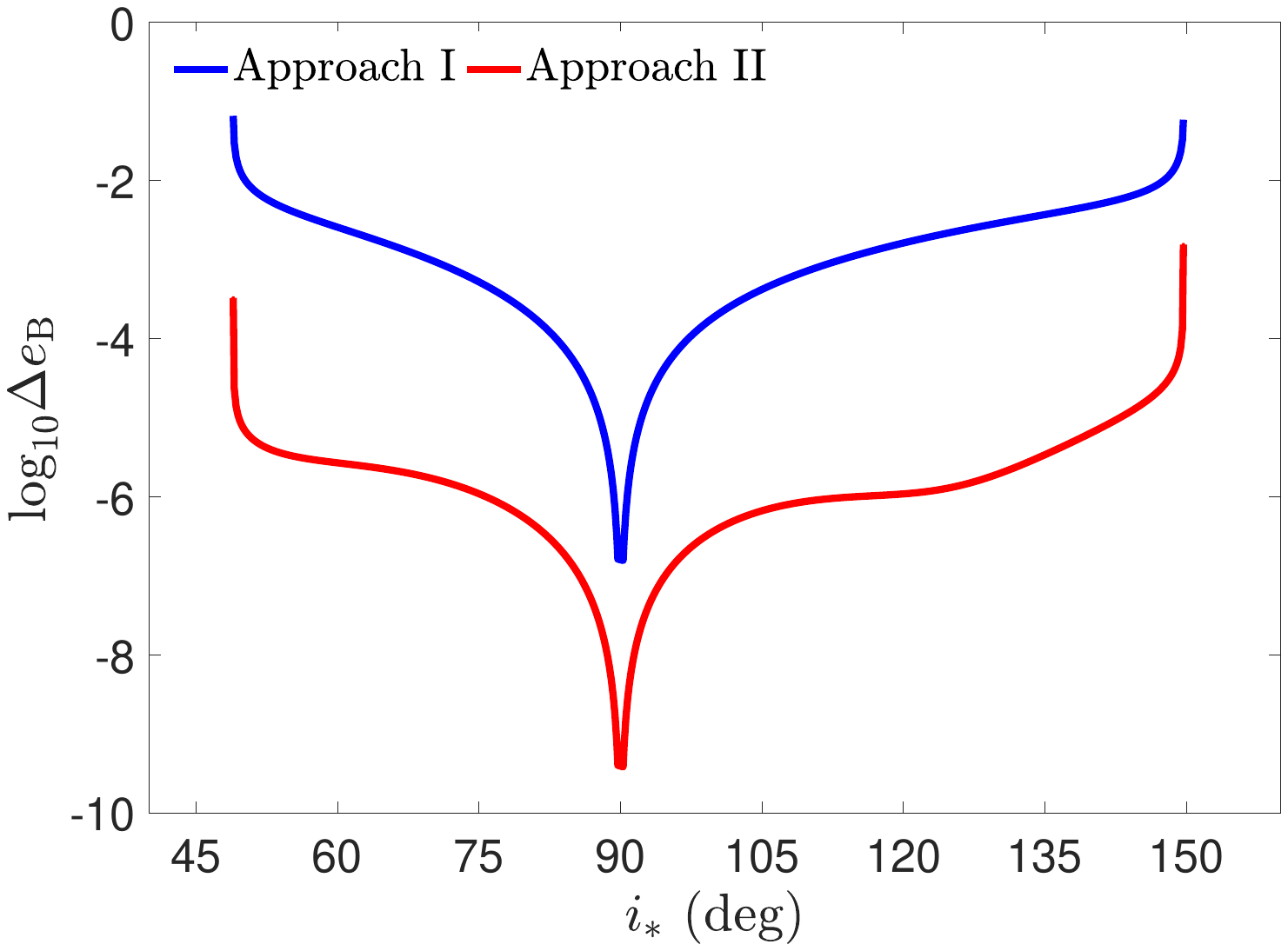}
\caption{The ZLK boundary under the extended Brown Hamiltonian model, determined by means of numerical and analytical methods (\textit{left panel}) and error curves of $e_{\rm B}$ of analytical expressions (\textit{right panel}), in the configuration of $a = 0.15\, {\rm au}$ (corresponding to $\alpha_{\rm H} = 0.423$). Note that the ZLK boundary $(e_{\rm B},i_{\rm B})$ is measured at $2\omega=\pi$ (the argument of ZLK center). Similar to Fig. \ref{Fig2}, level curves of the motion integral $H$ are presented as background in the left panel.}
\label{Fig3}
\end{figure*}

\begin{figure*}
\centering
\includegraphics[width=\columnwidth]{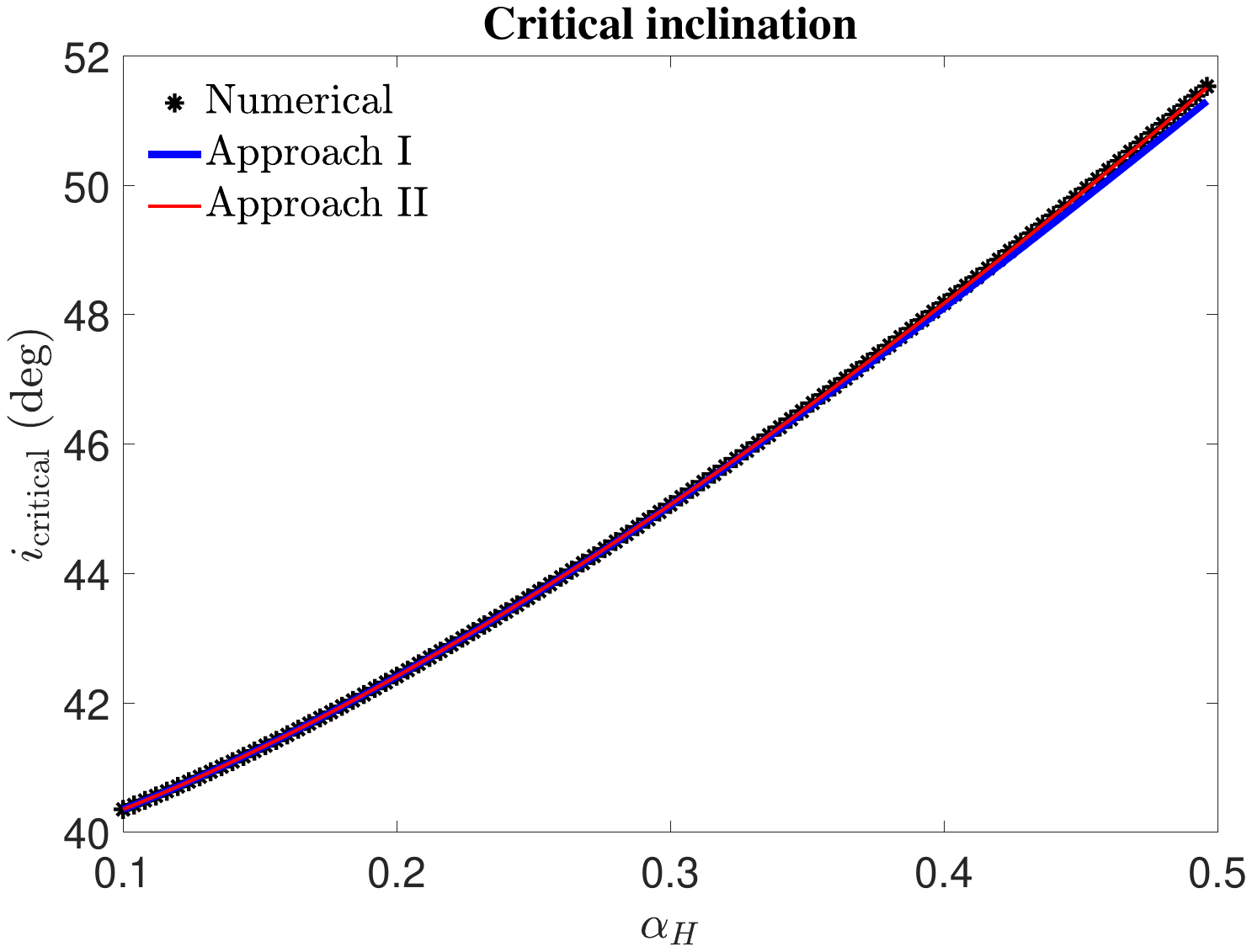}
\includegraphics[width=\columnwidth]{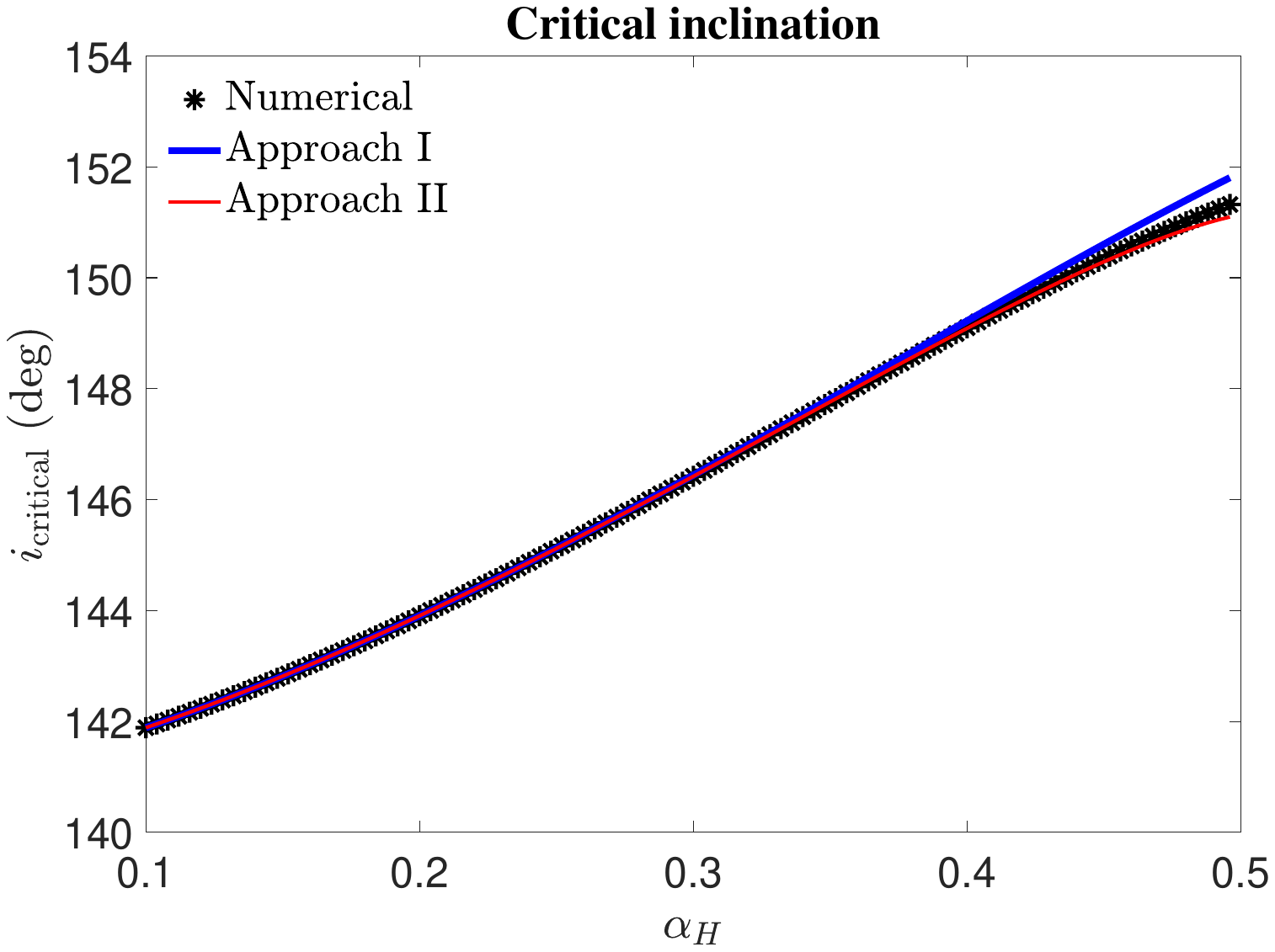}
\caption{Critical inclination of ZLK resonance as a function of $\alpha_{\rm H}$ under the extended Brown Hamiltonian model, determined by numerical and analytical methods, in the prograde configuration (\textit{left panel}) and in the retrograde configuration (\textit{right panel}).}
\label{Fig4}
\end{figure*}

\subsection{The maximum eccentricity (ZLK boundary)}
\label{Sect3-2}

When the ZLK resonance occurs, the eccentricity of the inner test particle may be excited from zero to a maximum value, denoted by $e_{\rm B}$. This eccentricity excitation phenomenon is usually considered as a consequence of the ZLK effect (or the ZLK mechanism). According to the phase portraits shown in Figure \ref{Fig1}, the maximum eccentricity of separatrix is located at $2\omega = \pi$, corresponding to the argument of ZLK center. At the point of maximum eccentricity, the associated inclination is denoted by $i_{\rm B}$, which is determined by the motion integral,
\begin{equation*}
H = \sqrt{1-e_{\rm B}^2}\cos{i_{\rm B}}.
\end{equation*} 
Additionally, we can see that the maximum eccentricity triggered by ZLK resonance is related to the ZLK separatrix, which plays a boundary role that divides the phase space into librating and circulating regions. According to level curves of Hamiltonian in phase portraits (see Figure \ref{Fig1}), the equation of ZLK separatrix can be formulated from the viewpoint of Hamiltonian,
\begin{equation}\label{Eq22}
{\cal F}_{\rm sep} = {\cal F}\left(e=0,2\omega=0\right) = {\cal F}\left(e=e_{\rm B},2\omega=\pi\right),
\end{equation}
or from the viewpoint of $C_{\rm ZLK}$,
\begin{equation}\label{Eq22-1}
{C_{\rm ZLK}^{\rm sep}} = {C_{\rm ZLK}}\left(e=0,2\omega=0\right) = {C_{\rm ZLK}}\left(e=e_{\rm B},2\omega=\pi\right).
\end{equation}
Substituting the expression of ${\cal F}$ into equation (\ref{Eq22}) or (\ref{Eq22-1}), we can get the equation of the maximum eccentricity $(e_{\rm B})$ as follows:
\begin{equation}\label{Eq23}
\begin{aligned}
&\left( {320{H^2} + 360{\varepsilon _{21}}{H^3} + 952{\varepsilon _{22}}{H^2} + 840{\varepsilon _{22}}{H^4}} \right)z\\
& - \left( {192 + 320{H^2} - 216{\varepsilon _{21}}H + 360{\varepsilon _{21}}{H^3} - 397{\varepsilon _{22}}} \right.\\
&\left. { + 1584{\varepsilon _{22}}{H^2} + 555{\varepsilon _{22}}{H^4}} \right){z^2} - 285{\varepsilon _{22}}{H^4} - 5{\varepsilon _{22}}{z^4}\\
& + \left( {192 - 216{\varepsilon _{21}}H - 392{\varepsilon _{22}} + 632{\varepsilon _{22}}{H^2}} \right){z^3} = 0,
\end{aligned}
\end{equation}
where $z = 1 - e_{\rm B}^2$ is the unknown variable to be solved. Similarly, we have two approaches in solving equation (\ref{Eq23}).

\textbf{Approach I: The ${\cal F}_{20}$ model as the starting point.}

Under the ${\cal F}_{20}$ model, the maximum eccentricity $e_{\rm B}$ excited by the ZLK effect is determined by the Kozai inclination $i_*$ in the following manner \citep{kozai1962secular}:
\begin{equation}\label{Eq24}
z_0 = 1 - e_{\rm B}^2 = \frac{5}{3}{H^2} = \frac{5}{3} \cos^2{i_*}.
\end{equation}
The associated inclination is equal to $i_{\rm B} = 39.2^{\circ}$ in the prograde space and $i_{\rm B} = 140.8^{\circ}$ in the retrograde space ($i_{\rm B}$ is known as Kozai angle). It means that, under the classical ZLK model, the ZLK resonance can occur within the inclination space between $39.2^{\circ}$ and $140.8^{\circ}$ \citep{kozai1962secular,naoz2016eccentric}.

Under the extended Brown Hamiltonian model, the solution can be expressed in a small-parameter series form (up to order 3) as follows:
\begin{equation}\label{Eq25}
z = {z_0} + {\varepsilon _{21}}{z_1} + \varepsilon _{21}^2{z_2} + \varepsilon _{21}^3{z_3} + {\varepsilon _{22}}{z_4} + {\varepsilon _{21}}{\varepsilon _{22}}{z_5},
\end{equation}
where the coefficients $z_i (i=\{1,...,5\})$ are to be determined. Replacing the series solution of equation (\ref{Eq25}) in equation (\ref{Eq23}), we can obtain
\begin{equation}\label{Eq26}
\begin{aligned}
z =& \frac{5}{3}{H^2} + \frac{{15}}{4}{\varepsilon _{21}}{H^3} + \frac{1}{{216}}{\varepsilon _{22}}{H^2}\left( {1623 - 545{H^2}} \right) + \frac{{135}}{{32}}\varepsilon _{21}^2{H^4}\\
& + \frac{{1215}}{{256}}\varepsilon _{21}^3{H^5} + \frac{1}{{384}}{\varepsilon _{21}}{\varepsilon _{22}}{H^3}\left( {6993 - 5705{H^2}} \right),
\end{aligned}
\end{equation}
which explicitly shows how the classical and extended Brown corrections influence the maximum eccentricity excited by ZLK effects.

\textbf{Approach II: The ${\cal F}_{20} + {\cal F}_{21}$ model as the starting point.}

Under the classical Brown Hamiltonian, the maximum eccentricity is given by \citep{grishin2018quasi}
\begin{equation}\label{Eq27}
{z_0} = 1-e_{\rm B}^2 = \frac{{5\left( {8 + 9{\varepsilon _{21}}H} \right)}}{{3\left( {8 - 9{\varepsilon _{21}}H} \right)}}{H^2}.
\end{equation}
Considering equation (\ref{Eq27}) as the starting point, we can formulate the perturbation solution under the extended Brown Hamiltonian model in a series form (up to order 3 in $\varepsilon_{22}$) as follows:
\begin{equation}\label{Eq28}
z = {z_0} + {\varepsilon _{22}}{z_1} + \varepsilon _{22}^2{z_2} + \varepsilon _{22}^3{z_3},
\end{equation}
where $z_i (i=1,2,3)$ are unknown coefficients to be determined. Similarly, substituting the series solution of equation (\ref{Eq28}) into equation (\ref{Eq23}), we can obtain
\begin{equation}\label{Eq29}
\begin{aligned}
{z_1} =& \frac{1}{{8{d_0}}}\left( {{z_0} - 1} \right)\left[ {8{H^2}{z_0}\left( {119 - 79{z_0}} \right) + z_0^2\left( {397 + 5{z_0}} \right)} \right.\\
&\left. { + 15{H^4}\left( {37{z_0} - 19} \right)} \right],
\end{aligned}
\end{equation}
\begin{equation}\label{Eq30}
\begin{aligned}
{z_2} =& \frac{{{z_1}}}{{4{d_0}}}\left\{ {z_0^2\left( {588 + 10{z_0}} \right) + 15{H^4}\left( {37{z_0} - 28} \right)} \right.\\
& - {z_0}\left( {397 + 288{z_1}} \right) - 4{H^2}\left( {119 - 396{z_0} + 237z_0^2 - 40{z_1}} \right)\\
&\left. { + 12{z_1}\left( {8 + 15{\varepsilon _{21}}{H^3}} \right) + 108{\varepsilon _{21}}H\left( {3{z_0} - 1} \right){z_1}} \right\},
\end{aligned}
\end{equation}
and
\begin{equation}\label{Eq31}
\begin{aligned}
{z_3} =& \frac{1}{{8{d_0}}}\left\{ {z_1^2\left[ { - 397 + 555{H^4} + {z_0}\left( {1176 + 30{z_0}} \right)} \right.} \right.\\
&\left. { + 24{H^2}\left( {66 - 79{z_0}} \right)} \right] - 24z_1^3\left( {8 - 9{\varepsilon _{21}}H} \right)\\
& - 2{z_2}\left[ {15{H^4}\left( {28 - 37{z_0}} \right) + {z_0}\left( {397 - 588{z_0} - 10z_0^2} \right)} \right.\\
&\left. { + 4{H^2}\left( {119 - 396{z_0} + 237z_0^2} \right)} \right]\\
&\left. { + 16{z_1}{z_2}\left[ {5{H^2}\left( {8 + 9{\varepsilon _{21}}H} \right) + 3\left( {8 - 9{\varepsilon _{21}}H} \right)\left( {1 - 3{z_0}} \right)} \right]} \right\},
\end{aligned}
\end{equation}
where
\begin{equation}\label{Eq32}
{d_0} = {z_0}\left[ {5{H^2}\left( {8 + 9{\varepsilon _{21}}H} \right) - 3\left( {8 - 9{\varepsilon _{21}}H} \right)} \right]
\end{equation}
and $z_0$ is provided by equation (\ref{Eq27}).

To summarize, the maximum eccentricity excited by ZLK effects (denoted by $e_{\rm B}$) and the associated inclination (denoted by $i_{\rm B}$) are expressed by
\begin{equation}\label{Eq33}
e_{\rm B} = \sqrt{1-z},\quad i_{\rm B} = \arccos{\left(\frac{\cos{i_*}}{\sqrt{z}}\right)},
\end{equation}
where $z$ is provided by equations (\ref{Eq25}) and (\ref{Eq28}), corresponding to approaches I and II, respectively. In particular, the distribution of $(e_{\rm B},i_{\rm B})$ represents the ZLK boundary in the eccentricity--inclination space, dividing the entire space into ZLK librating and circulating regions. Recall that the ZLK boundary $(e_{\rm B},i_{\rm B})$ is measured at the argument of ZLK center $(2\omega=\pi)$ and it corresponds to the line of $C_{\rm ZLK}=0$, as shown in Figure \ref{Fig1-1}.

Figure \ref{Fig3} presents the ZLK boundaries under the extended Brown Hamiltonian model, determined by numerical technique and analytical methods (including approaches I and II), together with the error of $e_{\rm B}$ as a function of the Kozai inclination $i_*$ for perturbation solutions. Similarly, the base 10 logarithm of the deviation between the analytical and numerical results is taken for the error analysis. Please refer to the caption for the detailed setting of model parameters. It is observed that (a) analytical curves can agree well with numerical curves for ZLK boundaries in the eccentricity--inclination space, and (b) the perturbation solution of approach II holds a higher (by three orders of magnitude) precision compared to that of approach I. 

\subsection{The critical Kozai inclination}
\label{Sect3-3}

Recall that the Kozai inclination is defined as the inclination at zero eccentricity, namely $H = \cos{i_*}$. According to the ZLK center shown in Figure \ref{Fig2}, we can see that there exists a critical value of Kozai inclination, denoted by $i_{\rm critical}$, above which the ZLK resonance begins to bifurcate. The critical inclination $i_{\rm critical}$ is different in the prograde and retrograde spaces. In particular, when $i_* > i_{\rm critical}$ in the prograde space (or $i_* < i_{\rm critical}$ in the retrograde space), the ZLK resonance occurs. The critical inclination has been discussed in \citet{beauge2006high} under nonlinear Hamiltonian model based on the Lie-series transformation and in \citet{lei2019semi} under the nonlinear Hamiltonian model based on the elliptic expansion of disturbing function. The purpose of this subsection is to formulate a perturbation solution of the critical Kozai inclination $i_{\rm critical}$ in an explicit form under our novel extended model. 

According to the stationary condition at the point of $(e=0,i=i_{\rm critical})$, we can get the equation of the critical Kozai inclination as follows: 
\begin{equation}\label{Eq34}
96 - 201{\varepsilon _{22}} - 108{\varepsilon _{21}}w - 160\left( {1 + {\varepsilon _{22}}} \right){w^2} - 180{\varepsilon _{21}}{w^3} - 135{\varepsilon _{22}}{w^4} = 0,
\end{equation}
where $w = \cos\left({i_{\rm critical}}\right)$ is the unknown variable to be determined. Similarly, we have two choices to construct a perturbation solution of $w$.

\textbf{Approach I: The ${\cal F}_{20}$ model as the starting point.}

Under the ${\cal F}_{20}$ model, the classical ZLK theory tells us that the critical inclination reads \citep{kozai1962secular}
\begin{equation}\label{Eq35}
w_0 = \cos\left({i_{\rm critical}}\right) =  \pm \sqrt {\frac{3}{5}}
\end{equation}
which provides the well-known critical inclinations at $i_{\rm critical} = 39.2^{\circ}$ in the prograde space and $i_{\rm critical} = 140.8^{\circ}$ in the retrograde space. With the solution of the ${\cal F}_{20}$ model as the starting point, the perturbation solution of $w$ can be written as (up to order 3)
\begin{equation}\label{Eq36}
w = {w_0} + {\varepsilon _{21}}{w_1} + \varepsilon _{21}^2{w_2} + \varepsilon _{21}^3{w_3} + {\varepsilon _{22}}{w_4} + {\varepsilon _{21}}{\varepsilon _{22}}{w_5},
\end{equation}
where $w_i (i=\{1,...,5\})$ are unknown variables to be determined. Replacing the series solution of equation (\ref{Eq36}) in equation (\ref{Eq34}), we can get the third-order solution of $w$ as follows:
\begin{equation}\label{Eq37}
w =  \pm \sqrt {\frac{3}{5}} \left( {1 + \frac{{729}}{{640}}\varepsilon _{21}^2 - \frac{9}{5}{\varepsilon _{22}}} \right) - \frac{{27}}{{40}}{\varepsilon _{21}}\left( {1 + \frac{{729}}{{320}}\varepsilon _{21}^2 - \frac{{61}}{{16}}{\varepsilon _{22}}} \right),
\end{equation}
where the upper sign is for the critical inclination in the prograde space and the lower sign for the one in the retrograde space. The perturbation solution given by equation (\ref{Eq37}) explicitly shows how the classical and extended Brown corrections influence the critical inclination of ZLK resonance.

\textbf{Approach II: The ${\cal F}_{20} + {\cal F}_{21}$ model as the starting point.}

Under the ${\cal F}_{20} + {\cal F}_{21}$ model, the equation of critical inclination can reduce to
\begin{equation}\label{Eq38}
8\left(3- 5{w_0^2}\right) - 9{\varepsilon _{21}} w_0 \left(3  + 5{w_0^2}\right) = 0,
\end{equation}
where the perturbation solution can be formulated up to order 6 in $\varepsilon_{21}$ as follows:
\begin{equation}\label{Eq39}
\begin{aligned}
{w_0} =&  \pm \sqrt {\frac{3}{5}} \left(1 + \frac{{729}}{{640}} \varepsilon _{21}^2 + \frac{{649539}}{{163840}} \varepsilon _{21}^4 + \frac{{10144677249}}{{524288000}} \varepsilon _{21}^6\right)\\
&- \frac{{27}}{{40}} {\varepsilon _{21}} \left(1  + \frac{{729}}{{320}}\varepsilon _{21}^2
  + \frac{{1003833}}{{102400}}\varepsilon _{21}^4\right),
\end{aligned}
\end{equation}
where the upper sign is for the critical inclination in the prograde space and the lower sign is for the one in the retrograde space. Then, taking the solution of the ${\cal F}_{20} + {\cal F}_{21}$ model as the starting point, we can write the perturbation solution of $w$ (up to order 3 in $\varepsilon_{22}$) as follows:
\begin{equation}\label{Eq40}
w = {w_0} + {\varepsilon _{22}}{w_1} + \varepsilon _{22}^2{w_2} + \varepsilon _{22}^3{w_3},
\end{equation}
where $w_i(i=1,2,3)$ are unknown coefficients to be determined. Replacing the perturbation solution of equation (\ref{Eq40}) in equation (\ref{Eq34}), we can obtain
\begin{equation}\label{Eq41}
{w_1} =  - \frac{1}{{4{h_0}}}\left( {201 + 160w_0^2 + 135w_0^4} \right),
\end{equation}
\begin{equation}\label{Eq42}
{w_2} =  - \frac{5}{{{h_0}}}{w_1}\left( {16{w_0} + 27w_0^3 + 8w_1 + 27{\varepsilon _{21}}{w_0}w_1} \right),
\end{equation}
and
\begin{equation}\label{Eq42-1}
\begin{aligned}
{w_3} =&  - \frac{5}{{2{h_0}}}\left( {16w_1^2 + 81w_0^2w_1^2 + 32{w_0}{w_2} + 54w_0^3{w_2}} \right.\\
&\left. { + 32{w_1}{w_2} + 18{\varepsilon _{21}}w_1^3 + 108{\varepsilon _{21}}{w_0}{w_1}{w_2}} \right),
\end{aligned}
\end{equation}
where ${h_0} = 80{w_0} + 27{\varepsilon _{21}}\left(1 + 5w_0^2\right)$.

In summary, the critical Kozai inclination can be determined by
\begin{equation}\label{Eq43}
i_{\rm critical} = \arccos{w},
\end{equation}
where $w$ is provided by equations (\ref{Eq36}) and (\ref{Eq40}), corresponding to approaches I and II, respectively.

Figure \ref{Fig4} shows the critical inclination determined by numerical and analytical methods, as a function of $\alpha_{\rm H}$ in the prograde and retrograde spaces. Here $\alpha_{\rm H}$ stands for the semimajor axis ratio to the Hill radius of the central object, defined by 
\begin{equation}\label{Eq44}
\alpha_{\rm H} = \frac{a}{r_{\rm H}},
\end{equation}
where $r_{\rm H}$ is defined in the limit of $m_0 \ll m_{\rm p}$ by \citep{tremaine2023dynamics}
\begin{equation}\label{Eq44-1}
\quad r_{\rm H} = a_{\rm p} \left(\frac{m_0}{3 m_{\rm p}}\right)^{1/3}.
\end{equation}

In practical simulations, the upper limit of $\alpha_{\rm H}$ is fixed at $0.5$ in order to ensure the stability of dynamical system \citep{hamilton1991orbital}\footnote{It should be noted that Hill stability separation can be reduced to as low as 0.4 due to the ZLK effects in the high-inclination regime \citep{grishin2017generalized,grishin2024irregularI}.}. From Figure \ref{Fig4}, we can see that (a) analytical results can agree well with the numerical curves and (b) in both the prograde and retrograde spaces the critical inclination is an increasing function of the semimajor axis in units of $r_{\rm H}$. It is noticed that the increasing relations between $i_{\rm critical}$ and $\alpha_{\rm H}$ shown in Figure \ref{Fig4} are consistent with the results of \citet{beauge2006high} and \citet{lei2019semi}.

\begin{figure*}
\centering
\includegraphics[width=\columnwidth]{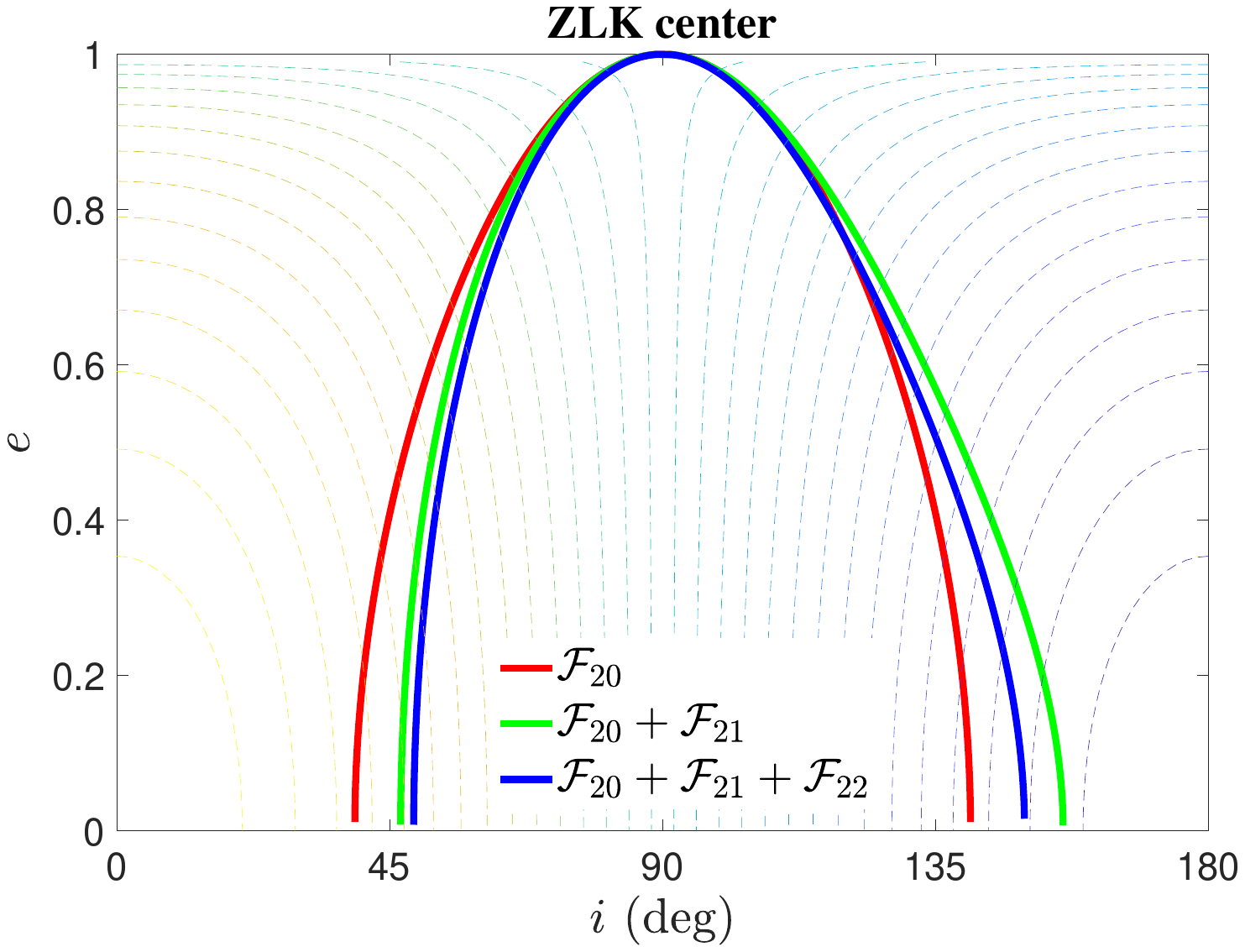}
\includegraphics[width=\columnwidth]{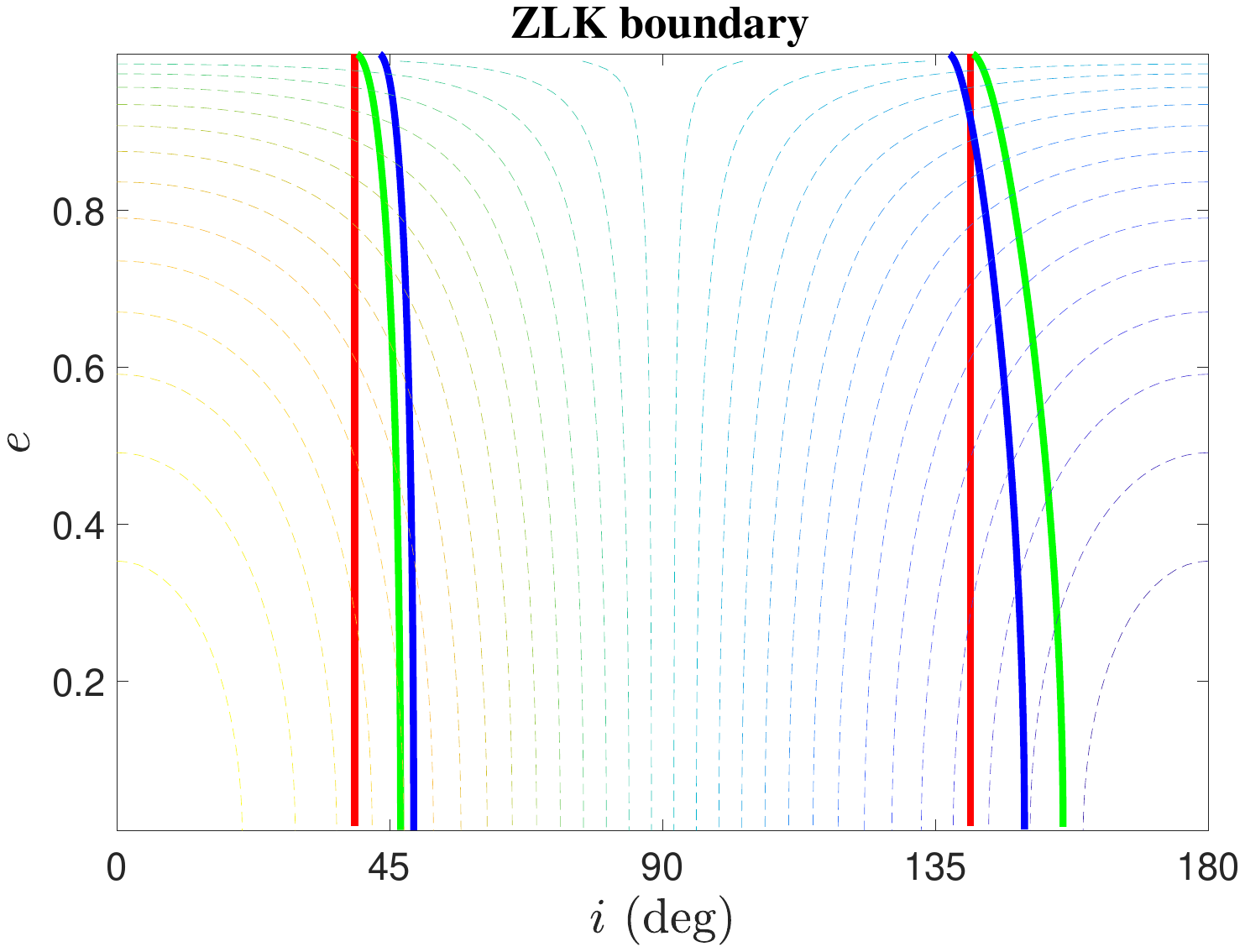}\\
\includegraphics[width=\columnwidth]{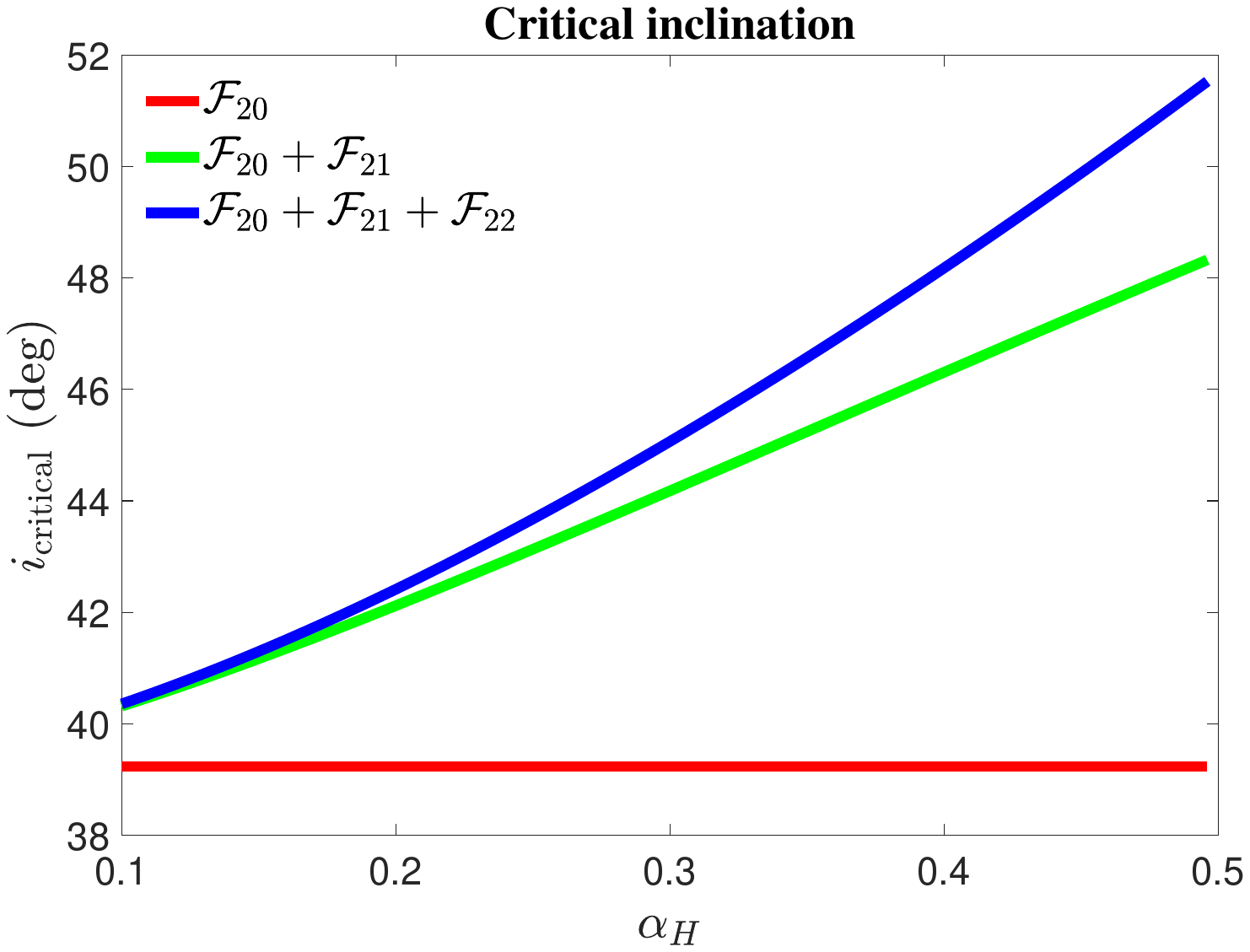}
\includegraphics[width=\columnwidth]{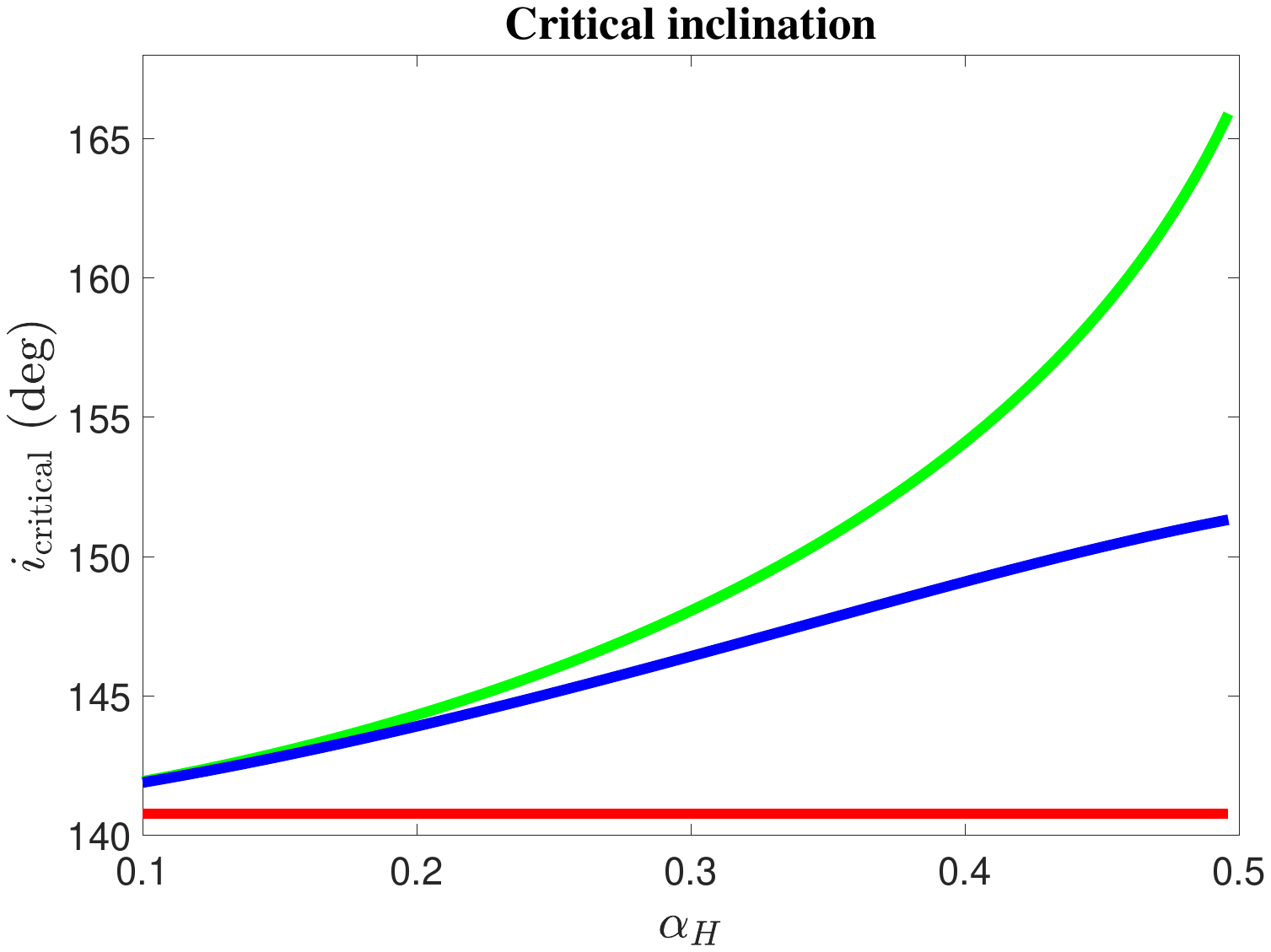}
\caption{Distribution of ZLK center (\textit{top-left panel}) and ZLK boundary (\textit{top-right panel}) in the case of $a = 0.15\, {\rm au}$, and critical inclination as a function of $\alpha_{\rm H}$ in the prograde configurations (\textit{bottom-left panel}) and in the retrograde configurations (\textit{bottom-right panel}) under different Hamiltonian models, including ${\cal F}_{20}$, ${\cal F}_{20} + {\cal F}_{21}$ and ${\cal F}_{20} + {\cal F}_{21} + {\cal F}_{22}$. To produce the ZLK characteristics, the perturbation solutions of approach II are utilized. In the top two panels, level curves of the motion integral $H$ are presented as background. %\eg{make the color consistent in all panels. The bottom left panel has different colors for different expansions.}
}
\label{Fig5}
\end{figure*}

\begin{figure*}
\centering
\includegraphics[width=\columnwidth]{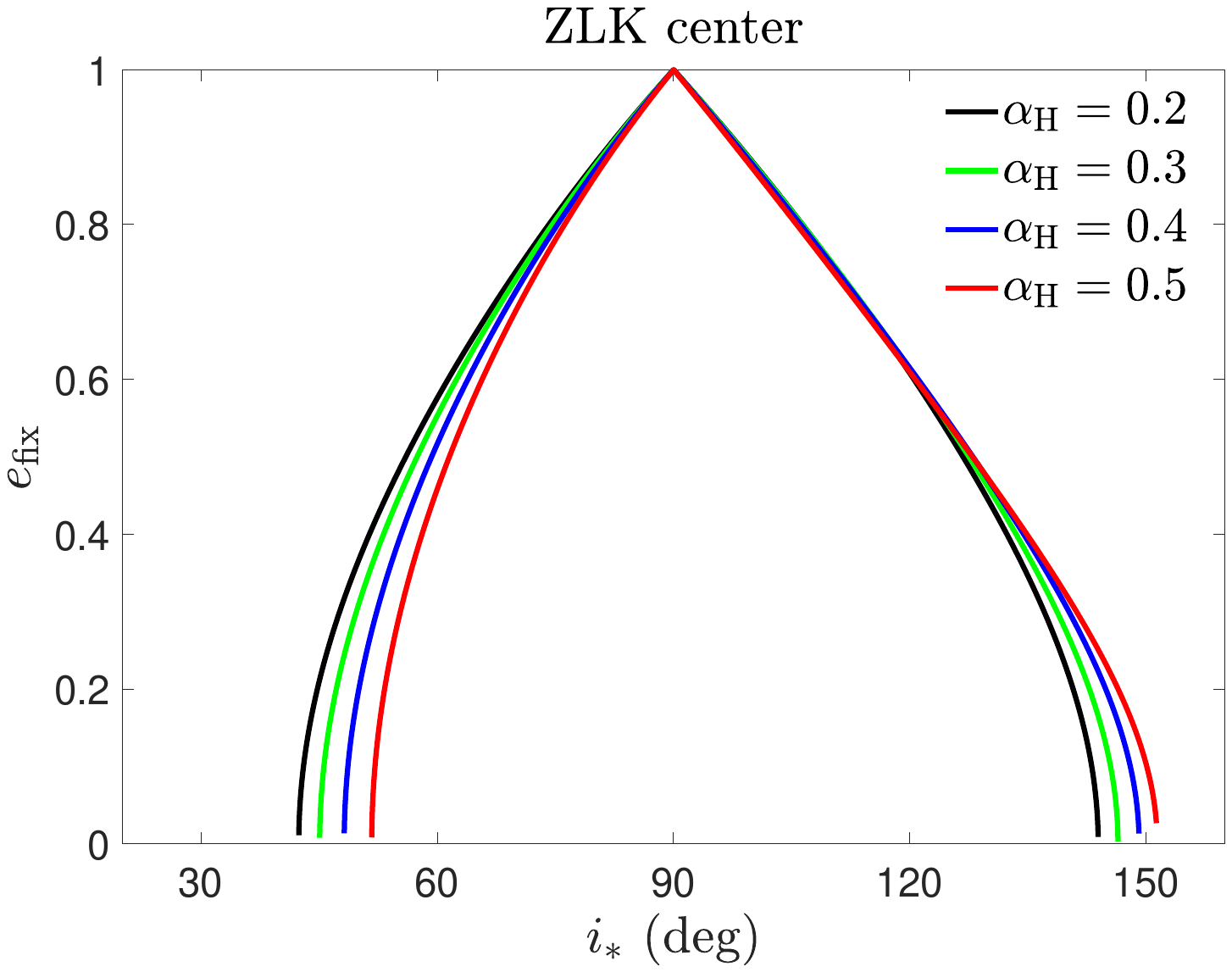}
\includegraphics[width=\columnwidth]{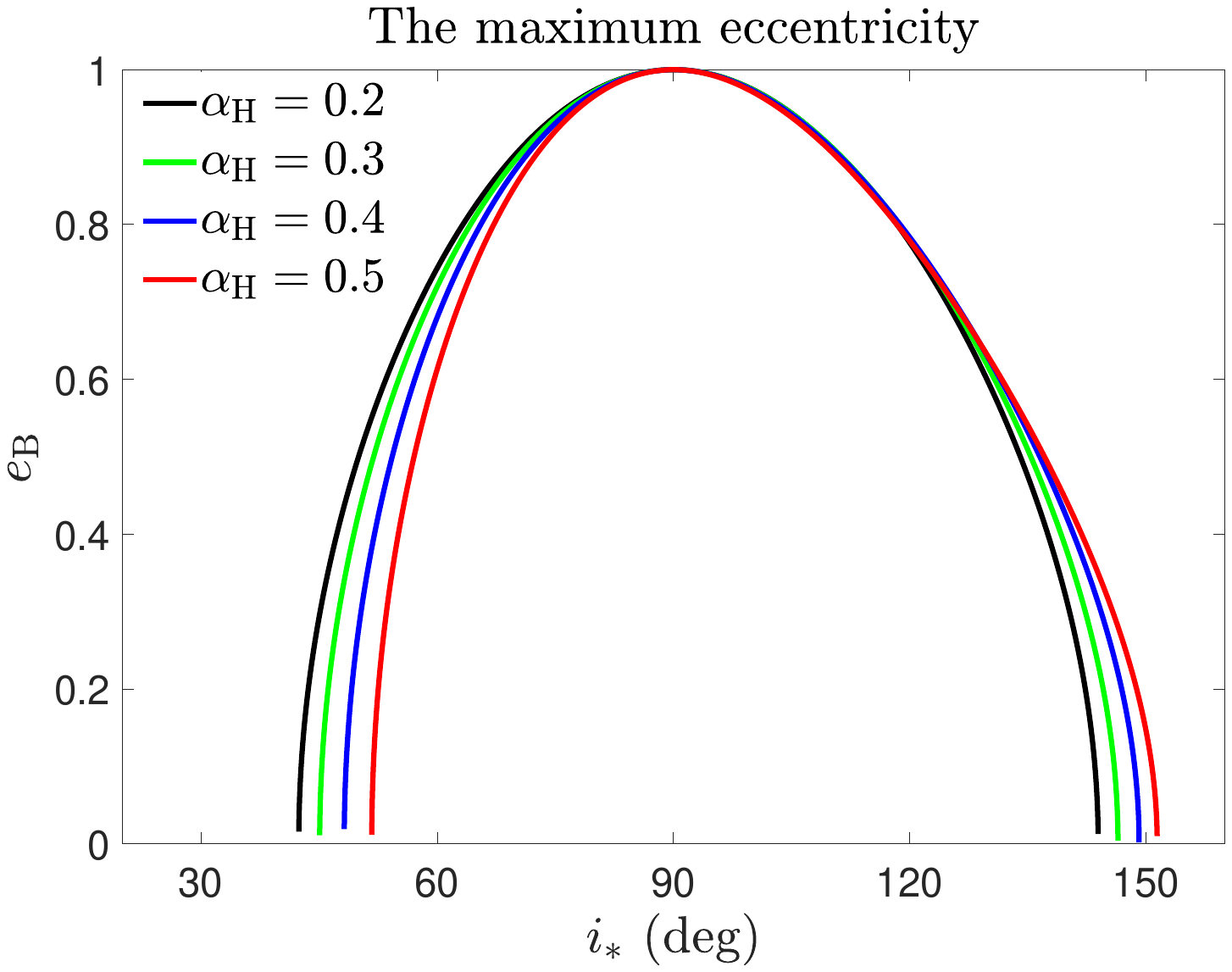}
\caption{The eccentricity of ZLK center (\textit{left panel}) and the maximum eccentricity excited by ZLK effects (\textit{right panel}) are shown as functions of the Kozai inclination $i_*$ for different levels of semimajor axis ratio to the Hill radius $\alpha_{\rm H}$. The curves are produced by the perturbation solution derived from approach II.}
\label{Fig5-1}
\end{figure*}

\subsection{Comparison of ZLK characteristics}
\label{Sect3-4}

Here, the perturbation solutions of ZLK characteristics derived from approach II are adopted because of its high precision. Figure \ref{Fig5} shows a comparison for the analytical results of ZLK characteristics, including the ZLK center, the ZLK boundary, and the critical inclination, under Hamiltonian models with different levels of correction. Under the extended Brown Hamiltonian, Figure \ref{Fig5-1} shows a comparison of analytical results of ZLK center and the maximum eccentricity excited by ZLK effects as functions of the Kozai inclination $i_*$ for different levels of $\alpha_{\rm H}$. Notice that the hierarchy level of triple systems reduces with an increase of $\alpha_{\rm H}$, because the single-averaging parameter $\epsilon_{\rm SA}$ increases with $\alpha_{\rm H}$.

It is observed from Figure \ref{Fig5} that (a) the distributions of ZLK center and ZLK boundary under the ${\cal F}_{20}$ model are symmetric with respect to the line of $i=90^{\circ}$, but they are no longer symmetric under the remaining two models, showing that Brown Hamiltonian corrections are responsible for breaking the symmetry of ZLK characteristics, (b) the inclination of ZLK boundary in the ${\cal F}_{20}$ model is independent on the eccentricity, while it is weakly dependent on the eccentricity under the remaining two models, (c) the critical inclinations under the ${\cal F}_{20}$ model are independent on the separation characterized by $\alpha_{\rm H}$, and it is equal to $39.2^{\circ}$ in the prograde space and equal to $140.8^{\circ}$ in the retrograde space, but they are increasing functions of $\alpha_{\rm H}$ under the remaining two models, and (d) the ${\cal F}_{20} + {\cal F}_{21}  + {\cal F}_{22}$ model holds a narrower space of inclination for triggering ZLK resonance, compared to the ${\cal F}_{20} + {\cal F}_{21}$ model. It is mentioned that the asymmetry was noticed by \citet{grishin2017generalized} in the context of the orbital distributions of irregular satellites.

From Figure \ref{Fig5-1}, we can see that (a) all curves corresponding to different values of $\alpha_{\rm H}$ coincide at the point of $i_*=90^{\circ}$, (b) the asymmetry of ZLK characteristics, including distributions of ZLK center and the maximum eccentricity, increases with the ratio $\alpha_{\rm H}$, (c) the ZLK center moves to a higher-inclination space when the ratio $\alpha_{\rm H}$ increases, and (d) the maximum eccentricity excited by ZLK effects is suppressed in the prograde space, while it may be enhanced in the retrograde space with an increase of $\alpha_{\rm H}$.

\begin{figure*}
\centering
\includegraphics[width=\columnwidth]{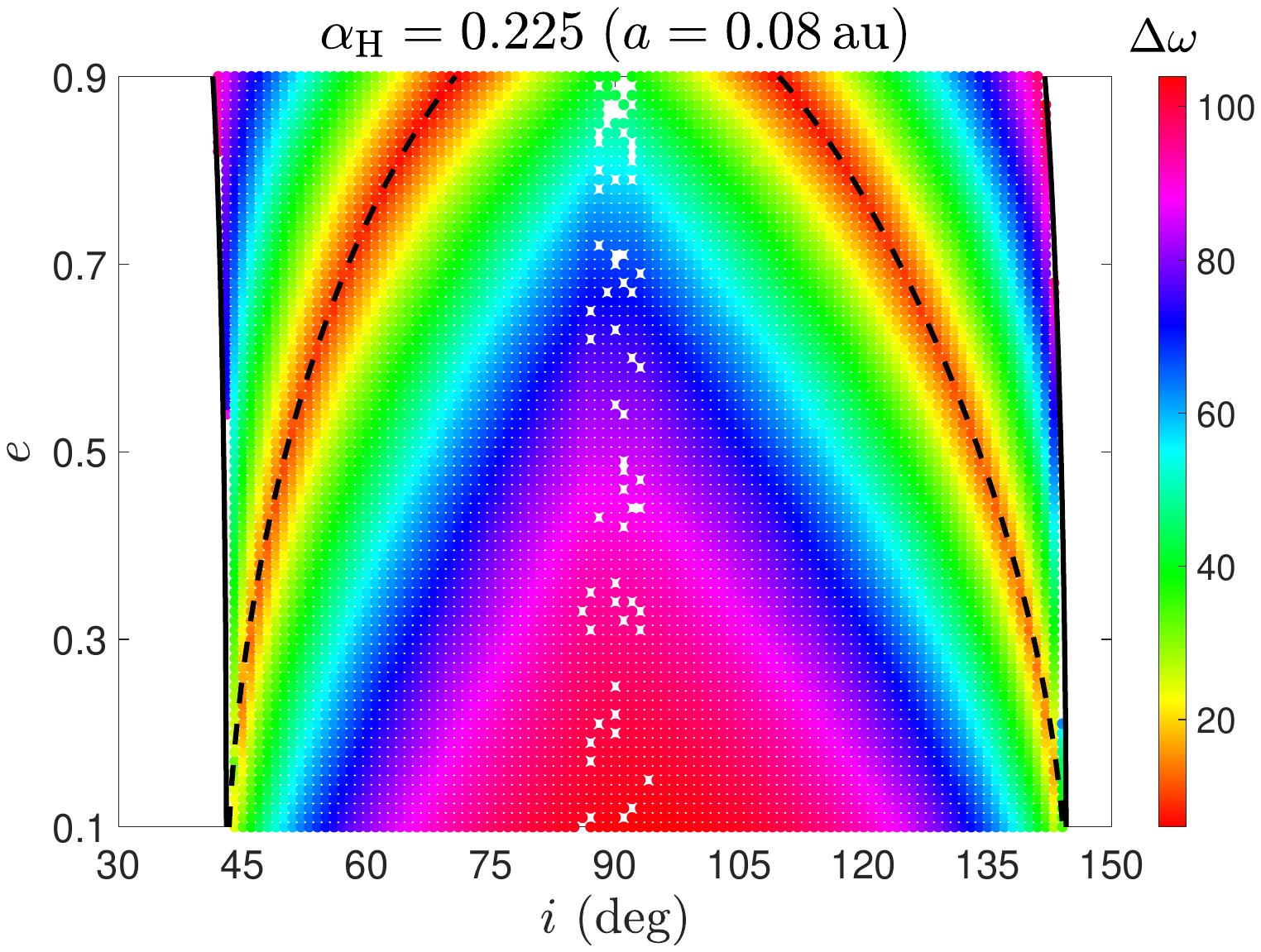}
\includegraphics[width=\columnwidth]{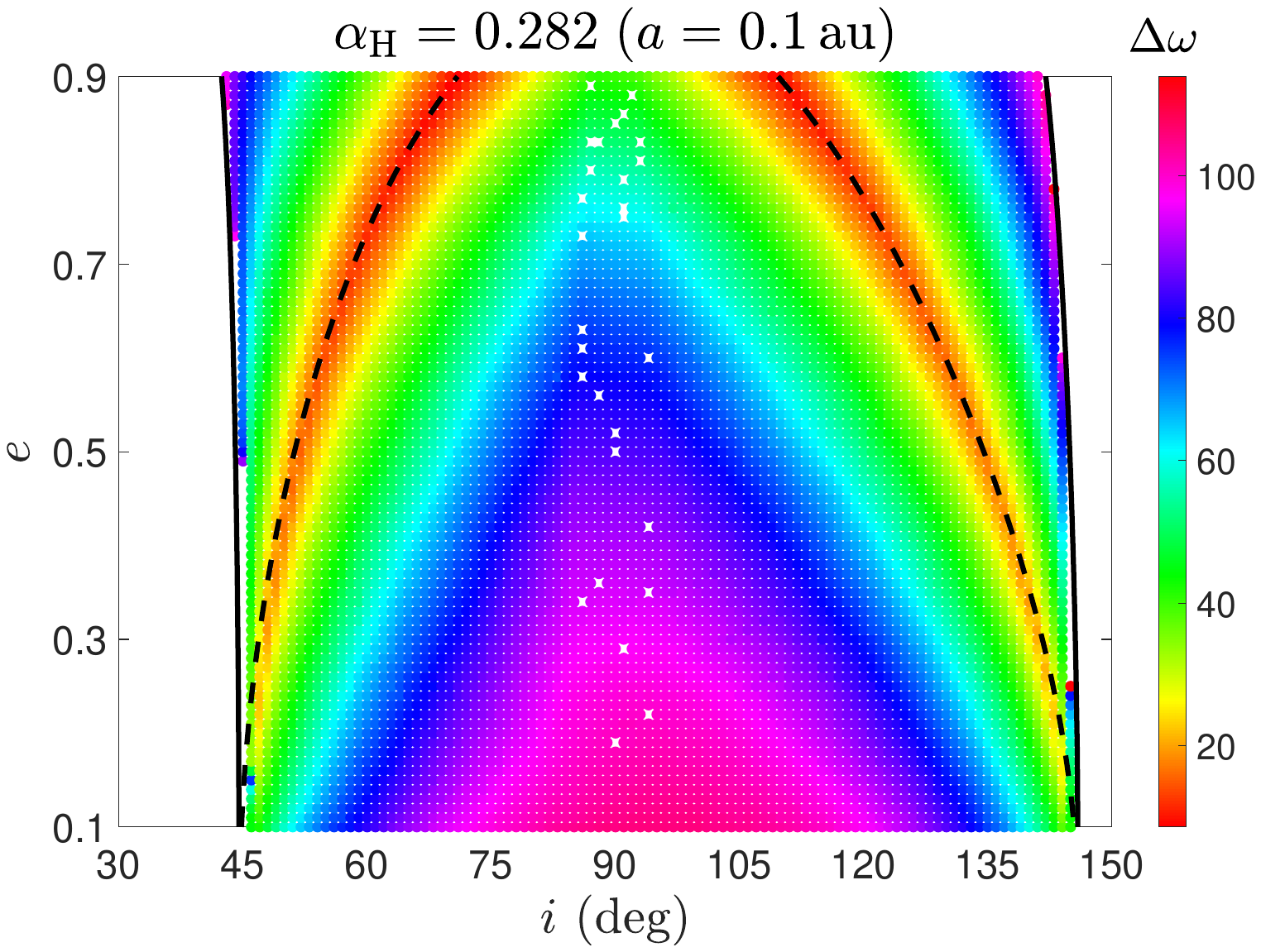}\\
\includegraphics[width=\columnwidth]{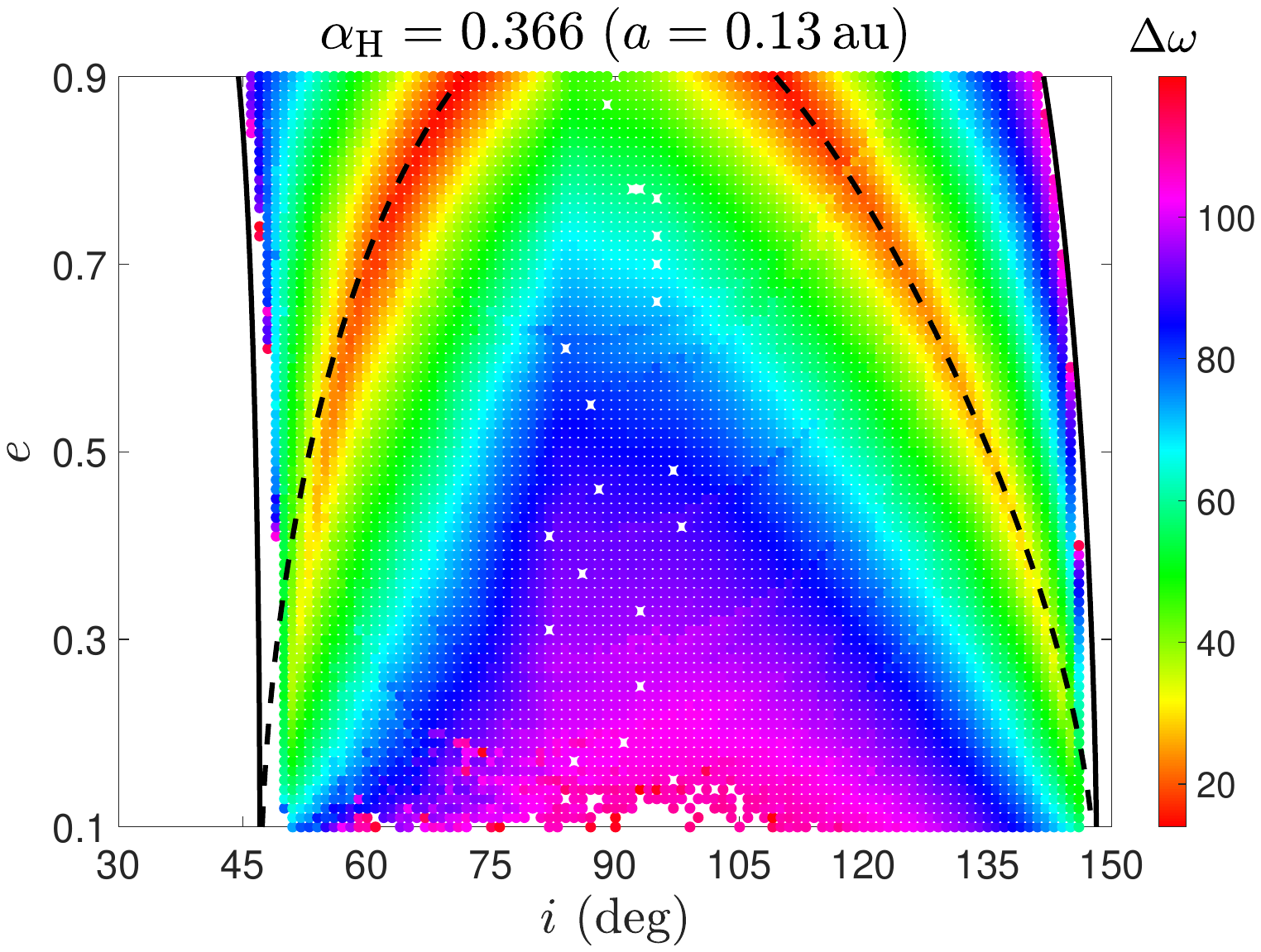}
\includegraphics[width=\columnwidth]{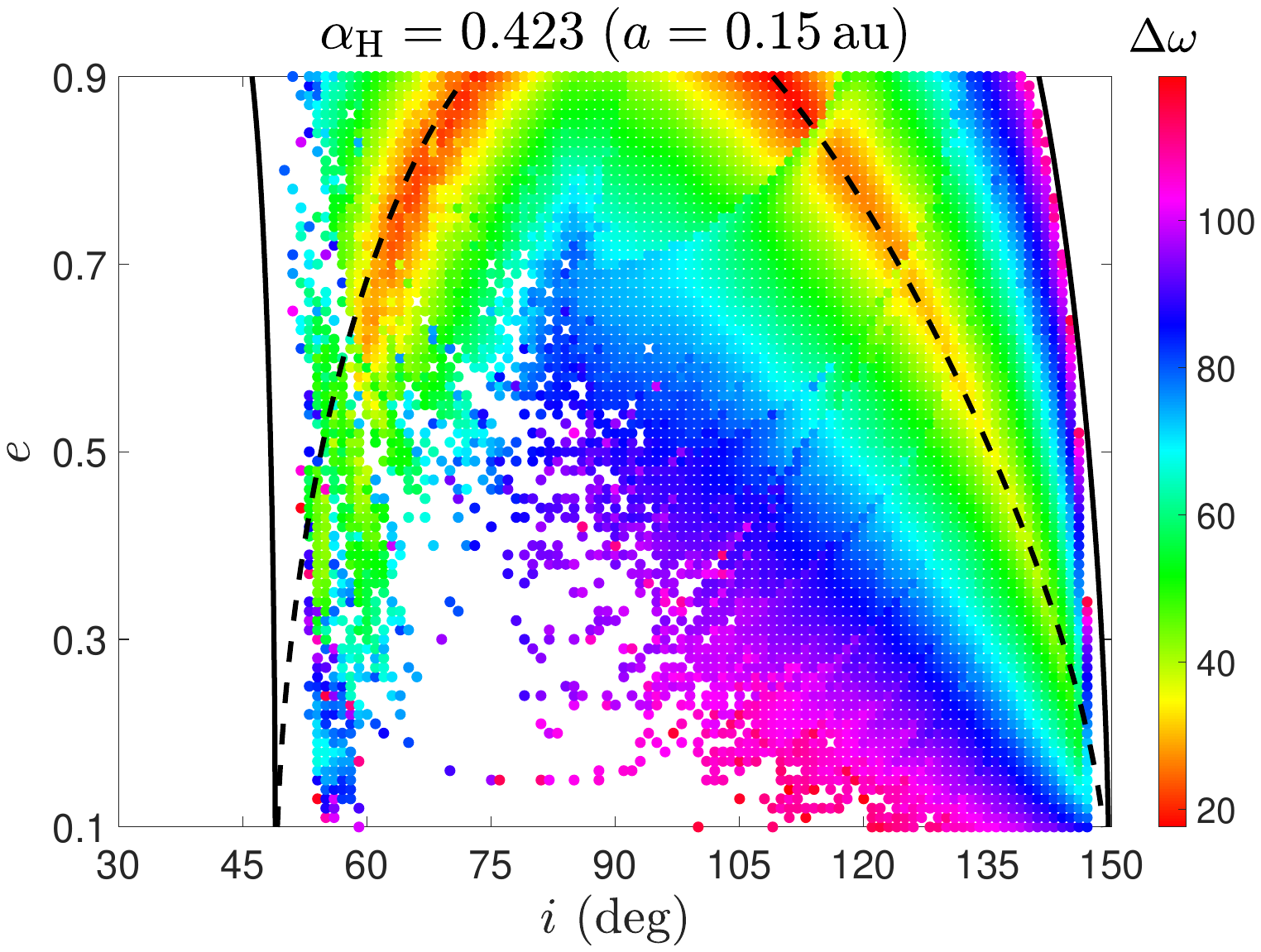}
\caption{Distribution of ZLK librating orbits in the $(i,e)$ space of initial conditions produced by performing $N$-body integration during the period of 300 $t_{\rm ZLK}$, together with analytical distribution of ZLK center (black dashed lines) and ZLK boundary (black solid lines) derived from the extended Brown Hamiltonian model, at different levels of $\alpha_{\rm H}$. The color represents the variation amplitude of the argument of ZLK resonance denoted by $\Delta \omega$ in units of degree. Orbits in the white space are either unstable or circulating.}
\label{Fig6}
\end{figure*}

\begin{figure*}
\centering
\includegraphics[width=\columnwidth]{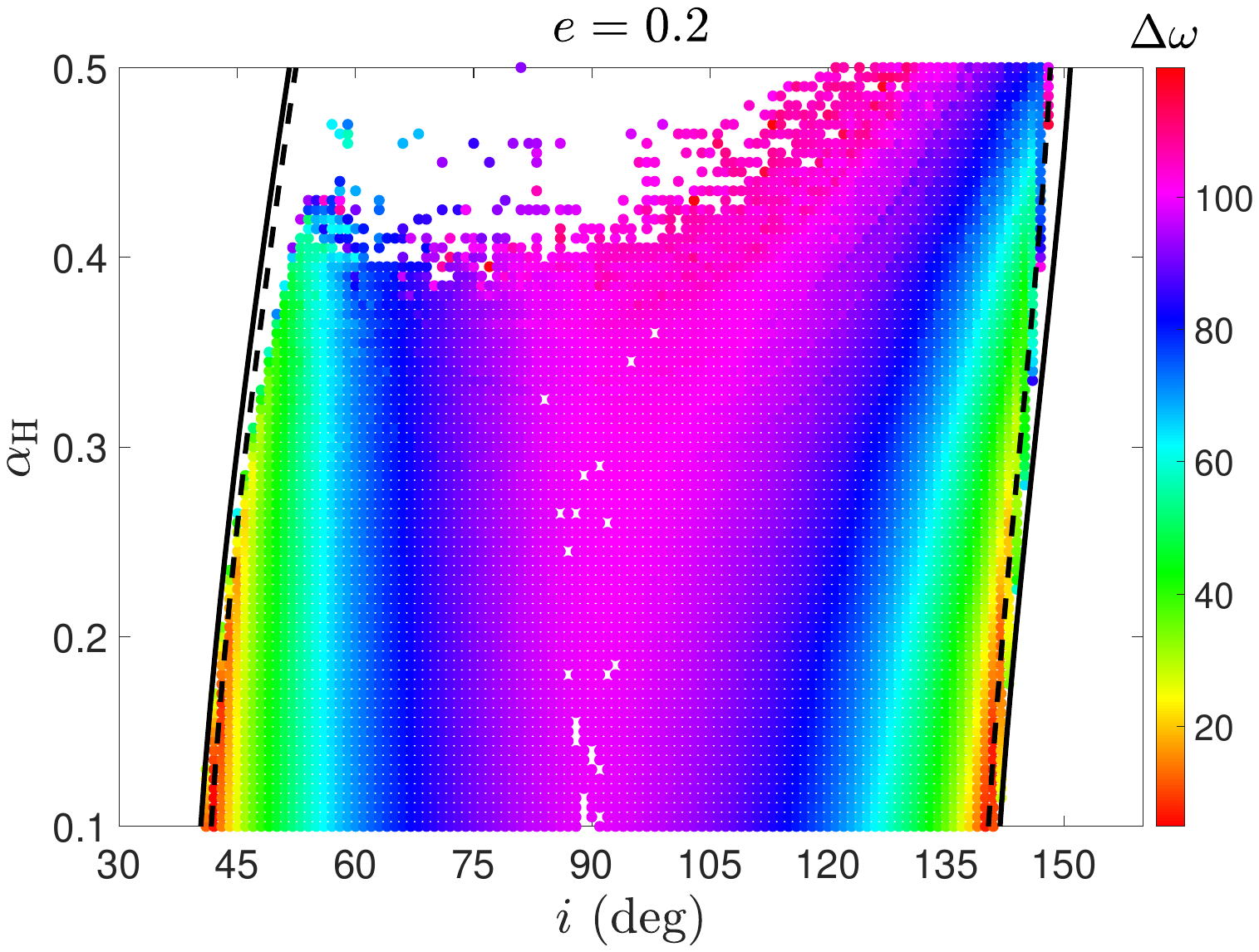}
\includegraphics[width=\columnwidth]{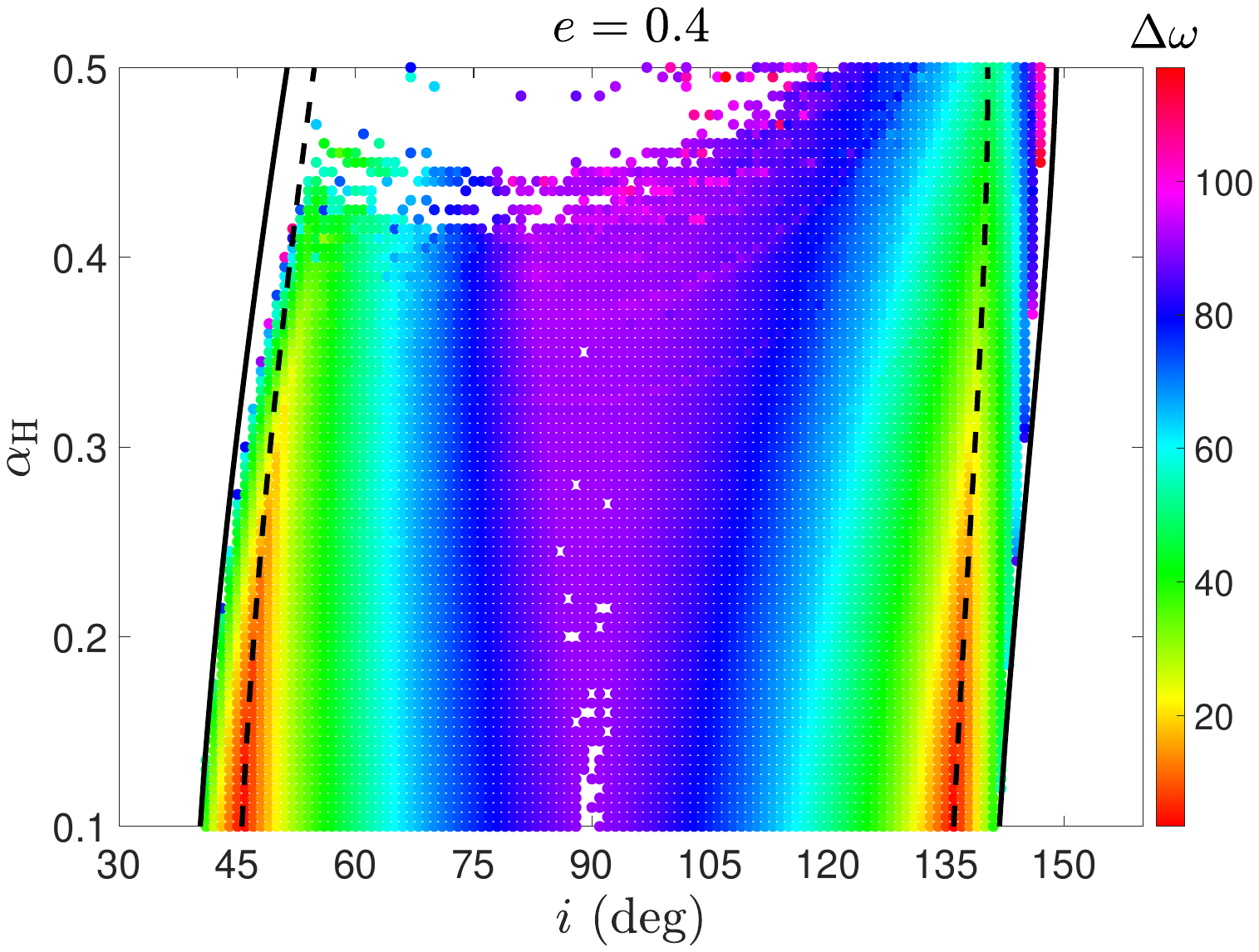}\\
\includegraphics[width=\columnwidth]{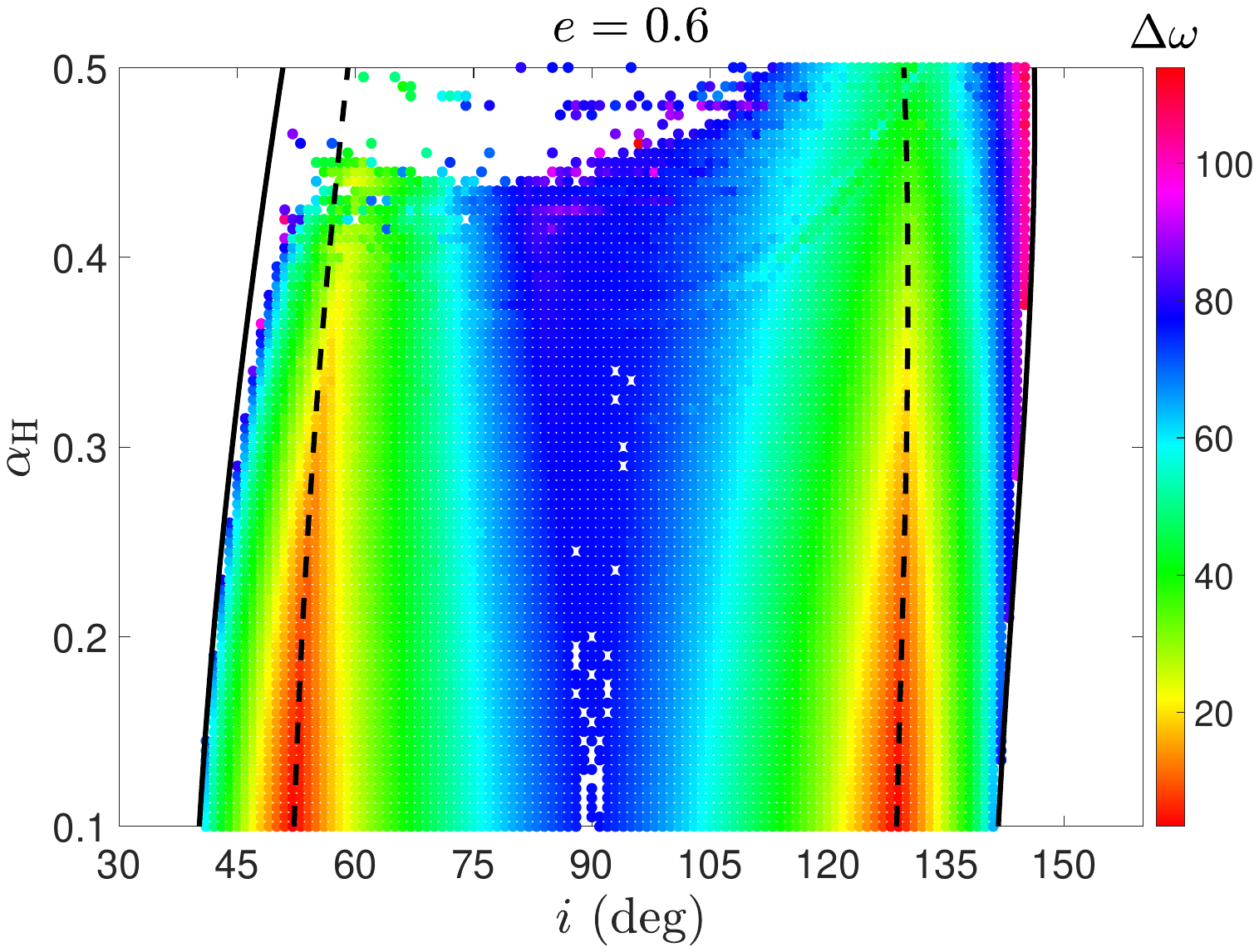}
\includegraphics[width=\columnwidth]{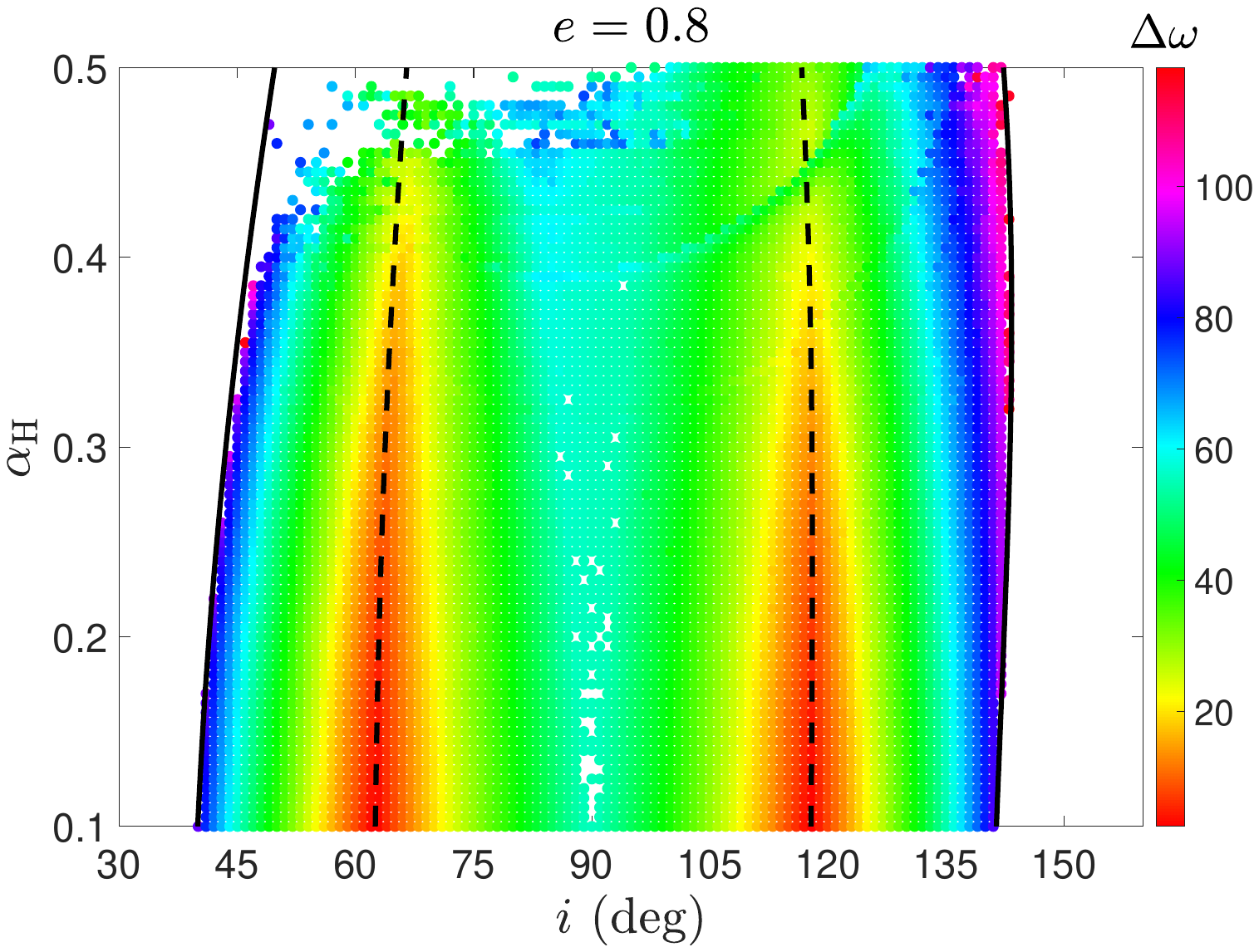}
\caption{Same as Figure \ref{Fig6} but for ZLK librating orbits shown in the $(i,\alpha_{\rm H})$ space with different initial eccentricities at $e=0.2$, $e=0.4$, $e=0.6$ and $e=0.8$.}
\label{Fig7}
\end{figure*}

\begin{figure}
\centering
\includegraphics[width=\columnwidth]{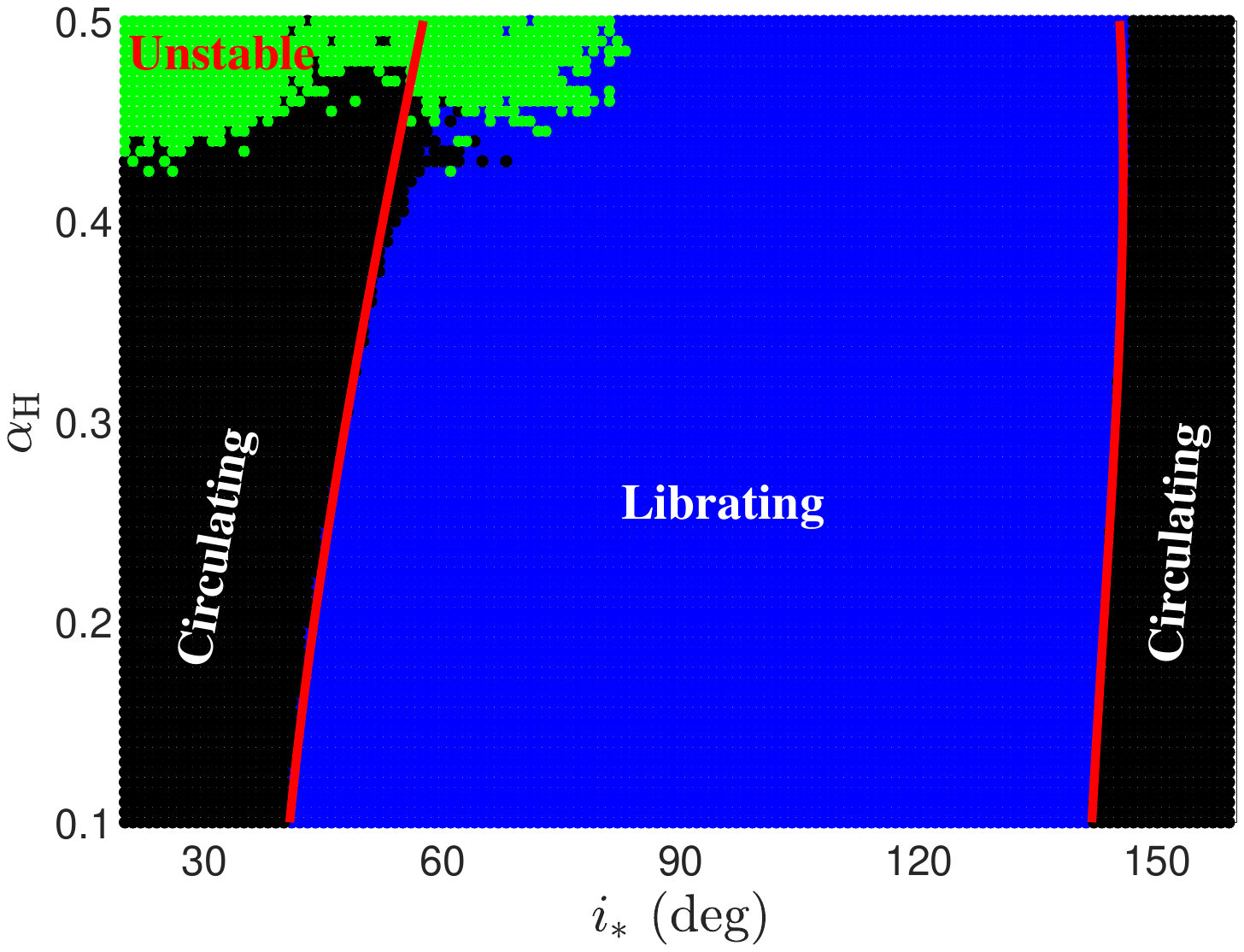}
\caption{Phase map of $N$-body integration in the $(i_*,\alpha_{\rm H})$ space. The ZLK librating and circulating regions are marked in blue and black, respectively. Unstable orbits are marked in green dots. Analytical boundaries corresponding to $e_{\rm fix} = e_{\rm c}$ are shown in red lines.}
\label{Fig8}
\end{figure}

\section{Numerical simulations}
\label{Sect4}

In the previous section, we have formulated perturbation solutions to describe the ZLK characteristics under the extended Brown Hamiltonian model. ZLK characteristics include the location of ZLK center (the eccentricity $e_{\rm fix}$ and the inclination $i_{\rm fix}$), the ZLK boundary (the maximum eccentricity $e_{\rm B}$ and the associated inclination $i_{\rm B}$), and the critical Kozai inclination ($i_{\rm critical}$). In this section, the approximate solutions derived from approach II are utilized to understand numerical structures associated with ZLK effects.

Initially, we consider dynamical maps associated with ZLK resonance in the spaces spanned by $\alpha_{\rm H}$, $e$, and $i$. In particular, the initial argument of pericenter is assumed at $\omega = 90^{\circ}$ corresponding to the argument of the ZLK center, and the remaining angles including $\Omega$ and $M$ are artificially taken as zero. It is mentioned that the initial condition $(\alpha_{\rm H},e,i,\Omega,\omega,M)$ is given in the form of ``averaged'' elements, which are suitable for our novel extended Hamiltonian model. 
To generate the associated initial condition for $N$-body simulations, we need to perform a mean-to-osculating transformation proposed in \citet{lei2025Extensions}. With a given osculating initial condition, the full $N$-body model is numerically propagated\footnote{In practical simulations, we adopted the classic eighth-order Runge–Kutta numerical integrator with a seventh-order method for automatic step-size control \citep{fehlberg1969classical}.} over a period of 300 $t_{\rm ZLK}$. A longer integration time is tested, and there is no qualitative change about the result.

Following the method of \citet{grishin2024irregularI}, orbits are generally classified into three types: (a) the orbit is unstable if $d > 3a$, where $d$ is the instantaneous distance from the central object and $a$ is the initial semimajor axis, (b) the orbit is circulating if $\sin{\omega}<-0.2$, corresponding to $\omega < -11.5^{\circ}$ or $\omega > 191.5^{\circ}$\footnote{This criterion strikes a balance between ensuring that librating orbits are not misclassified and minimizing the runtime for systems that undergo full circulation, as stated by \citet{grishin2024irregularI}.}, and (c) the remaining ones belong to librating orbits. It should be mentioned that the criterion $d > 3a$ for classifying unstable orbits is not mathematically rigorous. In this regard, some different criteria can be found in the literature \citep{grishin2017generalized,grishin2024irregularI}. Here we provide a simple explanation about their equivalence. During the long-term evolution, it is known that the semimajor axis has no significant change, so it holds $d < 2 a$ for almost all bounded orbits ($e<1$). Thus, the condition $d > 3a$\footnote{Considering the slight variation of semimajor axis during the long-term evolution, here we take $d>3a$ rather than $d>2a$. The adopted instability criterion implies that the inner binary orbit becomes unbound if its semimajor axis increases by less than 50\% of its initial value. This is a stricter condition than the instability criterion adopted by \citet{vynateya2022}.} adopted here means that the inner binary eccentricity exceeds unity (becoming unbound orbits). In this sense, the criterion of $d > 3a$ is approximately equivalent to the criterion of $e>1$, which has been adopted in \citet{grishin2017generalized}. Additionally, it has been numerically demonstrated that both the criteria of $e>1$ and $d > 3r_{\rm H}$ produce the same dynamical maps \citep{grishin2017generalized}. Thus, we can conclude that these three criteria $d>3a$, $e>1$ and $d > 3r_{\rm H}$ are approximately equivalent.

Figure \ref{Fig6} shows the distribution of librating orbits in the $(i,e)$ space with different levels of $\alpha_{\rm H}$, and Figure \ref{Fig7} presents the distribution of librating orbits in the $(i,\alpha_{\rm H})$ space with different values of initial eccentricity $e$. The index in the color bar stands for the variation amplitude of the argument of ZLK resonance, denoted by $\Delta \omega$. For comparison, the analytical distributions of ZLK center and ZLK boundary of the extended Brown Hamiltonian model are marked by black dashed and solid lines, respectively.

It is observed that (a) more and more orbits become unstable with an increase of $\alpha_{\rm H}$, (b) ZLK centers are distributed in the region closer to $i=90^{\circ}$ with an increase of eccentricity, (c) analytical curves of ZLK boundary provide good boundaries for the distribution of ZLK librating orbits, especially for the cases of low $\alpha_{\rm H}$, (d) analytical curves of the ZLK center agree well with the distribution of ZLK librating orbits with small amplitude of $\Delta \omega$, corresponding to the numerical distribution of ZLK center. Both Figures \ref{Fig6} and \ref{Fig7} show that, for those high-$\alpha_{\rm H}$ cases, some discrepancy between the analytical and numerical boundaries can be observed in the prograde regime. This is because chaotic orbits occupy the phase space in the vicinity of dynamical boundaries. In general, the range of chaotic sea is proportional to $\alpha_{\rm H}$.

Secondly, we construct a phase map by performing $N$-body simulations, analogous to Figure 7 in \citet{grishin2024irregularI}. For each 
$(i_*,\alpha_{\rm H})$ pair, the initial conditions are set at the ZLK center with $e=e_{\rm fix}$, $i=i_{\rm fix}$ and $\omega=\pi/2$, where $e_{\rm fix}$ and $i_{\rm fix}$ are determined by equation (\ref{Eq21}). If ZLK resonance does not occur, the initial eccentricity is set to zero, and the resulting orbit is either circulating or unstable \citep{grishin2024irregularI}. The longitude of ascending node $\Omega$ and mean anomaly $M$ are initialized at zero. The numerical integration period was set as 300 $t_{\rm ZLK}$, which is consistent with that of Figures \ref{Fig6} and \ref{Fig7}. The resulting phase map, shown in Figure \ref{Fig8}, categorizes the orbits into three types: ZLK librating orbits (blue dots), ZLK circulating orbits (black dots), and unstable orbits (green dots). In contrast to \citet{grishin2024irregularI}, we employ a mean-to-osculating transformation to provide the osculating initial conditions for $N$-body simulations, based on the method proposed in \citet{lei2025Extensions}. Thus, the resulting phase map is slightly different from Figure 7 of \citet{grishin2024irregularI}. 

According to Figure \ref{Fig8}, it is possible for us to understand the dynamical structures in the parameter space of $(i_*,\alpha_{\rm H})$ for those satellites of giant planets, which are located deeply inside ZLK resonance. We can see that, with an increase of $\alpha_{\rm H}$, the area of librating region in the prograde space reduces, while the area of libration region in the retrograde space increases. Additionally, in the prograde space, unstable orbits begin to appear when $\alpha_{\rm H}$ is higher than $\sim$$0.4$, which is in agreement with the Hill stability discussed in \citet{grishin2017generalized}. An intriguing feature is that there exist relatively smooth boundaries between the librating and circulating regions. It is known from \citet{grishin2024irregularI} that the boundaries can not be reproduced and understood from the classical Brown Hamiltonian model. Here we attempt to understand the structures based on our novel extended Hamiltonian model.

It is known that the zero-eccentricity point corresponds to the location of saddle points of long-term model. Thus, under the full $N$-body model, it is filled with a chaotic layer in the phase space near the zero-eccentricity point. Let us denote the thickness of the chaotic layer by $e_{\rm c}$. In general, numerical orbits initialized with $e_{\rm fix}<e_{\rm c}$ are inherently chaotic, whereas those with $e_{\rm fix}>e_{\rm c}$ may exhibit librating behavior. From Figure \ref{Fig6}, we can read the thickness of chaotic layer $e_{\rm c}$, which corresponds to the eccentricity of the intersection points between the numerical boundary of librating orbits and the lines of ZLK center. For example, in the case of $\alpha_{\rm H} = 0.366$ (see the bottom-left panel of Figure \ref{Fig6}), it is $e_{\rm c}=0.36$ in the prograde regime and it is $e_{\rm c} = 0.17$ in the retrograde regime. Consequently, within the framework of the extended Brown Hamiltonian model, we can delineate the boundaries between librating and circulating orbits by setting $e_{\rm fix} = e_{\rm c}$. Generally, the thickness of chaotic layer $(e_{\rm c})$ is proportional to the ratio $\alpha_{\rm H}$, as shown by Figure \ref{Fig6}. For simplicity, we adopt a linear relation between $e_{\rm c}$ and $\alpha_{\rm H}$. In particular, we take $e_{\rm c} = 1.0 \alpha_{\rm H}$ in the prograde regime and $e_{\rm c} = 0.45 \alpha_{\rm H}$ in the retrograde regime\footnote{The investigation of the specific relationship between $e_{\rm c}$ and $\alpha_{\rm H}$ is beyond the scope of this study.}. As a result, the analytical boundaries in the space of $(i_*,\alpha_{\rm H})$ can be determined by solving
\begin{equation}
{e_{\rm fix}} = \sqrt {1 - x}  = \left\{ \begin{array}{l}
1.0{\alpha _{\rm H}},\quad {\rm prograde}\\
0.45{\alpha _{\rm H}},\quad {\rm retrograde}
\end{array} \right.
\end{equation}
where $x$ is provided by equation (\ref{Eq17}).

In Figure \ref{Fig8}, the analytical boundaries between the librating and circulating orbits are marked by red lines. We can see that the analytical curves derived from the extended Brown Hamiltonian model agree well with the numerical boundaries between librating and circulating orbits in the retrograde regime and in the prograde space with $\alpha_{\rm H} < 0.4$. However, there is certain discrepancy between analytical and numerical boundaries in the high-$\alpha_{\rm H}$ prograde regime. This may be attributed to several possible reasons. Firstly, the linear relation between the thickness of chaotic layer and $\alpha_{\rm H}$ we adopt is not a good approximation in the prograde high-$\alpha_{\rm H}$ space (in fact it is a nonlinear relation). Secondly, it is due to strong instability in this region, as discussed in \citet{grishin2017generalized}. Lastly, our novel extended Hamiltonian model may be inadequate in this region.

\section{Conclusions}
\label{Sect5}

In this study, the ZLK effects are systematically studied within the framework of the extended Brown Hamiltonian model. The investigation focuses on key dynamical features, including phase-space structures arising in phase portraits, the location of ZLK center, the boundaries of ZLK librating region, the maximum eccentricity excited by ZLK effects, and the critical inclination for triggering the ZLK resonance. 

Two perturbative approaches are developed by treating $\varepsilon_{21}$ and $\varepsilon_{22}$ as small parameters: (I) using the solution of the classical ZLK model as the zeroth-order approximation, and (II) employing the solution of the classical Brown model as the starting point. In practice, perturbation solutions of approaches I and II are formulated up to order 3 and up to order 6 in $\varepsilon_{21}$, respectively. These explicit formulations enable a detailed assessment of how the classical and extended Brown corrections affect the ZLK dynamics. To verify the accuracy, the analytical descriptions are compared with numerical results, demonstrating excellent consistency between them. As desired, the analytical solution derived from approach II has a higher (by three orders of magnitude) precision than the ones from approach I.

Both analytical and numerical results show that (a) at a given $H$ the phase-space topologies are changed from the classical ZLK model, to the classical and extended Brown Hamiltonian models, (b) ZLK features are symmetric with respect to the line of $i=90^{\circ}$ under the ${\cal F}_{20}$ (the classical ZLK) model, while there is no symmetry within the framework of the Hamiltonian models with Brown corrections, (c) with inclusion of Brown corrections the critical Kozai inclination for triggering the ZLK resonance increases with the separation characterized by $\alpha_{\rm H}$, which is consistent with \citet{beauge2006high} and \citet{lei2019semi}, and (d) with the inclusion of the extended Brown correction the eccentricity excitation is suppressed in the prograde regime and it may be enhanced in the retrograde regime. Point (b) indicates that the Brown Hamiltonian corrections are responsible for breaking the symmetry of ZLK features. It shows that the level of asymmetry increases with the separation characterized by $\alpha_{\rm H}$.

Dynamical maps associated with ZLK resonance are numerically produced in the spaces of $(\alpha_{\rm H},e,i)$. The results show that those dynamical structures, including the boundaries of ZLK librating orbits and the distribution of ZLK center, can be well reproduced by the analytical solution formulated under the extended Brown Hamiltonian model. Similar to Figure 7 of \citet{grishin2024irregularI}, a phase map is made by means of $N$-body simulations for millions of orbits. It shows that there exist boundaries between librating and circulating regions in the $(i_*,\alpha_{\rm H})$ space, which cannot be reproduced from the classical Brown Hamiltonian model, as shown in \citet{grishin2024irregularI}. Under our novel extended model, it can be well captured by the analytical criterion of $e_{\rm fix} = e_{\rm c}$, where $e_{\rm c}$ is the thickness of the chaotic layer near the zero-eccentricity point.

In the forthcoming paper of this series, the extended Brown Hamiltonian developed in Paper I \citep{lei2025Extensions}, along with the analytical formulation of the ZLK characteristics presented in this work (Paper II), will be applied to the study of irregular satellites of giant planets, providing an analytical criterion for the onset of ZLK resonance.

Based on our extended model, future works can be conducted in the following topics: (a) relaxing the test-particle approximation to the general case \citep{mangipudi2022extreme}, (b) considering the combined influences of short-range forces, such as general relativity, tidal interactions and rotational bulges \citep{liu2015suppression,mangipudi2022extreme}, (c) designing science orbits of high-altitude lunar satellites under perturbations from the Earth \citep{carvalho2010some}, (d) revisiting the Hill stability criteria discussed in \citet{grishin2017generalized} within the framework of the extended Brown Hamiltonian model, (e) exploring stable regions of exomoons perturbed by a distant star \citep{quarles2021exomoons}, (f) analyzing the dynamics of exoplanets in stellar or black-hole binary systems \citep{roell2012extrasolar,li2015implications}, and (g) studying long-term evolution of stellar binaries under the perturbation of a distant supermassive black hole \citep{fang2019secular}.  

\section*{Acknowledgements}
Hanlun Lei would like to express gratitude to Prof. Scott Tremaine for helpful suggestions about the vectorial form of Brown Hamiltonian and to Dr. Hao Gao for his assistance in deriving the vectorial form of Brown corrections. In addition, we wish to thank the anonymous reviewer for his/her helpful comments. This work is financially supported by the National Natural Science Foundation of China (Nos. 12233003 and 12073011) and the China Manned Space Program with grant no. CMS-CSST-2025-A16. Evgeni Grishin acknowledges support from the ARC Discovery Program DP240103174 (PI: Heger).

%%%%%%%%%%%%%%%%%%%%%%%%%%%%%%%%%%%%%%%%%%%%%%%%%%
\section*{Data Availability}
The codes used in this article could be shared on reasonable request.

% The inclusion of a Data Availability Statement is a requirement for articles published in MNRAS. Data Availability Statements provide a standardised format for readers to understand the availability of data underlying the research results described in the article. The statement may refer to original data generated in the course of the study or to third-party data analysed in the article. The statement should describe and provide means of access, where possible, by linking to the data or providing the required accession numbers for the relevant databases or DOIs.

%%%%%%%%%%%%%%%%%%%% REFERENCES %%%%%%%%%%%%%%%%%%

% The best way to enter references is to use BibTeX:

\bibliographystyle{mnras}
\bibliography{mybib} % if your bibtex file is called example.bib

% Alternatively you could enter them by hand, like this:
% This method is tedious and prone to error if you have lots of references
%\begin{thebibliography}{99}
%\bibitem[\protect\citeauthoryear{Author}{2012}]{Author2012}
%Author A.~N., 2013, Journal of Improbable Astronomy, 1, 1
%\bibitem[\protect\citeauthoryear{Others}{2013}]{Others2013}
%Others S., 2012, Journal of Interesting Stuff, 17, 198
%\end{thebibliography}

%%%%%%%%%%%%%%%%%%%%%%%%%%%%%%%%%%%%%%%%%%%%%%%%%%

%%%%%%%%%%%%%%%%% APPENDICES %%%%%%%%%%%%%%%%%%%%%

% \appendix

% \section{Some extra material}

% If you want to present additional material which would interrupt the flow of the main paper,
% it can be placed in an Appendix which appears after the list of references.

%%%%%%%%%%%%%%%%%%%%%%%%%%%%%%%%%%%%%%%%%%%%%%%%%%

% Don't change these lines
\bsp	% typesetting comment
\label{lastpage}
\end{document}